\documentclass[aps,pra,twocolumn,footinbib,groupedaddress]{revtex4-1} 
	
\usepackage{amsmath}
\usepackage{amssymb}
\usepackage{graphicx} 
\usepackage[breaklinks=true,colorlinks,citecolor=blue,linkcolor=blue,urlcolor=blue]{hyperref}
\usepackage{physics}
\usepackage{bbold}
\usepackage{soul}
\usepackage{enumerate}
\usepackage{color}
\usepackage{ulem}

\begin{document}

\title{Thermal rectification through a nonlinear quantum resonator}

\author{Bibek Bhandari}
\affiliation{NEST, Scuola Normale Superiore and Istituto Nanoscienze-CNR, I-56126 Pisa, Italy}

\author{Paolo Andrea Erdman}
\affiliation{NEST, Scuola Normale Superiore and Istituto Nanoscienze-CNR, I-56126 Pisa, Italy}

\author{Rosario Fazio}
\affiliation{The Abdus Salam International Centre for Theoretical Physics , Strada Costiera 11, I-34151 Trieste, Italy}
\affiliation{Dipartimento di Fisica, Universit\`a di Napoli ``Federico II'', Monte S. Angelo, I-80126 Napoli, Italy}

\author{Elisabetta Paladino}
\affiliation{Dipartimento di Fisica e Astronomia Ettore Majorana, Universit\`a di Catania, Via S. Sofia 64, I-95123, Catania, Italy}
\affiliation{INFN, Sez. Catania, I-95123, Catania, Italy}
\affiliation{CNR-IMM, Via S. Sofia 64, I-95123, Catania, Italy}

\author{Fabio Taddei}
\affiliation{NEST, Istituto Nanoscienze-CNR and Scuola Normale Superiore, I-56126 Pisa, Italy}

\begin{abstract}

We present a comprehensive and systematic study of thermal rectification in a prototypical low-dimensional quantum system -- a non-linear resonator: we identify necessary conditions to observe thermal rectification and we discuss strategies to maximize it.
We focus, in particular, on the case where anharmonicity is very strong and the system reduces to a qubit.
In the latter case, we derive general upper bounds on rectification which hold in the weak system-bath coupling regime, and we show how the Lamb shift can be exploited to enhance rectification. We then go beyond the weak-coupling regime by employing different methods: i) including co-tunneling processes, ii) using the non-equilibrium Green's function formalism and iii) using the Feynman-Vernon path integral approach. We find that the strong coupling regime allows us to violate the bounds derived in the weak-coupling regime, providing us with clear signatures of high order coherent processes visible in the thermal rectification.
In the general case, where many levels participate to the system dynamics, we compare the heat rectification calculated with the equation of motion method and with a mean-field approximation. We find that the former method predicts, for a small or intermediate anharmonicity, a larger rectification coefficient.
\end{abstract}

\maketitle

\section{Introduction}
Thermal transport in quantum devices has garnered vast attention in the last decade fuelled by the incessant efforts in the miniaturization of electronic and 
thermal devices. Furthermore, the research in this field has been constantly growing thanks to advances in the experimental realization of nanoscale thermal 
devices that have sharpened our understanding on how energy/heat flows through small (quantum) systems~\cite{giazotto2006,giazotto2012, pekola2015, 
ronzani2018,maillet2019, maillet2020}. 

\begin{figure}[!htb]
	\centering
	\includegraphics[width=0.9\columnwidth]{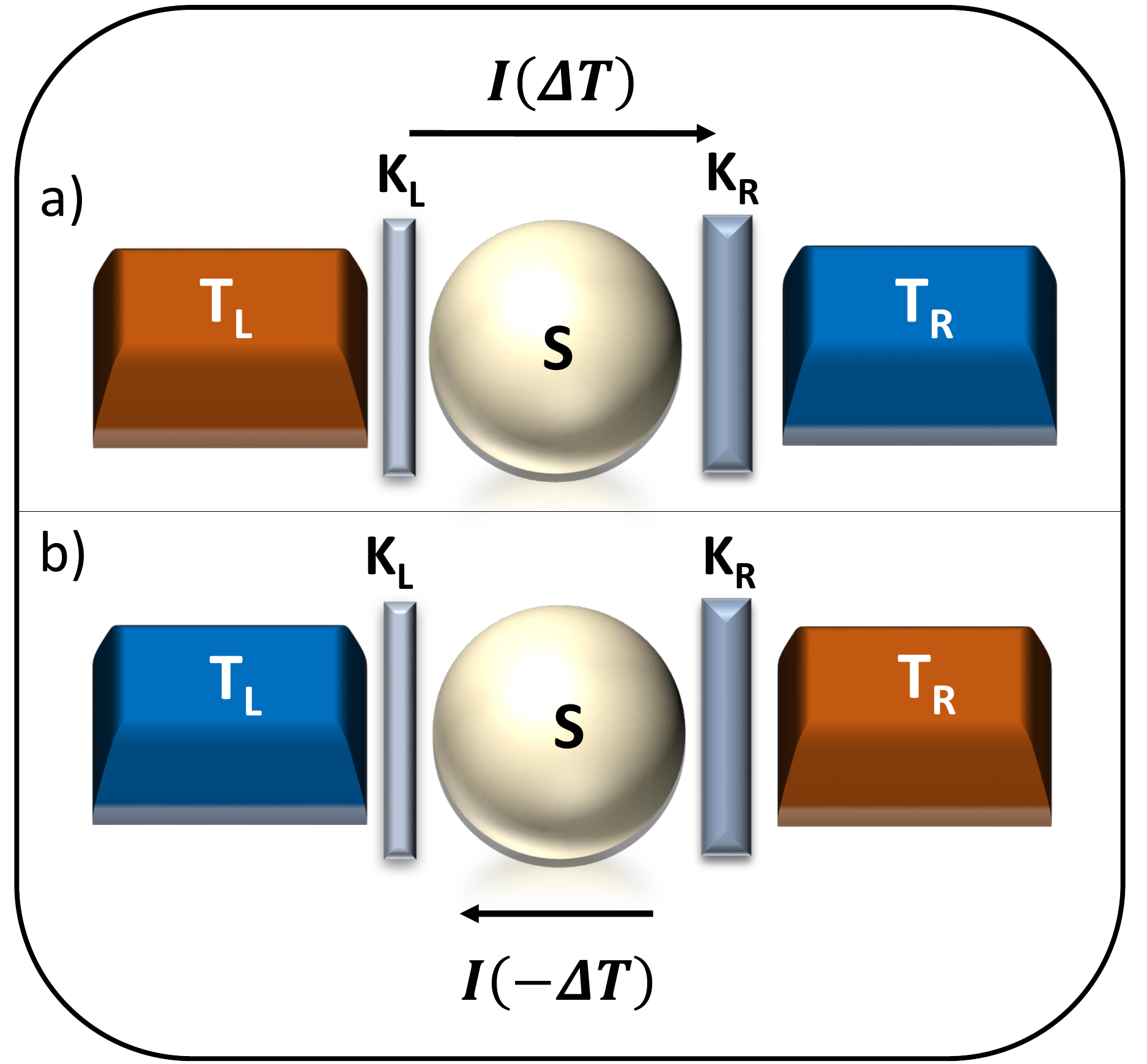}	
	\caption{Schematic representation of a quantum system S coupled to the two heat baths. The left and right baths are characterized, 
	respectively, by the temperatures $T_\text{L}$ and $T_\text{R}$. Panel~(a) represents the positive bias case, i.e. $T_\text{L}=T+\Delta T/2$ and 
	$T_\text{R}=T-\Delta T/2$ with $\Delta T>0$, while panel~(b) represents the negative bias configuration where the sign of $\Delta T$ is reversed. 
	In the presence of some asymmetry in the coupling to the baths (represented by the different thickness of the grey barriers), the magnitude of the 
	heat currents flowing through the device may depend on the sign of $\Delta T$, leading to thermal rectification.}
	\label{fig:setup1a}
\end{figure}

The phenomenon at the heart of our investigation is {\it thermal rectification}, an intriguing effect which may arise also at the nanoscale, where it may 
play a key role for heat management in small devices. It refers to the asymmetric conduction of heat, whereby the heat flow in one direction is different 
with respect to the heat flow in the opposite direction, see Fig.~\ref{fig:setup1a}. Thermal rectification, first observed experimentally by Starr in 1935~\cite{starr1935}, 
has been studied in a variety of setups since then, both theoretically~\cite{terraneo2002,li2004,segal2005,eckmann2006,zeng2008,ojanen2009,ruokola2009,archak,wu2009,wu2009b,kuo2010, otey2010,Zhang2010,Yang2018,roberts2011,ruokola2011, gunawardana2012, martinez2013, giazotto2013,liu2014,landi2014,jiang2015,sanchez2015, joulain2016,agarwalla,vicioso2018,giazotto2020} and experimentally~\cite{chang2006,scheibner2008,senior2019,schmotz2011,martinez2015}. 

From the practical viewpoint, the importance of thermal rectification stems from the fact that it can be used in a nanoscale device to divert heat from sensitive areas, while 
preventing it from flowing back in. On the other hand, it is interesting to understand what are the fundamental physical requirements for a system to exhibit 
thermal rectification, and what are the strategies to optimize it. In this paper we study thermal rectification through a multi-level quantum system (S) 
coupled to two thermal baths kept at different temperatures as schematically sketched in Fig.~\ref{fig:setup1a}. This is a paradigmatic situation that applies to 
several different experimentally available setups, such as the experiment of Ref.~\onlinecite{senior2019} where the effect was first observed in an artificial 
atom.  

Few ingredients are necessary for rectification to occur. As we shall see in the following of the paper, the baths must  be 
asymmetrically coupled to the system and inelastic scattering/interactions must necessarily be present. Indeed, in the absence of the latter, the current can be described 
by the Landauer-B{\"u}ttiker scattering approach \cite{landauer1957, buttiker1985}, expressed as an energy integral of a transmission function (which does not 
depend on temperature) multiplied by the difference of energy distribution of the baths. In this situation no rectification is possible, since the temperatures 
of the baths enter only through their distributions. Inelastic processes occur naturally in the presence of non-linearities, for example induced by interactions, 
or by time-dependent driving in the Hamiltonian describing the system \cite{campeny2019}. In the presence of interactions, at least when the spectral 
density of the baths have identical energy dependence, one can formally express the heat current analogous to the scattering theory with an effective 
transmission coefficient which now depends also on the temperatures of the baths \cite{ojanen,agarwalla}. If, in addition, the quantum system S is coupled 
asymmetrically to the two baths, thermal rectification can take place.

\begin{figure}[!t]
	\centering	\includegraphics[width=0.79\columnwidth]{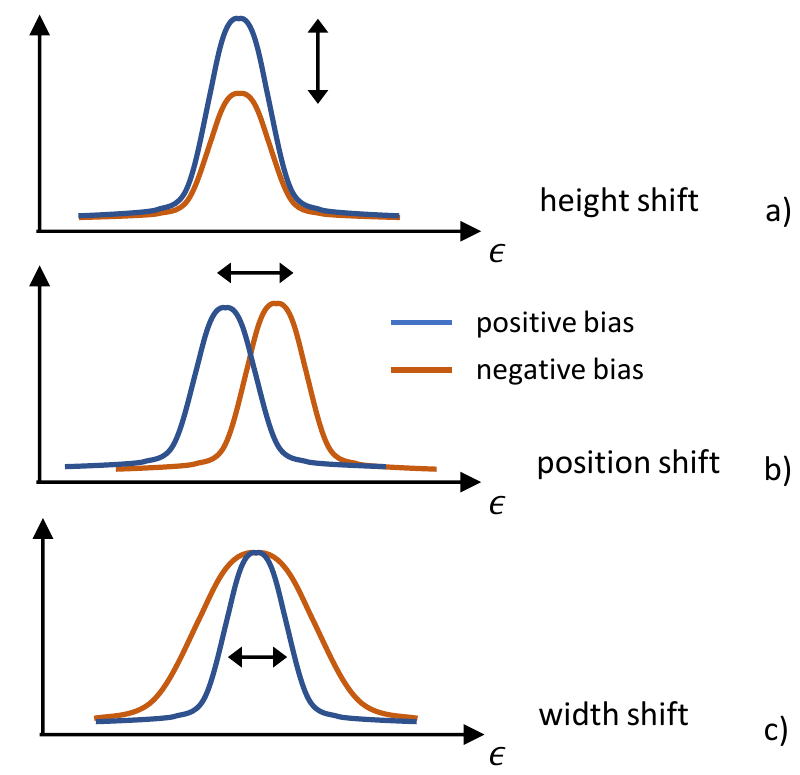}
	\caption{Schematic representation of the transmission function, as a function of the energy $\epsilon$, for the positive and negative bias case. 
	Each panel corresponds to a different variations of the transmission function which can give rise to heat rectification.}
	\label{fig:scheme}
\end{figure}

Within this framework, we can identify three possible ways the transmission probability can change upon inverting the temperature 
bias (from positive to negative). As schematically shown in Fig.~\ref{fig:scheme}, the transmission functions can change in height~(a), position~(b) and width~(c).
The height shift is the main mechanism that allows rectification even in the weak coupling regime, and it is present whenever one accounts for inelastic processes. 
The position shift is caused by the real part of the self energy, known as Lamb shift, which accounts for the renormalization of the system energy scales due to the 
system-bath coupling. Finally, the width of the transmission probability may change when the system is strongly coupled to the baths. In most cases we consider, 
the width and height shift occur together.

Rectification has been studied in different nanoscale devices, such as quantum dots~\cite{ruokola2011,kuo2010,vicioso2018}, spin-boson models~\cite{agarwalla,segal2005}, 
non-linear harmonic resonators~\cite{ruokola2009,archak}, and hybrid quantum devices~\cite{wu2009,wu2009b,martinez2013}, to name a few. In most cases, the weak coupling, 
wide band approximation has been employed. Asymmetric system-bath coupling and the presence of non-linearities are sufficient conditions to rectify~\cite{wu2009}. 
When studying the spin-boson model~\cite{segal2005,agarwalla} and the non-linear harmonic resonator~\cite{ruokola2009}, it has been observed that  rectification 
increases both as a function of the temperature difference and as a function of the asymmetry between the system-bath coupling strengths. The spin-boson model 
has been studied also beyond the weak coupling regime in Ref.~\onlinecite{agarwalla} 
using non equilibrium Green's functions or the Non-Interacting-Blip-Approximation (NIBA)~\cite{segal2005,boudjada2014}.

Although thermal rectification has been studied in various specific systems, strategies to \textit{maximize} rectification remain, to a large extent, unclear. 
Moreover, it is not known if there are any fundamental bounds to the maximum rectification that can be obtained, and what is the impact of quantum coherence 
on rectification. In this paper, we address these issues by considering, as a prototype model of a multi-level system, an anharmonic quantum oscillator.
In the limit of very strong anharmonicity the system reduces to a  a qubit (two-level systems) coupled to two different thermal baths. 
In our analysis we employed different formalisms to explore different regimes.
In the qubit case we used: (1) the Master equation (ME) taking co-tunneling into account, (2) non-equilibrium Green's 
functions (NEGF) and (3) exact calculations based on Feynman-Vernon path integral approach.
In the case of arbitrary anharmonicity we used the equation of motion (EOM) method. 

In the limit of very strong anharmonicity (qubit case), without assuming any specific model for the bath and system-bath Hamiltonian, we studied how to maximize the 
rectification and we derived general upper bounds valid within the weak coupling regime. Furthermore, we found that the rectification can be enhanced by exploiting the 
temperature dependence of the Lamb-shift, together with gapped density of states in the baths. Going beyond the weak-coupling regime we generalized the calculation 
of Ref.~\onlinecite{agarwalla} by addressing general spin couplings between the system and the baths, and by including the effect of the Lamb shift.  Furthermore, employing 
the Feynman-Vernon path integral approach, we were also able to study the strong coupling regime in an exactly solvable case. Thanks to the combination of all these 
approaches it was possible to see that many bounds and limitations emerging in the weak coupling regime can be overcome, and that rectification can be enhanced by 
higher order quantum coherent processes. The violation of such bounds provides a clear and simple, experimentally observable, strong-coupling signature of thermal rectification.

For smaller anharmonicity, the multi-level dynamics of the system comes into play and the qubit approximation breaks down. 
Ruokola et al.~\cite{ruokola2009} studied thermal rectification in a non-linear harmonic resonator using the mean-field Hartree approximation. Such approximation 
gives accurate results when the strength of anharmonicity is small compared to other energy scales of the system. In this paper, we go beyond the 
mean-field approximation employing the EOM method to study thermal rectification in the strong coupling and large-interaction regime.

The paper is organized as follows. In Section~\ref{sec:model} we introduce the model we are going to analyze in the rest of the paper. The  two baths, kept at different 
temperatures, can be of fermionic or bosonic nature. The Hamiltonian of the system is that of an anharmonic oscillator, more specifically with a Kerr-like $U$ non-linearity.
In the case of very large $U$, the model reduces to a two level system. Different types of coupling between the system and the baths are considered as well to see if 
different choices may lead to an enhancement of the rectification. The heat current and the rectification coefficient will be defined in Section~\ref{sec:heat_rect}. Here we will 
also introduce an expression for the current that will be used in the remainder of the paper when developing our approximation schemes. 
Sections~\ref{sec:weak_coupling}~-~\ref{sec:nonlin} contain the results of our analysis. We first start by analyzing the $U \to \infty$ case. In Section~\ref{sec:weak_coupling} we 
study the qubit case in the weak coupling regime, while in Section~\ref{sec:qubit_strong} we go beyond the weak coupling regime. The various approximation schemes here 
are also compared to an exact solution that we are able to derive for a specific choice of the couplings. In Section~\ref{sec:nonlin}, we relax the approximation of a two-level 
system and study thermal rectification in a non-linear resonator as a function of $U$. Finally, in Section~\ref{sec:conclusions} we draw the conclusions. Appendices contain 
several different details of the calculations, not inserted in the main text to favour the readability of the paper.

\section{The model}
\label{sec:model}
We consider a quantum system S arbitrarily coupled to two thermal baths denoted by $\text{L}$ (left) and $\text{R}$ (right) [see Fig.~\ref{fig:setup1a} for a sketch]. The  
Hamiltonian governing the dynamics of this setup is given by
\begin{equation}
	\mathcal{H} = \mathcal{H}_\text{L} + \mathcal{H}_\text{R}  + \mathcal{H}_{U} + \mathcal{H}_{\text{L,S}} + \mathcal{H}_{\text{R,S}},
\end{equation}
where $\mathcal{H}_\alpha$, for $\alpha=\text{L},\text{R}$, is the Hamiltonian of bath $\alpha$, $\mathcal{H}_{U}$ is the Hamiltonian of the system S 
and $\mathcal{H}_{\alpha,\text{S}}$ describes the coupling between bath $\alpha$ and S. Each of these components -- the baths, the system, and the couplings -- 
contribute in different ways to the thermal properties of the device. Below we describe in detail these different parts.

\subsection{Thermal baths}
The baths are assumed, as customary,  to be ``large'' quantum systems in equilibrium with a well defined temperature $T_\alpha$ (and equal chemical potential $\mu$ in the 
fermionic case).  The Hamiltonian of the bosonic (B) and fermionic (F) baths is given by
\begin{equation}
\begin{aligned}
	\mathcal{H}_\alpha^\text{(B)} &= \sum_k \epsilon_{\alpha k}\, b^\dagger_{\alpha k} b_{\alpha k}, \\
	\mathcal{H}_\alpha^\text{(F)} &= \sum_k \epsilon_{\alpha k} \, c^\dagger_{\alpha k} c_{\alpha k},
\end{aligned}
\label{eq:h_bath}
\end{equation}
where $b_{\alpha k}$ and $b_{\alpha k}^\dagger$ ($c_{\alpha k}$ and $c_{\alpha k}^\dagger$) are, respectively, the destruction and creation bosonic 
(fermionic) operators of an excitation  with energy $\epsilon_{\alpha k}$ in bath $\alpha$ and quantum number $k$. The operators satisfy the usual commutation and 
anticommutation relations: $[b_{\alpha k},b^\dagger_{\alpha^\prime k^\prime}] = \delta_{\alpha,\alpha^\prime}\delta_{k,k^\prime}$, $[b_{\alpha k},b_{\alpha^\prime k^\prime}]= 0$, 
$\{c_{\alpha k},c^\dagger_{\alpha^\prime k^\prime}\} = \delta_{\alpha,\alpha^\prime}\delta_{k,k^\prime}$ and $\{c_{\alpha k},c_{\alpha^\prime k^\prime}\}= 0$, where 
$[\dots,\dots]$ and $\{\dots,\dots\}$ denote, respectively, the commutator and anticommutator. In the following, for simplicity, we will generically use the symbol $d_{\alpha k}$ to 
denote both cases and will later specify the nature of the particles  forming the bath. Since the baths are at thermal equilibrium, the bosonic baths are
prepared in a thermal Gibbs state $\rho^\text{(B)}_\alpha = e^{-\mathcal{H}^\text{(B)}_\alpha/(k_\text{B}T_\alpha)}/Z^\text{(B)}_\alpha$, where $Z^\text{(B)}_\alpha= 
\Tr\, [e^{-\mathcal{H}^\text{(B)}_\alpha/(k_\text{B}T_\alpha)}]$ is the partition function of bath $\alpha$, while we assume the fermionic baths to be prepared in the 
state  $\rho^\text{(F)}_\alpha = e^{-(\mathcal{H}^\text{(F)}_\alpha-\mu\mathcal{N}_\alpha)/(k_\text{B}T_\alpha)}/Z^\text{(F)}_\alpha$, where $Z^\text{(F)}_\alpha
= \Tr[e^{-(\mathcal{H}^\text{(F)}_\alpha-\mu\mathcal{N}_\alpha)/(k_\text{B}T_\alpha)}]$ is the grand partition function of bath $\alpha$ and $\mathcal{N}_\alpha 
= \sum_k c^\dagger_{\alpha k} c_{\alpha k}$ is the particle number operator of bath $\alpha$.

\subsection{The System}
The system S  connecting the two reservoirs is a multi-level quantum system. As discussed in the introduction, as a paradigmatic case also relevant for experiments, we will 
consider   a non-linear   resonator whose Hamiltonian is given by
\begin{equation}
	\mathcal{H}_{U} = \Delta b^{\dagger}b + \frac{U}{2}b^{\dagger}b^\dagger bb,
	\label{eq:sys_res}
\end{equation}
where $\Delta$ determines the frequency of the harmonic resonator, $b$ ($b^\dagger$) is a bosonic destruction (creation) operator, and $U$ describes 
the strength of the non-linear term. The essential ingredient, we believe, are the multi-level structure of the spectrum and its non-harmonic nature. In this
perspective, the Kerr-like form represents a generic situation capturing all these features.
In fact it bridges between weakly anharmonic systems like the transmon \cite{koch2007} for small values of $U$ to multilevel qubits like the fluxonium \cite{manucharyan2009,koch2013} or the phase qubits \cite{martinis2002} for increasing values of the anharmonicity parameter.
In the limit of large $U$, only the two number states $| n=0 \rangle$ and $| n=1 \rangle$ are relevant for the dynamics and the corresponding Hamiltonian reduces to that 
of a qubit ($U=\infty$)

\begin{equation}
	\mathcal{H}_{\infty}  = \frac{\Delta}{2}\sigma_z,
	\label{eq:def_hq}
\end{equation}
(we dropped an irrelevant constant) where $\Delta$ is the energy spacing between the ground and excited state, and $\sigma_z$ denotes a Pauli matrix.  Physically, in a 
bosonic system the qubit may represent a non-linear harmonic oscillator where the interaction is so strong that only the first two states are energetically accessible.

\subsection{System-bath coupling}
The coupling allows energy exchange between the baths and S. In the $U=\infty$ (qubit) case, we can write the most general system-bath interaction as
\begin{equation}
{\cal H}_{\alpha, \text{S}}=\sigma^+\otimes  B_\alpha  +  \sigma^- \otimes B_\alpha^\dagger  +\sigma_z \otimes B_{\alpha z} ,
\label{eq:h_int_gen1}
\end{equation}
where $B_\alpha$ is an arbitrary operator (not necessarily Hermitian) acting on the Hilbert space of bath $\alpha$, while $B_{\alpha z}$ is an Hermitian operator 
acting on the space of bath $\alpha$ (see App.~\ref{app:qubit_master_eq} for details). This expression can be derived by expanding the operators acting on the 
tensor space of S and of the bath onto the product basis, and choosing the Pauli matrices and the identity as basis of Hermitian operators acting on the qubit space. 
Aside from deriving some general properties, throughout this paper we will mainly consider the ``linear coupling'' and the ``non-linear coupling'' cases, i.e. 
\begin{align}
	B_{\alpha}^\text{(lin)} &= \sum_k V_{\alpha k}\, d_{\alpha k}, \label{eq:qubit_lin_coupling} \\
	B_{\alpha}^\text{(non-lin)} &= \sum_k V_{\alpha k}\, d^2_{\alpha k},	\label{eq:nonlin_coupling}
\end{align}
respectively. Obviously the non-linear coupling [Eq.~(\ref{eq:nonlin_coupling})] only applies to the case of boson baths. The coupling strength is determined by  $V_{\alpha k}$. When 
assessing strong coupling effects, we will focus on the spin-boson  model, i.e. we will consider a bosonic bath coupled to the qubit via the following interaction
\begin{equation}
	\mathcal{H}_{\alpha,\text{S}} = \sum_{i=x,y,z} {u}_{\alpha,i}\sigma_i \otimes \sum_k V_{\alpha k}\, (b_{\alpha k} + b_{\alpha k}^\dagger ),
	\label{eq:h_int_generic2}
\end{equation}
where $\vec{u}_\alpha = (\sin\theta_\alpha\cos\phi_\alpha, \sin\theta_\alpha\sin\phi_\alpha, \cos\theta_\alpha)$ is a unit vector parametrized by the angles $\theta_\alpha$ and $\phi_\alpha$.

In the generic non-linear resonator case (finite $U$), we will consider bosonic baths. Interactions are already present in S, so we will focus on the following linear coupling
\begin{equation}
	\mathcal{H}_{\alpha,\text{S}} =  \sum_k V_{\alpha k}\, b_{\alpha k}^\dagger b + h.c.
	\label{eq:lin_coup}
\end{equation}

As we will see in the following, the system-bath interaction can be conveniently characterized by the spectral density 
\begin{equation}
	\Gamma_\alpha(\epsilon) = 2\pi \sum_k \delta(\epsilon - \epsilon_{\alpha k}) V_{\alpha k} V^*_{\alpha k}.
	\label{eq:spectraldensity}
\end{equation}
Taking the continuum limit for the energy spacing of the baths, and assuming that the coupling constants $V_{\alpha k}$ only depend on the energy $\epsilon_{\alpha k}$, we can rewrite Eq.~(\ref{eq:spectraldensity}) as
\begin{equation}
	\Gamma_\alpha(\epsilon) = 2\pi D_\alpha(\epsilon) |V_\alpha(\epsilon)|^2,
	\label{eq:spectral_continuum}
\end{equation}
where $D_\alpha(\epsilon)$ is the density of states of bath $\alpha$, and $V_\alpha(\epsilon_{\alpha k}) = V_{\alpha k}$. In the following, we will consider generic spectral 
densities for the two baths. In some cases, explicitly mentioned, we will focus on bosonic baths with Ohmic spectral densities and an exponential cut-off 
energy $\epsilon_\text{C}$, i.e.
\begin{equation}
	\Gamma_\alpha (\epsilon)= \pi K_\alpha
	\,\epsilon\, e^{-\epsilon/\epsilon_{\rm C}} \equiv K_\alpha J(\epsilon),
	\label{eq:negf_spectral}
\end{equation}
where $K_\alpha$ is the dimensionless Ohmic coupling strength~\cite{weiss}.

\section{Heat current and rectification coefficient}
\label{sec:heat_rect}
We are interested in studying the steady-state heat current flowing across the device when a temperature bias is imposed between the baths. Specifically, as 
depicted in Fig.~\ref{fig:setup1a}, we fix $T_\text{L} = T + \Delta T/2$ and $T_\text{R}=T-\Delta T/2$, where $T$ is the average temperature. Since no work is 
performed on the system (in the fermionic case we consider no chemical potential bias), the first 
principle of thermodynamics tell us that heat will flow from left to right if $\Delta T>0$ (positive bias case, see Fig.~\ref{fig:setup1a}a), otherwise it will flow from 
right to left (negative bias case, see Fig.~\ref{fig:setup1a}b). Furthermore, since we consider steady state currents, the heat flowing out of one bath is equal to 
the one flowing into the other bath. Therefore, for simplicity we define the heat flowing out of the left lead as
\begin{equation}
    I(\Delta T) \equiv -\lim_{t\to+\infty}\, \frac{d}{dt} \ev*{H_\text{L}}(t),
    \label{eq:j_gen_def}
\end{equation}
where $\ev{\dots}(t) = \Tr[\rho(t) \dots]$, $\rho(t)$ being the density matrix representing the state of the total system at time $t$. Notice that the time variation of the 
energy associated with the coupling Hamiltonian vanishes in steady state~\cite{ludovico2016}.
According to the definition~(\ref{eq:j_gen_def}), the heat current $I(\Delta T)$ is positive when $\Delta T>0$ and negative when $\Delta T<0$.

As discussed in the introduction, it is possible to construct devices where the magnitude of the heat current depends on the sign of the temperature bias. Specifically, 
if the left-right symmetry is broken, the magnitude of the heat current $I(\Delta T)$ induced by a positive bias may be different with respect to the magnitude of $I(-\Delta T)$, which is the 
heat current induced by a negative bias. We therefore define the rectification coefficient $R$ as
\begin{equation}
	R =  \frac{I(\Delta T)+I(-\Delta T)}{I(\Delta T)-I(-\Delta T)} ,
\label{eq:r_def}
\end{equation}
for $\Delta T>0$ [so that $I(-\Delta T)<0$ and the numerator represents the difference of the magnitudes of the currents]. The definition is such that $|R|\leq 1$. Furthermore, $R=0$ means that no rectification takes place, while $|R|=1$ means that we have 
perfect rectification (i.e. the heat current is finite in one direction, and null in the other). Positive (negative) values of $R$ indicate that the heat flow is 
greater for positive (negative) temperature biases.

For later convenience, it is useful to write the current in a Meir-Wingreen~\cite{meirwingreen} form which will make apparent the necessary ingredients for rectification. 
Starting from the formal definition of the heat current given in Eq.~(\ref{eq:j_gen_def}), we can simplify the calculation of the heat current using a standard 
procedure known as ``bath embedding''~\cite{stefanucci}, which is valid whenever the operators of the bath appear linearly in $\mathcal{H}_{\alpha, \text{S}}$. 
This approach applies to all models except for the qubit with non-linear coupling, Eq.~(\ref{eq:nonlin_coupling}), which will be treated in the weak coupling regime only.
Under such hypothesis, the formally exact Meir-Wingreen-type formula~\cite{meirwingreen} for the heat current can be written as~\cite{ojanen,velizhanin,wang2006,saito2008,segal2014} 
 \begin{equation}
	I(\Delta T) = \int \frac{d\epsilon}{2\pi\hbar} \, \epsilon \Tr\left[ G^<(\epsilon)\Sigma_\text{L}^>(\epsilon) - G^>(\epsilon)\Sigma^<_\text{L}(\epsilon) \right],
	\label{eq:j_green}
\end{equation}
where the integration is performed over $[0,+\infty]$ ($[-\infty,+\infty]$) for bosonic (fermionic) baths, and the $\Tr[...]$ runs over the internal degrees of freedom of the system S. In the previous expression $G^\lessgtr(\epsilon)$ is the 
Fourier transform of the lesser/greater Green's function of S, in the presence of the baths, defined as $G^<(t-t^\prime) =  \mp i \ev*{d^\dagger(t^\prime) d(t)}$ and $G^>(t-t^\prime) 
=  -i\ev*{d(t) d^\dagger(t^\prime)}$ (upper sign for bosons and lower sign for fermions). Moreover, $\Sigma_\text{L}^<(\epsilon) = \mp i\Gamma_\text{L}(\epsilon)n_\text{L}(\epsilon)$ and $\Sigma_\text{L}^>(\epsilon)
=-i\Gamma_\text{L}(\epsilon)(1 \pm n_\text{L}(\epsilon))$ are, respectively, the Fourier transform of the lesser and greater \textit{embedded} self energies induced by 
the left bath, and $n_{\rm L}(\epsilon)$ denotes the energy distribution of the left bath. Therefore, $n_\text{L}(\epsilon) = (e^{\epsilon/(k_\text{B}T_\text{L})} - 1)^{-1}$ for bosonic 
baths, while $n_\text{L}(\epsilon) = (e^{(\epsilon-\mu)/(k_\text{B}T_\text{L})} + 1)^{-1}$ for fermionic baths.
The only quantities which must be determined in Eq.~(\ref{eq:j_green}) are $G^\lessgtr(\epsilon)$. 

There is a typical situation in which Eq.~(\ref{eq:j_green}) can be written as a simpler and more transparent expression. Namely, if the spectral densities $\Gamma_\alpha(\epsilon)$ of the baths are proportional, i.e. $\Gamma_\text{L}(\epsilon) \propto \Gamma_\text{R}(\epsilon)$, we can write Eq.~(\ref{eq:j_green}) as \cite{jauho}
\begin{equation}
	I(\Delta T) = \int \frac{d\epsilon}{2\pi\hbar} \, \epsilon\, \mathcal{T}(\epsilon,T, \Delta T) \left[ n_\text{L}(\epsilon) - n_\text{R}(\epsilon) \right],
	\label{eq:j_green_prop}
\end{equation}
where
\begin{equation}
	 \mathcal{T}(\epsilon,T ,\Delta T) = i\Tr\left\{\frac{\Gamma_\text{L}(\epsilon)\Gamma_\text{R}(\epsilon)}{\Gamma_\text{L}(\epsilon)+\Gamma_\text{R}(\epsilon)}[ G^>(\epsilon) - G^<(\epsilon) ] \right\}
	 \label{eq:def_transmission}
\end{equation}
and $n_\alpha(\epsilon)$ denotes the energy distribution of bath $\alpha$. 
This formula was used in Ref. \cite{agarwalla} to study the spin-boson problem. The dependence of $\mathcal{T}(\epsilon,T,\Delta T)$ on the temperatures may arise from $G^\lessgtr(\epsilon)$, which are indeed correlation functions of
$S$ computed in the presence of the baths. We notice that, in the absence of this temperature dependence, the magnitude of the heat current would remain the same in the 
positive and negative bias cases, and there would be no thermal rectification. Indeed, this is the case for non-interacting systems, where Eq.~(\ref{eq:def_transmission}) reduces 
to the well known scattering formula with a transmission function that does not depend on the temperature of the baths. It is therefore crucial to introduce a non-linearity in the 
local system or in the coupling Hamiltonian to observe thermal rectification.

By computing the Green function in certain approximation schemes, rectification can be explored at different orders in the system-bath coupling. We will first consider the weak coupling 
approximation where the results can equivalently be derived by a Master Equation.

\section{$U=\infty$ - Weak coupling regime}
\label{sec:weak_coupling}
We start our analysis considering the case of two-level system (the $U=\infty$ case). Furthermore, in this Section, we derive general properties and upper bounds to the  rectification coefficient $R$ in the limit in which the baths are weakly coupled to the qubit. 
This regime is obtained by performing a leading order expansion in $\mathcal{H}_{\alpha,\text{S}}$. 
At this level the baths are effectively treated as Markovian~\cite{breuer2002} and transport is described by a Lindblad master equation for the reduced density matrix of the system. 
Heat transport takes place via sequential tunneling processes where the transition from the ground (excited) to the excited (ground) state involves single photon absorption (emission) in the bosonic case~\cite{boese2001,segal2005}.
The approximations considered in this Section
lead to height and position shifts in the transmission function (see Fig.~\ref{fig:scheme}), but width shifts are neglected. Indeed, in this regime the width of the transmission function is the smallest energy scale, so it is infinitesimal, see Eq.~(\ref{eq:heat_currrents_weak}).
Width shifts will appear beyond the weak coupling regime, as discussed in the following Sections.

Under weak coupling, the rectification ratio is found to be (see App.~\ref{app:rect_weak} for details)
\begin{equation}
	R = \frac{\Upsilon^{-1}_\text{L}(T_\text{L})+\Upsilon^{-1}_\text{R}(T_\text{R})-\Upsilon^{-1}_\text{L}(T_\text{R})-\Upsilon^{-1}_\text{R}(T_\text{L})}
	{\Upsilon^{-1}_\text{L}(T_\text{L})+\Upsilon^{-1}_\text{R}(T_\text{R})+\Upsilon^{-1}_\text{L}(T_\text{R})+\Upsilon^{-1}_\text{R}(T_\text{L})},
\label{eq:r_general}
\end{equation}
where
$\Upsilon_\alpha(T) = [1 +  e^{- \Delta /(k_B T)}] \Upsilon^-_\alpha (T) $ is the total dissipation rate induced by bath $\alpha$~\cite{footnote1}.
The rate $\Upsilon_\alpha^-(T)$ is associated to the 
transition of the qubit from the excited to the ground state by exchanging energy with bath $\alpha$. Since bath $\alpha$ is prepared in a thermal state, the other rate 
$\Upsilon^+_\alpha(T)$ (from the ground to the excited state) is related by the detailed balance.
The dissipation rate can then be calculated by evaluating 
\begin{equation}
	\Upsilon^-_\alpha(T) = \frac{1}{\hbar^2}\int\nolimits_{-\infty}^{+\infty} dt\, e^{i\Delta t/\hbar} \ev{B_\alpha(t)B_\alpha^\dagger(0)},
	\label{eq:gamma_def}
\end{equation}
where the expectation value is taken with respect to the \textit{equilibrium} thermal state $\rho_\alpha$ of the bath, $\Delta$ is the energy spacing of the qubit,
$B_\alpha(t)$ and $B^\dagger_\alpha(t)$ are interaction picture operators
[see App.~\ref{app:tunn_rates} for a derivation of Eq.~(\ref{eq:gamma_def})]. 
Notice that Eq. (\ref{eq:r_general}) holds for arbitrary spectral densities of the baths. The role of the Lamb-shift, which is neglected in Eq. (\ref{eq:r_general}), will be discussed in Section~\ref{sec:lamb_weak}.  

We can now study $R$ for any weakly coupled system using Eqs.~(\ref{eq:r_general}) and (\ref{eq:gamma_def}), which are generally easy to compute (we will 
consider various models explicitly in the following subsections). As a consequence of the weak coupling assumption, the coupling term $\sigma_z \otimes B_{\alpha z}$ 
[see Eq.~(\ref{eq:h_int_gen1})] does not contribute to $I(\Delta T)$, therefore neither to $R$. This is due to the fact that, in the weak coupling regime, the heat current is mediated by 
transitions in S. Therefore, the heat current only depends on the population of the qubit, which in turn is solely determined by the coupling 
terms proportional to $\sigma^+$ and $\sigma^-$ (see App.~\ref{app:rect_weak} for more details). There is nevertheless ample room for optimizing the rectification by considering different coupling Hamiltonians.
We decompose the dissipation rate as
\begin{equation}
	\Upsilon_\alpha(\Delta,T) = \Gamma_\alpha(\Delta) g_\alpha (\Delta,T),
	\label{eq:similar_rates}
\end{equation}
where $\Gamma_\alpha(\Delta)$ is the spectral density of bath $\alpha$, given by Eq.~(\ref{eq:spectral_continuum}), and 
$g_\alpha (\Delta,T)\geq 0$ are arbitrary non-negative functions.
In App.~\ref{sec:tunneling_rates} we explicitly evaluate $g_\alpha (\Delta, T)$ for various models.

The rest of the section is organized as follows: in Section~\ref{sec:similar_bath} we consider the case in which the two functions $g_{\rm L}(\Delta,T)$ and $g_{\rm R}(\Delta,T)$ are equal.
Under such a weak condition on the baths, we derive general bounds on $R$.
We then study the impact of linear and non-linear coupling to the baths.
In Section~\ref{sec:different_bath} we show that rectification can be enhanced in the case in which $g_{\rm L}\neq g_{\rm R}$.
In Section~\ref{subsubsec:weak_sigma}, we study a generic spin coupling to the bath. At last, in Section \ref{sec:lamb_weak} we show 
how the Lamb-shift can be exploited to further enhance rectification.

\subsection{$g_{\rm L}=g_{\rm R}$ case}
\label{sec:similar_bath}
In this section we assume that, in Eq.~(\ref{eq:similar_rates}), $g_{\rm L}(\Delta,T)=g_{\rm R}(\Delta,T)\equiv g(\Delta,T)$.
This implies that the dissipation rates of the two baths, as a function of temperature, are equal up to a prefactor. 
However, the rates may have \textit{any dependence} on the gap of the qubit through the spectral densities. 
 
Having defined the asymmetry coefficient $\lambda$ as 
\begin{equation}
	\lambda =\frac{\Gamma_\text{L}(\Delta)-{\Gamma}_\text{R}(\Delta)}{\Gamma_\text{L}(\Delta)+{\Gamma}_\text{R}(\Delta)},
	\label{eq:l_def}
\end{equation}
such that $|\lambda| \leq 1$, and using Eq.~(\ref{eq:similar_rates}), one can cast Eq.~(\ref{eq:r_general}) into the simple form
\begin{equation}
	R =\lambda \, \frac{ g(\Delta,T_\text{C}) -  g(\Delta,T_\text{H})}{ g(\Delta,T_\text{C}) +  g(\Delta,T_\text{H}) }.
	\label{eq:r_similar}
\end{equation}
Without specifying the precise model, we can derive the following general properties of $R$: 
\begin{itemize}
	\item If $\lambda>0$, $R$ is a decreasing function of $g(\Delta,T_\text{H})$, and an increasing function of $g(\Delta,T_\text{C})$ (the monotonicity 
	is inverted if $\lambda<0$). Therefore, if  $g(\Delta,T)$ is monotonous with respect to $T$, then $R$ is monotonous with respect to $\Delta T$.
	
	\item $R$ is linear, therefore monotonous, with respect to $\lambda$.	
	\item Given the first property, we can maximize the possible rectification by taking the limits where $g(\Delta,T_\text{H})$ and $g(\Delta,T_\text{C})$ 
	respectively tend to zero and infinity. This yields the following bound
	\begin{equation}
			|R| \leq |\lambda|.
		\label{eq:r_similar_bound}
	\end{equation}
	As a consequence, the maximum rectification is severely limited by the asymmetry ratio $\lambda$. As expected, for $\lambda=0$ we find that there is no 
	rectification, and the only way to obtain perfect rectification is to have a vanishingly small coupling to one bath.
\item Given the second property, $|R|$ is bounded by $|(g(\Delta,T_\text{C})-g(\Delta,T_\text{H}))/(g(\Delta,T_\text{C})+g(\Delta,T_\text{H}))|$. We therefore have 
stronger rectification when $g(\Delta,T)$ has a strong temperature dependence.
\end{itemize}

\subsubsection{Linear system-bath tunnel couplings}
\label{subsec:linear_coup}
In this subsection we study heat rectification through a qubit where the coupling to the baths is linear, i.e. defined by Eq.~(\ref{eq:qubit_lin_coupling}). For 
fermionic baths weakly coupled to the qubit, we have that (see App.~\ref{subsec:fermionic_rates}, for details)
\begin{equation}
	g(\Delta,T) = 1.
	\label{eq:fermionic_rates}
\end{equation}
Plugging this value into Eq.~(\ref{eq:r_similar}) shows us that no rectification is possible~\cite{senior2019}. This is indeed expected, since a qubit coupled to fermionic reservoirs 
can be described by a non-interacting fermionic Hamiltonian, where the Landauer-B{\" u}ttiker formula can be used to compute the heat current. Next, we 
consider bosonic baths. In this case, as shown in App.~\ref{subsec:bosonic_tunnel_rates}, we have that
\begin{equation}
	g(\Delta,T) = \coth{[\Delta/(2k_{\rm B}T)]},
	\label{eq:bosonic_tunnel_rates}
\end{equation}
so rectification is possible. In particular, we find the following properties:
\begin{itemize}
	\item Since $g(\Delta,T)$ is a monotonous increasing function of $T$, $\lambda$ and $R$ have opposite signs. This means that more heat flows out of the lead which is more weakly coupled.
	\item Since $g(\Delta,T)$ is a monotonous function of $T$, the rectification increases with $\Delta T$. 
	\item Since $g(\Delta,T)$ is never zero, but it diverges for $T \to \infty$, the bound in Eq.~(\ref{eq:r_similar_bound}) is saturated only in the limit of infinitely hot reservoir ($T_\text{H} \to \infty$).
	\item It can be explicitly seen that $R$ is a decreasing function of the gap $\Delta$, so it is maximum in the limit $\Delta \to 0$. In this limit, we can expand the hyperbolic cotangent, finding the following bound
	\begin{equation}
		|R| \leq \lambda\, \frac{T_\text{H}-T_\text{C}}{T_\text{H}+T_\text{C}}.
		\label{eq:r_boson_t}
	\end{equation}
\end{itemize}

\subsubsection{Bosonic baths with non-linear coupling}
\label{subsec:boson_nonlin}
For the non-linear coupling 
$g(\Delta, T)$ is defined through the relation $\Upsilon_\alpha(\Delta,T) = \Gamma_\alpha(\Delta/2) g(\Delta,T)/2$.
By replacing  $\Gamma_\alpha(\Delta)$ with $\Gamma_\alpha(\Delta/2)$ in the definition of
$\lambda$, Eq.~(\ref{eq:l_def}),
 the function  $g(\Delta, T)$  reads [see App.~\ref{app:bosonic_nonlin}]
\begin{equation}
	g(\Delta,T) = 1 +\coth^2[\Delta/(4k_{\rm B}T)].
	\label{eq:rates_nonlin_bosonic}
\end{equation}
The properties listed below follow:
\begin{itemize}
\item $g(\Delta,T)$, as a function of $T$, has the same monotonicity as the bosonic case with linear coupling, it only diverges for $T\to \infty$, and it is never zero. 
Therefore, it has the same first 3 properties of Subsection~\ref{subsec:linear_coup}.
\item Also in this case, $g(\Delta,T)$ is monotonous with respect to $\Delta$, such that $R$ is maximized in the limit $\Delta \to 0$. Performing an expansion for 
small $\Delta$, we find the following bound
\begin{equation}
	|R| \leq \lambda\, \frac{ T_\text{H}^2 - T_\text{C}^2 }{ T_\text{H}^2 +  T_\text{C}^2 }.
\end{equation}
Comparing with Eq.~(\ref{eq:r_boson_t}), we see that the non-linear coupling may be more effective in rectifying the heat current. This can be explicitly verified by 
comparing the exact expressions of $R$ using Eq.~(\ref{eq:r_similar}).  
\end{itemize}

\subsection{$g_{\rm L}\neq g_{\rm R}$ case}
\label{sec:different_bath}
It may be useful to check if allowing arbitrary functions $g_{\rm L}(\Delta,T)$ and $g_{\rm R}(\Delta,T)$ in Eq.~(\ref{eq:similar_rates}) results in a stronger rectification.
Notice that $g_{\rm L}$ and $g_{\rm R}$ are different whenever 
the correlation function in Eq.~(\ref{eq:gamma_def}) is different for the two baths. This may happen considering different bath Hamiltonians, and/or different coupling 
Hamiltonians.

As an example, we take the two bosonic baths, but we consider two different coupling Hamiltonians. We assume the left bath to be linearly coupled to the qubit 
(as in Section~\ref{subsec:linear_coup}) and the right bath to be non-linearly coupled to the qubit (as in Section~\ref{subsec:boson_nonlin}). 
Plugging the respective rates for the left and the right lead into Eq.~(\ref{eq:r_general}) yields an exact expression for $R$ which in the limit $\Delta \to 0$ is simply given by
\begin{equation}
	R =\frac{T_\text{H}-T_\text{C}}{T_\text{H}+ T_\text{C}}
	\label{eq:r_t_nonlin_tun}
\end{equation}
\textit{regardless of $\lambda$} [under the obvious assumption that $\Gamma_\alpha(\Delta\to 0)$ does not diverge]. Notably, a general property of  the regime 
analyzed in Section~\ref{sec:similar_bath} is that $|R| \leq |\lambda|$, 
while here we have rectification even for $\lambda=0$, and it can be made arbitrarily large simply by choosing larger and larger temperature differences. 

This shows that, in general, an asymmetry in the form of the system-bath couplings can produce large rectification coefficients.

\subsection{Generic spin coupling}
\label{subsubsec:weak_sigma}
In this section, we investigate what happens when the qubit is coupled to the baths through the same arbitrary bath operators, but through different Pauli spin matrices. 
As an example, we consider the coupling Hamiltonian given in Eq.~(\ref{eq:h_int_generic2}) with $V_{\text{L}k} = V_{\text{R}k}$, although also non-linear couplings can 
be treated on the same footing. As shown in App.~\ref{app:different_sigma}, this system can be mapped into the case discussed in Section~\ref{sec:similar_bath}
 with an effective ${\Gamma}_\alpha(\Delta) \propto \sin^2{\theta_\alpha} $. 
Therefore, all the properties derived in Section~\ref{sec:similar_bath} hold in this case, where the asymmetry coefficient is given by
\begin{equation}
	\lambda = \frac{\sin^2\theta_\text{L}-\sin^2\theta_\text{R}}{\sin^2\theta_\text{L}+\sin^2\theta_\text{R}},
\end{equation}
while the function $g(\Delta,T)$ depends on the bath and system-bath Hamiltonian [in the specific case of Eq.~(\ref{eq:h_int_generic2}), we have 
$g(\Delta,T) = \coth{[\Delta/(2k_{\rm B}T)]}$, see Eq.~(\ref{eq:bosonic_tunnel_rates})]. The rectification does not depend on the angle $\phi_\alpha$; 
as we will show, this property does not hold beyond the weak coupling regime thanks to coherent transport effects. The only relevant parameter is $\theta_\alpha$, 
which is the angle between the coupling term and the qubit Hamiltonian (proportional to $\sigma_z$). Since the rectification is linear in $\lambda$, it reaches  the 
maximum when $\theta_\text{L}$ is $0$ and $\theta_R$ is $\pi/2$, or viceversa. 

\subsection{Role of the Lamb-Shift}
\label{sec:lamb_weak}
Until now we have ignored the Lamb shift, i.e. the renormalization of the energy gap of the qubit induced by the presence of the baths. The renormalization of 
the qubit splitting depends on both ($\text{L/R}$) temperatures, thus it may influence the rectification properties of the device. This allows us to have rectification not subject 
to the bounds derived in the previous sections.

As shown in App.~\ref{sec:lamb_shift}, if the system-bath Hamiltonian does not contain terms proportional to $\sigma_z$ (i.e. $B_{\alpha z} = 0$), the Lamb shift 
Hamiltonian (which has to be summed to the bare Hamiltonian $\mathcal{H}_{\infty}$) takes the following form~\cite{breuer2002}:

\begin{equation}
\tilde{\mathcal{H}} = \left[ \delta\Delta_\text{L}(\Delta,T_\text{L}) + \delta\Delta_\text{R}(\Delta,T_\text{R}) \right]\sigma_z,
	\label{eq:hls}
\end{equation}
where
\begin{equation}
	\delta\Delta_\alpha(\epsilon,T) = \frac{1}{2\pi}\mathcal{P}\int\nolimits_{-\infty}^{+\infty} \frac{\Upsilon_\alpha(\epsilon^\prime,T)}{\epsilon - \epsilon^\prime}d\epsilon^\prime.
		\label{eq:s}
\end{equation}
In Eq.~(\ref{eq:s}), $\mathcal{P}$ indicates a Cauchy principal value integration. We recall that the $\Delta$ appearing in Eq.~(\ref{eq:hls}) is the \textit{bare gap}. 
The renormalized gap is therefore given by
\begin{equation}
	\tilde{\Delta}(\Delta T) = \Delta +  \delta\Delta_\text{L}(\Delta,T+\Delta T/2) + \delta\Delta_\text{R}(\Delta,T-\Delta T/2),
	\label{eq:delta_tilde}
\end{equation}
and it may change upon inverting the temperature bias ($\Delta T \to -\Delta T)$. 
In the presence of a Lamb shift, $R$ is still given by Eq.~(\ref{eq:r_general}) provided that we replace $\Delta \to \tilde{\Delta}(\Delta T)$.

In general, we notice that the renormalization terms $ \delta\Delta_\alpha(\Delta,T_\alpha)$ is of the same order in the coupling strength as the rates
 $\Upsilon_\alpha(\epsilon,T_\alpha)$ (which are evaluated at leading order in the coupling). Therefore, if the rates $\Upsilon_\alpha(\epsilon,T_\alpha)$ are 
 smooth functions of $\epsilon$, their variation due to the Lamb shift will be beyond leading order in the coupling strength. The effect of the Lamb shift on rectification 
 is thus negligible in the weak coupling regime when the spectral density of the baths is a smooth function of the energy (on the $\hbar\Upsilon_\alpha$ scale). 
On the contrary, the Lamb shift may become relevant for rectification whenever there is a strong energy dependence in $\Upsilon_\alpha(\epsilon,T_\alpha)$, 
for example, if the density of states of the baths has a gap. As we will show in detail in the following, even a small renormalization of the gap can have a large 
impact on the current.

\begin{figure}[!tb]
	\centering
	\includegraphics[width=0.99\columnwidth]{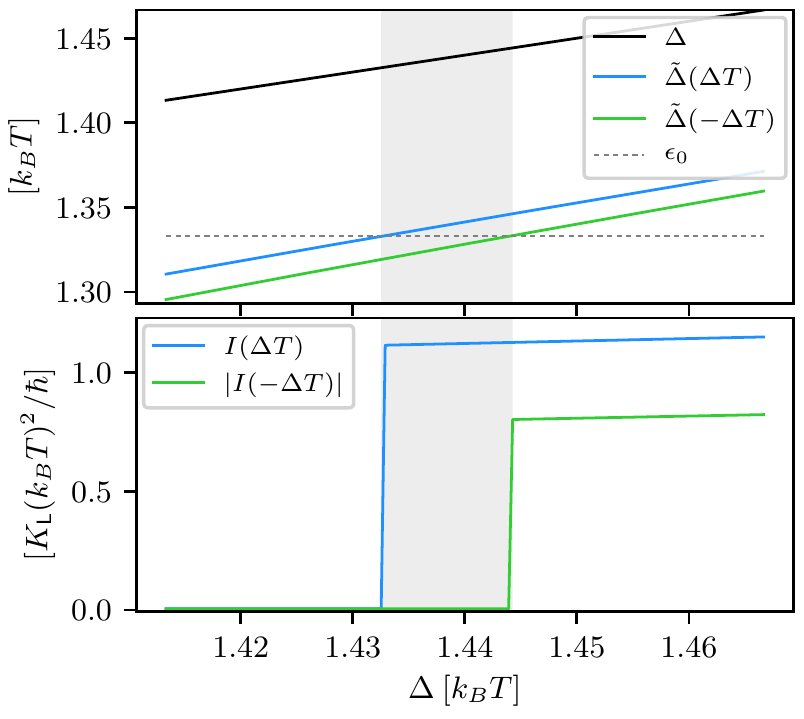}
	\caption{Upper panel: the bare gap $\Delta$, the renormalized gap $\tilde{\Delta}(\Delta T)$ in the positive bias case and the renormalized gap 
	$\tilde{\Delta}(-\Delta T)$ in the negative bias case, as a function of the bare gap $\Delta$. The dashed gray line corresponds to the gap $\epsilon_0$ 
	in the density of states of the baths, while the region highlighted in gray shows where the renormalized gaps are respectively larger and smaller than 
	$\epsilon_0$. Lower panel: the heat currents $I(\Delta T)$ and $|I(-\Delta T)|$. In the highlighted region we have perfect rectification (up to higher 
	order corrections in the coupling strength). The parameters are:  $K_\text{R}/K_\text{L} = 5$, $\epsilon_{\rm C} = (20/3)\,k_{\rm B}T$, $\epsilon_0 = (4/3)\, k_{\rm B}T$ 
	and $\Delta T/T = 2/3$. Note that the current is plotted in units of $(k_BT)^2/\hbar$ and, once the ratio $K_\text{R}/K_\text{L}$ is fixed, scales with $K_{\rm L}$.}
\label{fig:lamb_shift}
\end{figure}

We consider two bosonic  baths with a cutoff frequency $\epsilon_{\rm C}$ and a gap in the density of states  $\epsilon_0$,
\begin{equation}
	\Upsilon_\alpha(\epsilon,T_\alpha) =
	 \frac{\pi}{\hbar}\, K_\alpha\, \theta(\epsilon-\epsilon_0)\, \epsilon \, e^{-\epsilon/\epsilon_{\rm C}} \coth{[\epsilon/(2k_{\rm B}T_\alpha)]},
\end{equation}
where  
$\theta(\epsilon)$ is the Heaviside function and $K_\alpha$ is the dimensionless Ohmic coupling strength introduced in Eq.~(\ref{eq:negf_spectral}).
In the upper panel of Fig.~\ref{fig:lamb_shift} we show the bare gap $\Delta$ (black curve), the 
renormalized gap $\tilde{\Delta}(\Delta T)$ for the positive bias case (blue curve), and the renormalized gap $\tilde{\Delta}(-\Delta T)$ for the negative 
bias case (green curve), as a function of the bare gap $\Delta$. The renormalized gaps are different in the positive and negative bias cases. In particular, in the 
highlighted region $\tilde{\Delta}(-\Delta T)$ is inside the gap, i.e. it is smaller than $\epsilon_0$ (dashed gray line), while $\tilde{\Delta}(\Delta T)$ is outside the gap. 
We therefore expect a finite heat current in the latter case, and a zero heat current in the former. This is confirmed in the lower panel of Fig.~\ref{fig:lamb_shift} 
where $I(\Delta T)$ and $|I(-\Delta T)|$ are plotted as a function of the bare gap. The heat currents are computed using Eq.~(\ref{eq:heat_currrents_weak}) with 
$\Delta \to \tilde{\Delta}(\Delta T)$ to account for the Lamb shift. The highlighted region is the one in which perfect rectification is possible. 

This is however an ideal situation. The inclusion of higher order effects in the coupling strength (for example co-tunneling effects) will reduce the rectification. 
Indeed, the perfect rectification visible in the grey region in the lower panel of Fig.~\ref{fig:lamb_shift} is a consequence of the current $I(-\Delta T)$ being directly 
proportional to the density of states, therefore exactly zero for $\tilde\Delta(-\Delta T)<\epsilon_0$. On the other hand, higher order effects create small yet finite 
currents even in this parameter range. Nonetheless, it is useful to have identified  a mechanism to enhance rectification exploiting the Lamb shift.

\section{$U=\infty$ - Beyond weak coupling}
\label{sec:qubit_strong}
As shown in the previous section, in the weak-coupling regime there are bounds to the rectification coefficient $R$. In this section, we show that some of 
these bounds can be overcome by increasing the coupling between the system and the baths. We will see that quantum coherence can be beneficial for rectification. 
From the point of view of the transmission function, see Fig.~\ref{fig:scheme}, going beyond the weak coupling regime allows us to consider also the effect of a width shift, which was neglected in the sequential tunneling regime. 
For concreteness we focus on bosonic baths and consider three different approaches.
First, we include co-tunneling contributions to the heat current in the sequential tunneling regime derived from the master equation.
The importance of considering co-tunneling resides not only  in improving the analysis compared to the weak-coupling regime. It is also an important guide to interpret the results derived with other two methods we employ.
These consist first of all in an approach based on non-equilibrium Green's function theory (NEGF); secondly
in a formal exact solution for the heat current valid for general spectral densities and coupling strengths 
derived within the Feynman-Vernon path-integral approach. 
For  Ohmic baths we consider the special strong coupling condition characterized by  $K_R+K_L=1/2$ where the heat current is derived in closed form.
This solution, which extends to the non-equilibrium case the Toulouse limit of the spin-boson model, 
also provides a benchmark for the non-perturbative non-equilibrium Green's function results. In addition, it
holds beyond the non-adiabatic regime treated in the NIBA~\cite{segal2005,agarwalla,boudjada2014}.

We will mainly consider two different couplings: 
the ``XX coupling'', where both left and right baths are coupled to the system through $\sigma_x$, i.e. 
$\theta_\text{L}=\theta_\text{R}=\pi/2,\;\phi_{\rm L}=0,\; \phi_{\rm R}=0$, and the ``XY coupling'', i.e. $\theta_\text{L}=\theta_\text{R}=\pi/2,\;\phi_{\rm L}=0,\; \phi_{\rm R}=\pi/2$. 
 Since the XX and XY couplings only differ by the angle $\phi_\alpha$ [see Eq.~(\ref{eq:h_int_generic2})], both cases display identical rectification within 
 the weak coupling regime (see Section~\ref{subsubsec:weak_sigma}). As we will see, this property is violated when going beyond the weak coupling regime, 
 signaling the effect of higher order coherent quantum effects. 
Heat transport in the   $\Delta\rightarrow 0$ limit will be studied for 
 arbitrary spin coupling as defined in Eq.~(\ref{eq:h_int_generic2}).
 This limiting case exhibits  vanishing heat current in the sequential tunneling limit. Hence, thermal current and thermal rectification becomes solely due to higher order processes.

We proceed by considering first co-tunneling processes  (in Section~\ref{subsec:me_cot}), and then applying the NEGF method in Section~\ref{subsec:negf}.
The exact results, based on Feynman-Vernon path integral approach, will be introduced in Section~\ref{subsec:exact}, while its impact on rectification will be discussed in 
Section~\ref{subsec:r_strong}. The results for arbitrary couplings in the $\Delta\rightarrow 0$ limit are presented at the end of the Section.

\subsection{Co-tunneling processes}
\label{subsec:me_cot}
Both for the XX and XY couplings, only elastic co-tunneling processes contribute. These are processes that coherently transfer an excitation from one bath to 
the other (via a virtual state) without changing the state of the qubit, and thus without affecting the ME itself. The reason why co-tunneling processes can 
only be elastic comes from the fact that in a co-tunneling process the coupling Hamiltonian (which changes the state of the qubit) is applied twice. Therefore the 
two-level system is brought back to the initial state.

The heat current, including both sequential and co-tunneling processes,
can be expressed as (see App.~\ref{app:cotunn} for details of the calculation)
\begin{equation}
	I(\Delta T) = I^\text{seq}(\Delta T) + I^\text{cot}(\Delta T),
	\label{eq:j_cot}
\end{equation}
where $I^\text{seq}(\Delta T)$ is the heat current relative to the sequential regime, given by Eq.~(\ref{eq:heat_currrents_weak}), and 
\begin{multline}
	I^\text{cot}(\Delta T) = \int_0^\infty \frac{d\epsilon}{2\pi\hbar}\epsilon\;\Gamma_\text{L}(\epsilon)\Gamma_\text{R}(\epsilon)\\
\times\left|\frac{1}{\Delta+\epsilon+i\eta}\pm \frac{1}{\Delta-\epsilon+i\eta}\right|^2 \left[n_\text{R}(\epsilon)-n_\text{L}(\epsilon)\right]
\label{eq:cot}
\end{multline}
is the contribution due to co-tunneling, where $\eta$ is an infinitesimal positive quantity, and $n_\alpha(\epsilon)$ is the Bose-Einstein distribution 
relative to bath $\alpha$. The plus sign in Eq.~(\ref{eq:cot}) refers to the XX coupling, while the minus sign to the XY coupling. 

Eq.~(\ref{eq:cot}) diverges logarithmically in the limit $\eta\to 0^+$, but the co-tunneling rates can be regularised as discussed extensively in the literature, see 
Refs.~\onlinecite{turek,Kaasbjerg2016,bibek,paolo}. Assuming that the two-level system is in the ground state, the first term inside the square modulus of Eq.~(\ref{eq:cot}) 
arises from virtual transitions of an excitation from one bath to the qubit, and then from the qubit to the other bath (see App.~\ref{app:cotunn}). The second term instead 
arises from the (virtual) process in which excitations are created both in one bath and in the qubit, and then by destroying an excitation in the qubit and in the other bath. 
The choice of the  XX and XY couplings produce opposite interference effects between these two processes. If we had neglected the ``counter rotating'' terms in 
$\mathcal{H}_{\alpha ,\text{S}}$ [Eq.~(\ref{eq:h_int_generic2})], the second term inside the square modulus would have vanished and the co-tunneling 
rates would have become the same in the XX and XY cases. 

Crucially, since in Eq.~(\ref{eq:cot}) the temperatures only enter through the Bose-Einstein distributions, $I^\text{cot}(\Delta T)$  is an \textit{anti-symmetric} 
function, i.e. $I^\text{cot}(-\Delta T) = -I^\text{cot}(\Delta T)$. Therefore the contribution of co-tunneling to the heat current is the same both for the positive and negative bias case. 
The impact of co-tunneling on rectification can be easily appreciated by plugging Eq.~(\ref{eq:j_cot}) into Eq.~(\ref{eq:r_def}):
\begin{equation}
    R = \frac{I^\text{seq}(\Delta T)+I^\text{seq}(-\Delta T)}{I^\text{seq}(\Delta T)-I^\text{seq}(-\Delta T) +2 I^\text{cot}(\Delta T)},
    \label{eq:r_cot}
\end{equation}
where we fixed $\Delta T >0$, so that $I(\Delta T)>0$ and  $I(-\Delta T)<0$.
Notably, co-tunneling only appears in the denominator of Eq.~(\ref{eq:r_cot}). 
Defining $R^\text{seq} = [I^\text{seq}(\Delta T)+I^\text{seq}(-\Delta T)]/[I^\text{seq}(\Delta T)-I^\text{seq}(-\Delta T)]$, we see that if $I^\text{cot}(\Delta T) < 0$, then 
\begin{equation}
    |R| > |R^\text{seq}|,
\end{equation}
($|R|<|R^\text{seq}|$ if $I^\text{cot}(\Delta T) > 0$). 
Therefore, in the presence of sequential tunneling, co-tunneling can enhance rectification despite being an elastic process 
which would induce no rectification on its own.

Interestingly, as discussed below, the co-tunneling contribution 
$I^\text{cot}(\Delta T)$ is usually \textit{negative} when sequential tunneling dominates. 
In the weak coupling regime (i.e. when $\hbar\Upsilon$ is the smallest energy scale), sequential tunneling dominates when $\Delta$ is of the order of $k_BT$. 
In the presence of sequential tunneling processes only, the transmission function ${\cal T}(\epsilon,T,\Delta T)$ in Eq.~(\ref{eq:j_green_prop}) can be thought of as an infinitely narrow function in $\epsilon$ peaked at the resonant condition.
The co-tunneling contributions broadens the transmission function, as qualitatively illustrated in Fig.~\ref{fig:scheme}(c). 
As the width of the transmission function increases, the weight of ${\cal T}(\epsilon,T,\Delta T)$ moves from its peak to its tails. 
Therefore, where sequential processes dominate, i.~e. around the peak, co-tunneling contributions \textit{decrease} the heat flow. On the other hand, where sequential tunneling is suppressed, 
i.~e. in the tails of the transmission function, co-tunneling \textit{increases} the heat flow.

\subsection{Non-equilibrium Green's function method}
\label{subsec:negf}
In this section we will employ the NEGF method to compute heat currents. It is convenient to first express spin operators in a Majorana representation~\cite{liu,schad,schad2}
\begin{equation}
	\sigma_a = -\frac{i}{2}\sum_{bc= x,y,z}\epsilon_{abc}\, \eta_b\eta_c,
	\label{spintomaj}
\end{equation}
where $\epsilon_{abc}$ is the Levi-Civita symbol, and $\eta_a$ denotes three Majorana fermion operators (they satisfy the anti-commutation relation 
$\{\eta_a,\eta_b\} = 0$ for $a\neq b$, $\eta_a^2 = 1$ and $\eta_a = \eta_a^\dagger$). The system and coupling Hamiltonians [see Eqs.~(\ref{eq:def_hq}) 
and (\ref{eq:h_int_generic2})] therefore become (up to an irrelevant additive constant)
\begin{equation}
\begin{aligned}
	\mathcal{H}_{\infty} &= -i\frac{\Delta}{2}\eta_{x}\eta_{y}, \\
	\mathcal{H}_{\alpha, \text{S}} &= -\frac{i}{2}\sum_{abc} u_{\alpha,a} \epsilon_{abc} \,\eta_b\eta_c \otimes \sum_k V_{\alpha k}(b_{\alpha k}+ b^\dagger_{\alpha k}),
\end{aligned}
\label{hamilmaj0}
\end{equation}
where the indices  $a$, $b$ and $c$ run over $x$, $y$ and $z$ in the sum.
The advantage is that in this representation the system Hamiltonian is quadratic, while the non-linearity is transferred to the coupling term. In the Majorana representation 
the system-bath coupling gives us the non-linear effects that, as we discussed, are necessary to observe rectification.

Assuming that the spectral densities of the two baths are proportional,  the heat current is given by Eq.~(\ref{eq:j_green_prop}) where $\mathcal{T}(\epsilon,T,\Delta T)$, 
in general, must be computed numerically. However, we are able to find an analytic expression for the transmission function by solving the Dyson equation for the 
Green's functions with an expression for the self energy expanded to leading order in the coupling Hamiltonian $\mathcal{H}_{\alpha,\text{S}}$ in Eq.~(\ref{hamilmaj0})
(see App.~\ref{app:negf} for details). In the XX coupling case, this method leads to
\begin{multline}
	\mathcal{T}_\text{XX}(\epsilon, T, \Delta T)=
	\\
	\frac{4\,\Delta^2\Gamma_\text{L}(\epsilon)\Gamma_\text{R}(\epsilon)}{\left(\epsilon^2-2\epsilon\left(\delta {\Delta}_\text{L}(\epsilon,T_\text{L})+
	\delta{ \Delta}_\text{R}(\epsilon,T_\text{R})\right)-\Delta^2\right)
	^2+\xi^2(\epsilon)}
\label{xxtra}
\end{multline}
where $\xi (\epsilon)=\epsilon\sum_\alpha{\Gamma}_\alpha(\epsilon)[1+2n_\alpha(\epsilon)]$, and $\delta\Delta_\alpha(\epsilon,T_\alpha)$, which describes the Lamb shift 
induced by bath $\alpha$, is defined in Eq.~(\ref{eq:s}) with $\Upsilon_\alpha(\epsilon^\prime,T_\alpha) = \Gamma_\alpha(\epsilon^\prime)\coth{[\epsilon^\prime/(2k_\text{B}T_\alpha)]}$ [Eq.~(\ref{eq:similar_rates}) with Eq.~(\ref{eq:bosonic_tunnel_rates})]
\cite{footnote:majorana}. 

\begin{figure}[t]
	\centering
	\includegraphics[width=0.9\columnwidth]{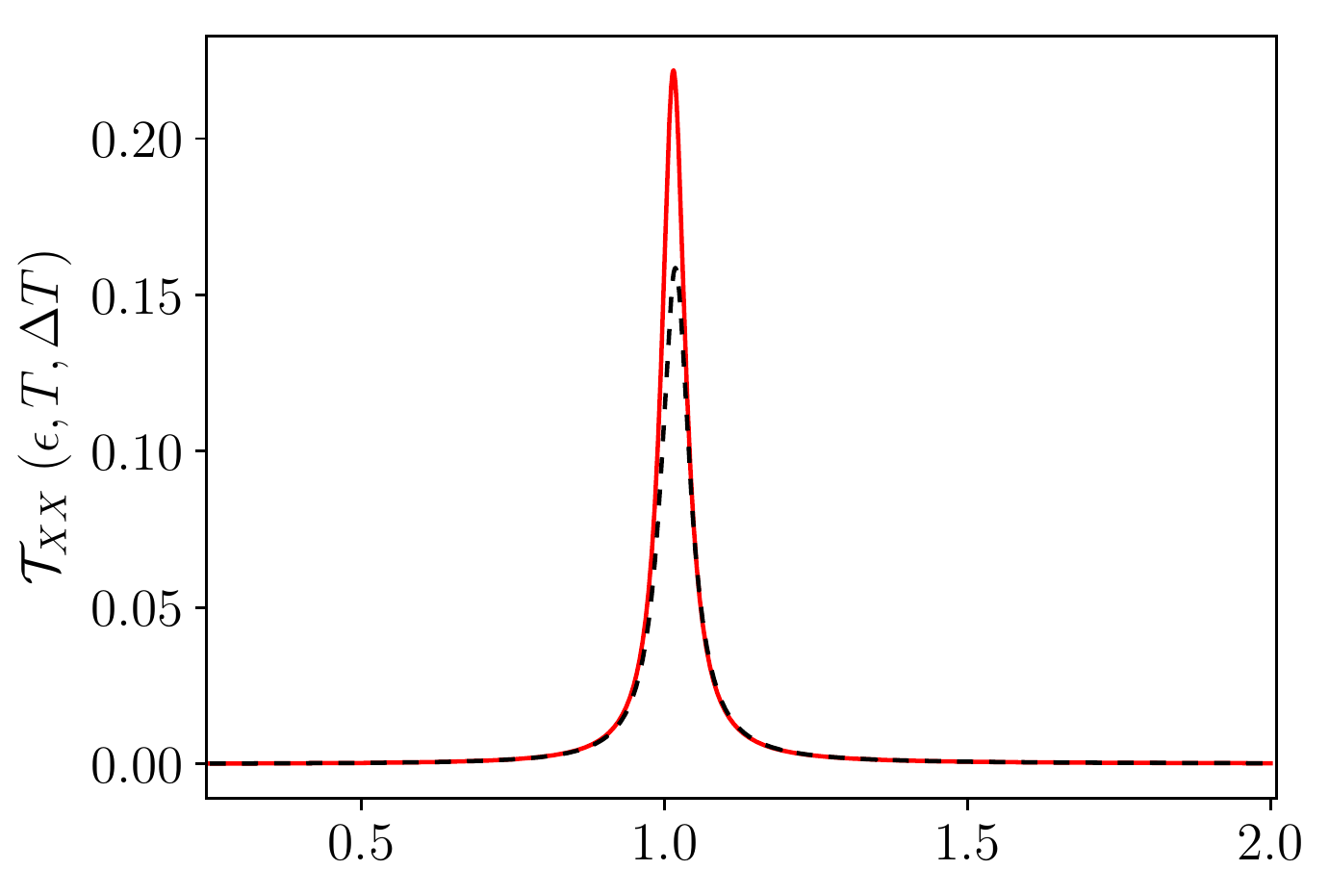}
	\includegraphics[width=0.9\columnwidth]{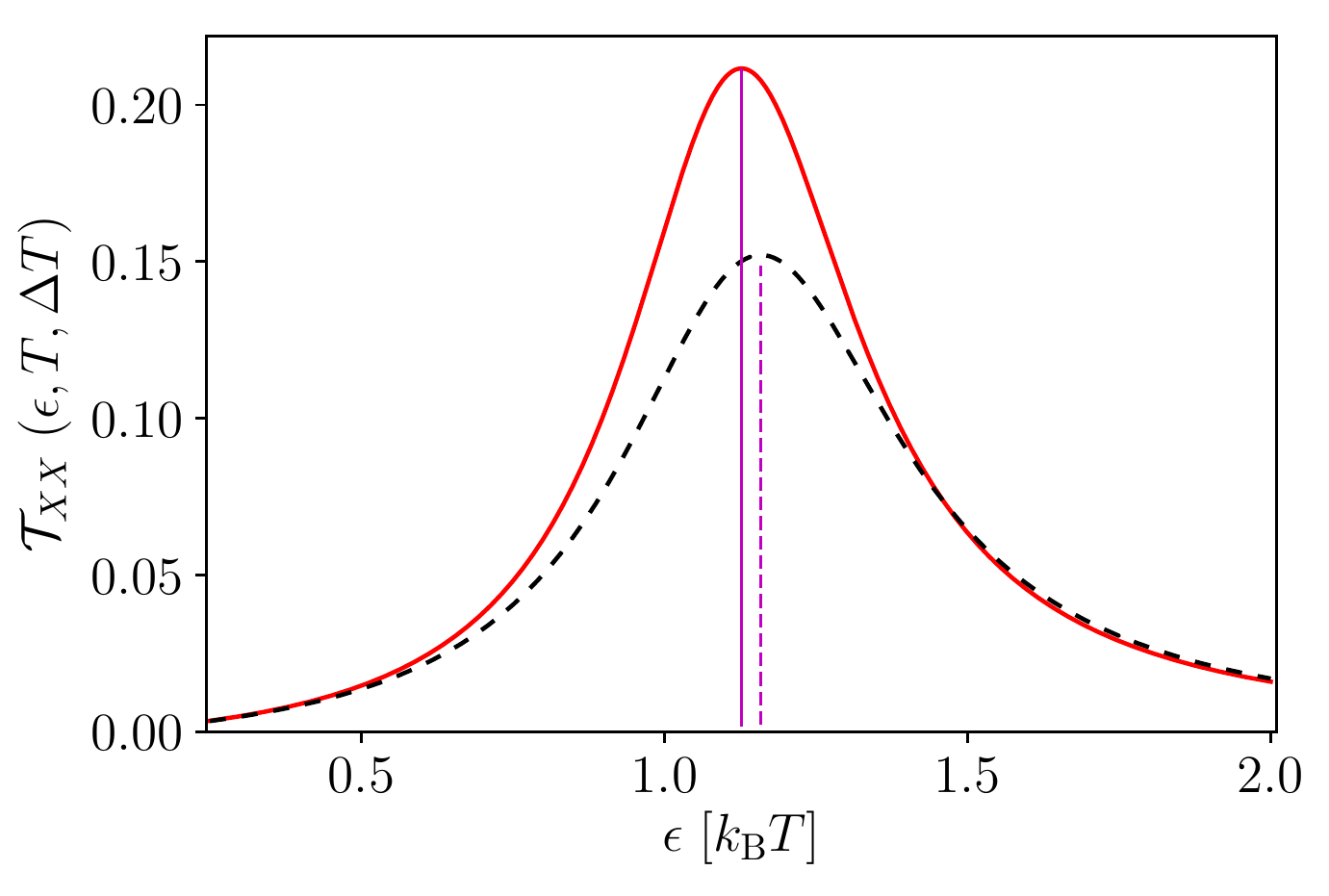}
\caption{Transmission function $\mathcal{T}_{XX}(\epsilon,T,\Delta T)$ as a function of energy $\epsilon$ for $K_L=0.006$ and $K_R=0.003$ (top panel); $K_L=0.06$ and $K_R=0.03$ (bottom panel).
Red solid (black dashed) curves are for positive (negative) bias, while the vertical lines are guides for the eye.
The other parameters are $\Delta = k_{\rm B}T$, $\epsilon_{\rm C}=6 k_{\rm B}T$ and the temperature difference is $\Delta T = 0.6 T$.}
	\label{figtrans}
\end{figure}

Instead, in the XY coupling case we find
\begin{equation}
\mathcal{T}_\text{XY}(\epsilon, T, \Delta T)=\frac{4\,\epsilon^2\,\Gamma_\text{L}(\epsilon)\Gamma_\text{R}(\epsilon)}{\left(\epsilon^2-\mathcal{X}(\epsilon)-\Delta^2\right)^2+\mathcal{Y}^{2}(\epsilon)},
\label{xytra1}
\end{equation}
where 
\begin{multline}
\mathcal{X}(\epsilon)=2\epsilon\left[\delta{\Delta}_\text{L}(\epsilon, T_\text{L})+\delta{\Delta}_\text{R}(\epsilon, T_\text{R})\right)+\left(1+2n_\text{L}(\epsilon)\right]\\
\times \left[1+2n_\text{R}(\epsilon)\right]\Gamma_\text{L}(\epsilon)\Gamma_\text{R}(\epsilon)-4\delta{\Delta}_\text{L}(\epsilon,T_\text{L})\delta{\Delta}_\text{R}(\epsilon,T_\text{R}),
\label{x1}
\end{multline}
and
\begin{equation}
\mathcal{Y}(\epsilon)=\sum_{\substack{\alpha,\beta = \text{L},\text{R} \\ \alpha\neq \beta}}\left[2\delta{\Delta}_{ \alpha}(\epsilon,T_\alpha)-\epsilon\right]\left[1+2n_{ \beta}(\epsilon)\right]\Gamma_{\beta}(\epsilon).
\label{x2}
\end{equation}
As shown in App.~\ref{app:negf}, this approach provides results which are more accurate with respect to the ME approach 
also including co-tunneling processes, since it 
contains higher order processes beyond sequential and co-tunneling thanks to the implicit re-summation performed by solving the Dyson equation.
As shown in Fig.~\ref{figtrans}, indeed the transmission function (\ref{xxtra}) includes height (top panel) and position (bottom panel) shifts effects bridging between the
sequential and sequential-plus-co-tunneling regimes illustrated in the previous sections. Red solid curves refer to the positive bias, while black dashed curves refer to negative bias. The vertical lines in the bottom panel are guides to the eye to highlight the Lamb-shift.
A discussion of the results deriving 
from this formulation will be deferred to Section~\ref{subsec:r_strong} in order to allow for a comparison between the different approaches.

\subsection{Exactly solvable case - $K_\text{R}+K_\text{L} =1/2$}
\label{subsec:exact}

By applying the Feynman-Vernon path-integral approach to the spin-boson problem~\cite{weiss},
rectification can be computed exactly for the XX coupling condition.
This case 
will also serve as a benchmark for approximate studies in the non-perturbative regimes.
The exact formal expression for the heat current Eq.~(\ref{eq:j_green}) for generic spectral densities 
of the two baths
having the same energy dependence, i.e. $\Gamma_\text{L} (\epsilon) \propto \Gamma_\text{R}(\epsilon)$,
takes the form of Eq.~(\ref{eq:j_green_prop}) with $\mathcal{T}(\epsilon,T,\Delta T)$ replaced by
\begin{equation}
\mathcal{T}_\text{XX}^{\rm (ex)}(\epsilon,T,\Delta T)=  2 \frac{\Gamma_\text{L}(\epsilon) \Gamma_\text{R}(\epsilon)}{ \Gamma_\text{L}(\epsilon) + 
\Gamma_\text{R}(\epsilon)} {\rm Im}\left[\chi(\epsilon)\right]
\label{eq:transmission_exact}
\end{equation}
(see Appendix~\ref{app:exact} for details).
In Eq.~(\ref{eq:transmission_exact}), $\chi(\epsilon)$ is the Fourier transform of the qubit dynamical susceptibility in the presence of the two baths, 
$\chi(t)= (i/\hbar) \Theta(t) \langle [ \sigma_x(t), \sigma_x(0)] \rangle $, given in Eq.~(\ref{chi-exact}).

Here we focus on Ohmic spectral densities, defined as in Eq.~(\ref{eq:negf_spectral}).
The dimensionless   coupling strength $K_\alpha$ enters  the exact expression of the dynamical susceptibility in a form which allows the path 
summation in analytic form when $K_\text{R}+K_\text{L} =1/2$, 
analogously  to the $K=1/2$ regime of the spin-boson model, corresponding to the Toulouse limit of the anisotropic Kondo problem  \cite{weiss,sassetti1990}, see Eq.~(\ref{chi-ohmic}). 
We obtain
\begin{multline}
	\chi(t)=\frac{2\Delta^2}{\hbar^3\gamma}\Theta(t)e^{-\gamma t/2}\int_0^\infty d\tau P(\tau) \\
	\times \left[e^{-\gamma |t-\tau|/2}-e^{-\gamma |t+\tau|/2}\right],
	\label{eq:chiTou}
\end{multline}
where $\gamma=\pi\Delta^2/(2\hbar\,\epsilon_{\rm C})$ and 
\begin{equation}
	P(\tau) = \prod_\alpha\left(\frac{\epsilon_{\rm C}}{\pi k_{\rm B}T_\alpha}\sinh\left(\frac{\pi|\tau|k_{\rm B}T_\alpha}{\hbar}\right)\right)^{- 2 K_\alpha}.
\end{equation}
We note that $\chi(t)$ takes the same form of the spin-boson model at $K=1/2$ with the only difference that
the bath-induced (dipole or intra-blip, see App.~\ref{app:exact}) interactions involving the two baths enter $P(\tau)$
in factorized form. When $K_\text{L}= K_\text{R} =1/4$ and $\Delta T=0$ we recover the susceptibility  at the Toulouse point and the heat current trivially vanishes. 

In order to evaluate the rectification, the heat current (\ref{eq:j_green_prop}), is more
conveniently written by substituting Eqs.~(\ref{eq:transmission_exact}) and (\ref{eq:chiTou}), and reads
\begin{equation}
I = \frac{1}{\hbar}\frac{K_\text{L} K_\text{R}}{K_\text{L}+ K_\text{R}} \Im \int_{-\infty}^{+\infty}dt\, \chi(t)F(-t),
\label{eq:gm_t_1}
\end{equation}
where
\begin{multline}
	F(-t)=  (k_{\rm B} T_\text{R})^3 \psi^{(2)}\left(1 + \frac{k_{\rm B} T_\text{R}}{\epsilon_\text{C}} \left(1-i \frac{ \epsilon_\text{C}\,t}{\hbar}\right)\right) \\
	 -(k_{\rm B} T_\text{L})^3 \psi^{(2)}\left(1 + \frac{k_{\rm B} T_\text{L}}{\epsilon_\text{C}} \left(1-i \frac{ \epsilon_\text{C}\,t}{\hbar}\right)\right)
\label{eq:gm_t}
\end{multline}
and $\psi^{(2)}(z)$ denotes the second derivative of the digamma function.
The resulting exact form of the heat rectification  includes all possible heat transfer processes. 
In the following we will evaluate it explicitly by numerical integration  of the current expressed as in Eq.~(\ref{eq:gm_t_1}).

\subsection{Rectification coefficient}
\label{subsec:r_strong}
In this subsection we show that the general properties and bounds derived in Section~\ref{sec:weak_coupling} can be overcome in the strong-coupling regime, allowing the system to enhance rectification.
Furthermore, we will also identify the effect of higher order coherent transport processes on rectification. We will consider Ohmic spectral density, as in 
Eq.~(\ref{eq:negf_spectral}), for both baths. 
\begin{figure}[t]
	\centering
	\includegraphics[width=0.9\columnwidth]{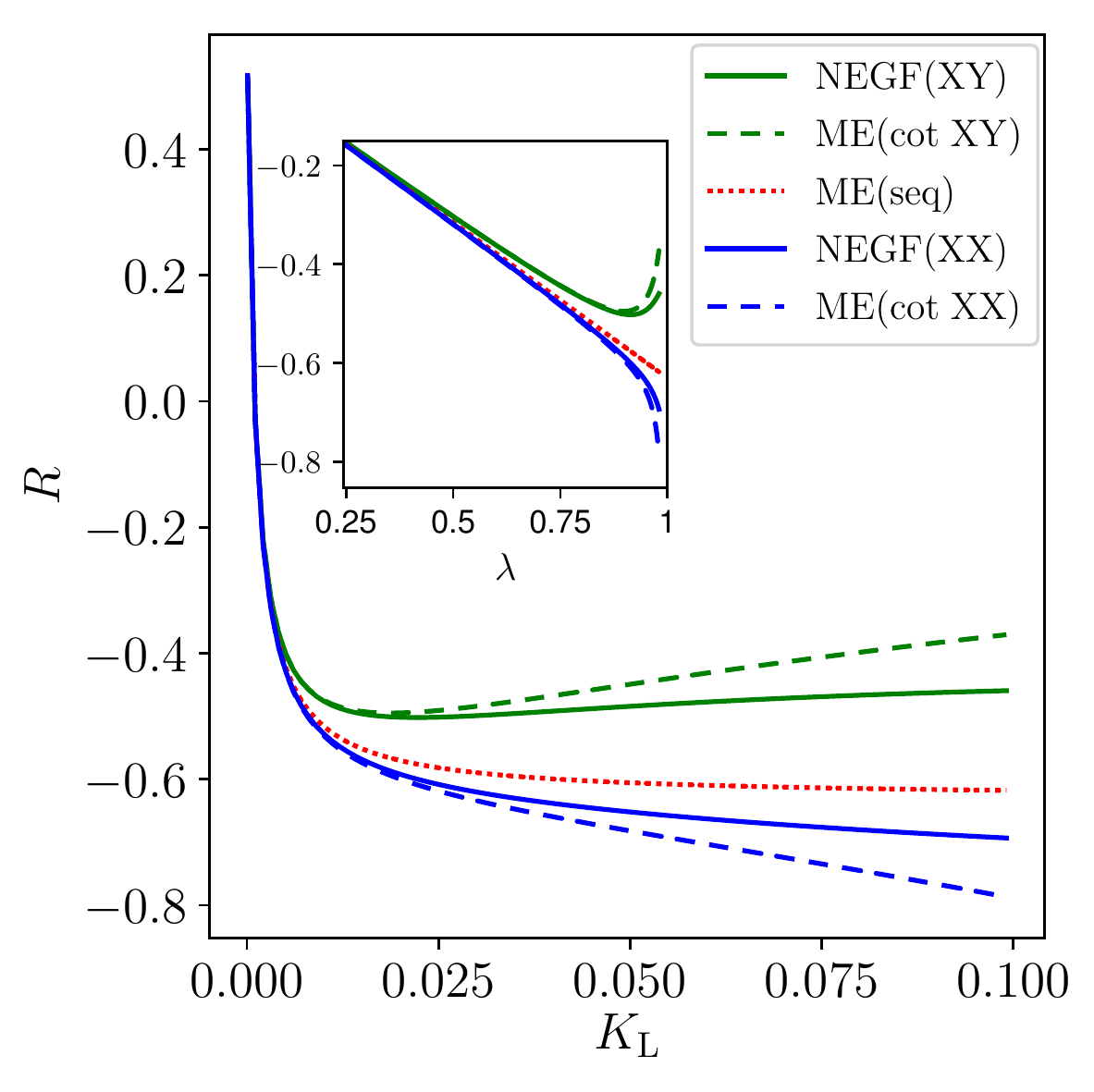}
\caption{Rectification coefficient $R$, computed with the three methods described in the legend, plotted as a function of $K_\text{L}$ both for the XX and XY couplings [the ME(seq) case is the same for both couplings]. The parameters are $K_\text{R} = 0.005$, $\Delta = 0.8k_\text{B}T$, $\epsilon_\text{C} = 10 k_\text{B}T$ and $\Delta T/T = 8/5$. We denote with ``NEGF'' the calculations performed with the non-equilibrium Green's function method described in Section~\ref{subsec:negf}, with ``ME(cot)'' those performed with the master equation which includes co-tunneling described in Section~\ref{subsec:me_cot}, and with ``ME(seq)'' the calculations performed in the weak coupling limit as described in Section~\ref{sec:weak_coupling}. The inset shows the same points plotted as a function of $\lambda$ for $\lambda \in [0.25,1]$. We neglect the Lamb shift in this plot.}
	\label{fig:r_gamma}
\end{figure}

\begin{figure}
	\centering
	\includegraphics[width=\columnwidth]{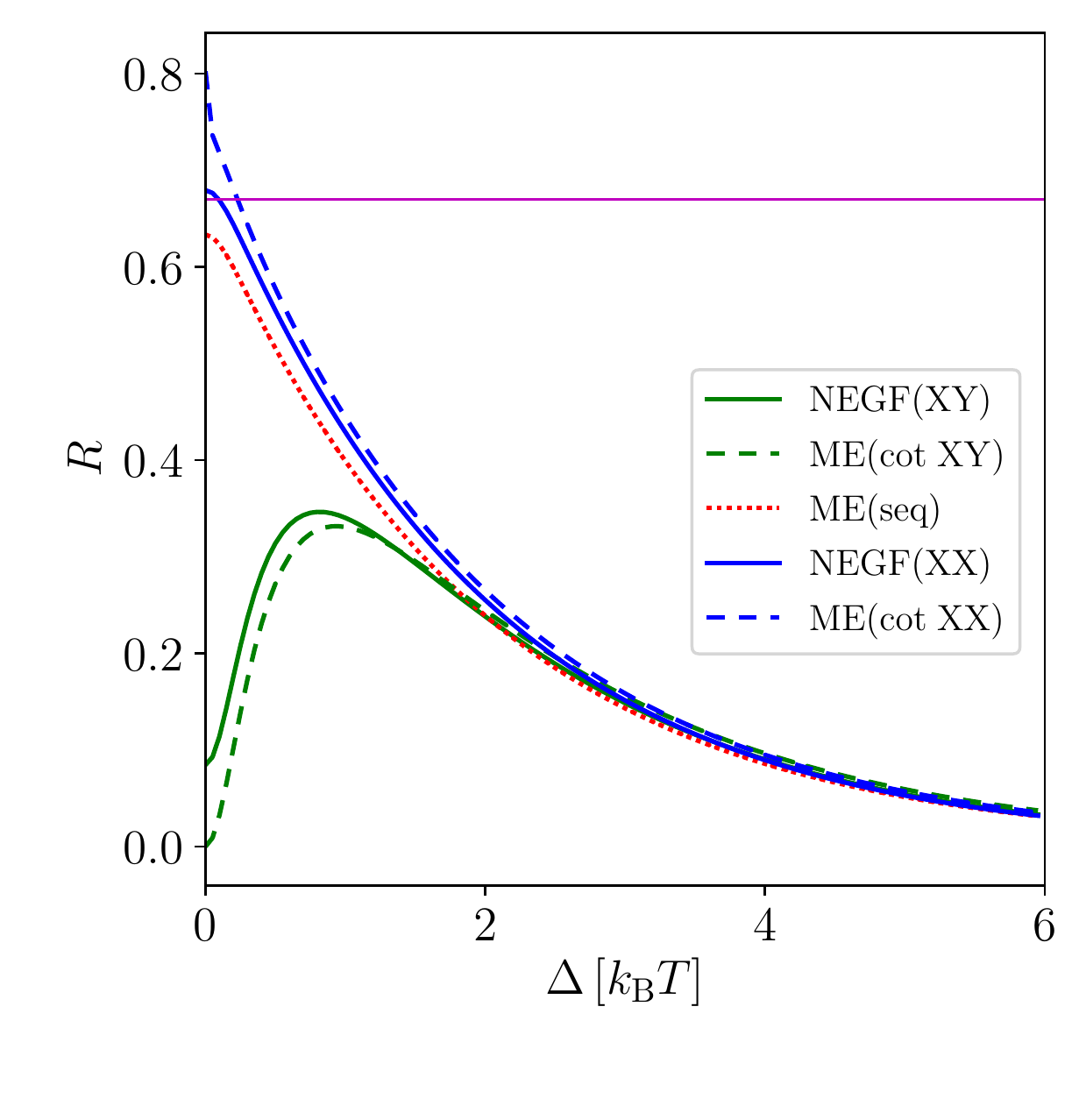}
\caption{Rectification coefficient $R$, computed with the three methods described in the legend, plotted as a function of the qubit gap $\Delta$. The parameters are $K_\text{L} = 0.006$, $K_\text{R}=0.03$, $\epsilon_\text{C} = 10 k_\text{B}T$ and $\Delta T/T = 1.9$. 
The horizontal magenta line shows $|\lambda| =\left|( K_{\rm L}-K_{\rm R})\big/( K_{\rm L}+K_{\rm R})\right|=0.67$. We neglect the Lamb shift in this plot. }
	\label{fig:r_delta}
\end{figure}

In Fig.~\ref{fig:r_gamma} we plot $R$ as a function of $K_\text{L}$ in the XX and XY case comparing the NEGF calculation, the ME calculation including co-tunneling 
effects [ME(cot)] and the ME calculation in the weak coupling regime [ME(seq)]. The coupling constant $K_\text{R}$ is set to 0.005 and the temperatures are fixed. 
First we notice that for small values of $ K_\text{L}$, i.e. in the weak coupling limit, all curves coincide, as expected.
As $K_\text{L}$ increases, we notice 
that the NEGF and ME(cot) curves nicely agree up to $K_\text{L}\approx 0.025$, and then we see some deviations. Interestingly, we notice that the rectification 
obtained using NEGF and ME(cot) method is \textit{different} in the XX and XY cases, contrary to what is obtained using the ME(seq) method. Indeed, in 
Section~\ref{subsubsec:weak_sigma} we showed that, in the weak coupling regime, rectification only depends on the angle between the qubit ($\sigma_z$) and 
the coupling term. Higher order coherent processes, instead, are able to distinguish these different couplings, as they produce different interference effects 
[see the $\pm$ in Eq.~(\ref{eq:cot})]. Rectification is \textit{enhanced} in the XX coupling case thanks to higher order processes, while it 
is suppressed in the XY case (we will explain this behavior describing Fig.~\ref{fig:r_delta}).  In the inset of Fig.~\ref{fig:r_gamma} we plot the same points as a 
function of the asymmetry coefficient $\lambda = (K_\text{L}-K_\text{R})/(K_\text{L}+K_\text{R})$ [see Eq.~(\ref{eq:l_def})]. We recall that, in the weak coupling 
regime, we proved that $R$ is linear in $\lambda$ (see Section~\ref{sec:similar_bath}). Indeed, for small values of $\lambda$, the behavior is linear. Interestingly, 
the behaviour becomes non-linear for larger values of $\lambda$, which correspond to larger values of the coupling constant $K_\text{L}$. This non-linearity is yet 
another signature of higher order coherent processes.

In Fig.~\ref{fig:r_delta} we plot $R$, computed with the three methods described above, as a function of $\Delta$. The choice of  the values of the coupling constants 
has been made in order to show that  the NEGF and ME(cot) results, although qualitatively agreeing with each other, present quantitative deviations with respect to the 
ME(seq) calculation.
First of all we notice that for $\Delta > 2 k_{\rm B}T$ all methods predict similar values of $R$.
On the other hand, for $\Delta\leq 2 k_{\rm B}T$, rectification is stronger in the XX case (blue curves), while it is weaker in the XY case (green curves) 
as compared to the sequential tunneling result, consistently with Fig.~\ref{fig:r_gamma}.
This means that coherent processes can decrease (XY case) or increase (XX case) rectification.
This different behavior can be understood by recalling the discussion in Sec.~\ref{subsec:me_cot} and that the co-tunneling contributions depend on the coupling.
In particular, in the XY case $I^{\rm cot}(\Delta T)$ is positive and, according to Eq.~(\ref{eq:r_cot}), $R$ is suppressed with respect to the sequential result, while in the XX case $I^{\rm cot}(\Delta T)$ is negative and $R$ is increased.
We also notice that, for small $\Delta$, in the XX case the two terms inside the square modulus of Eq.~(\ref{eq:cot}) tend to cancel each other, resulting in a small co-tunneling contribution, while in the XY case co-tunneling remains finite.
At the same time, the heat current due to only sequential processes tends to zero as $\Delta/(k_\text{B}T)\to 0$ [see Eq.~(\ref{eq:heat_currrents_weak})].
These observations explain the large deviation between the ME(seq) curve and the XY case for small $\Delta$.
Moreover, we notice that in the XY case, thanks to higher order processes, the NEGF and ME(cot) are non-monotonous with respect to $\Delta$ (as discussed in 
Section~\ref{subsec:linear_coup}, $R$ is monotonous in $\Delta$ in the weak coupling regime).
Interestingly, the value of $R$ computed using the NEGF and ME(cot) methods in the XX case 
shows a violation of the general weak-coupling bound of Eq.~(\ref{eq:r_similar_bound}), i.~e. we find that $|R|>|\lambda| = 0.67$ (denoted with a horizontal magenta line 
in Fig.~\ref{fig:r_delta}).
Finally, according to expectations, we mention that the contributions of co-tunneling on the heat current gets smaller for decreasing $\Delta$ (corresponding to increasing temperature).
This, however, does not prevent $R$ from largely deviating with respect to $R^{\rm seq}$, since such deviations depend on the ratio between co-tunneling and sequential contributions to the current [see Eq.~(\ref{eq:r_cot})].
\begin{figure}[!htb]
	\centering
	\includegraphics[width= 0.9\columnwidth]{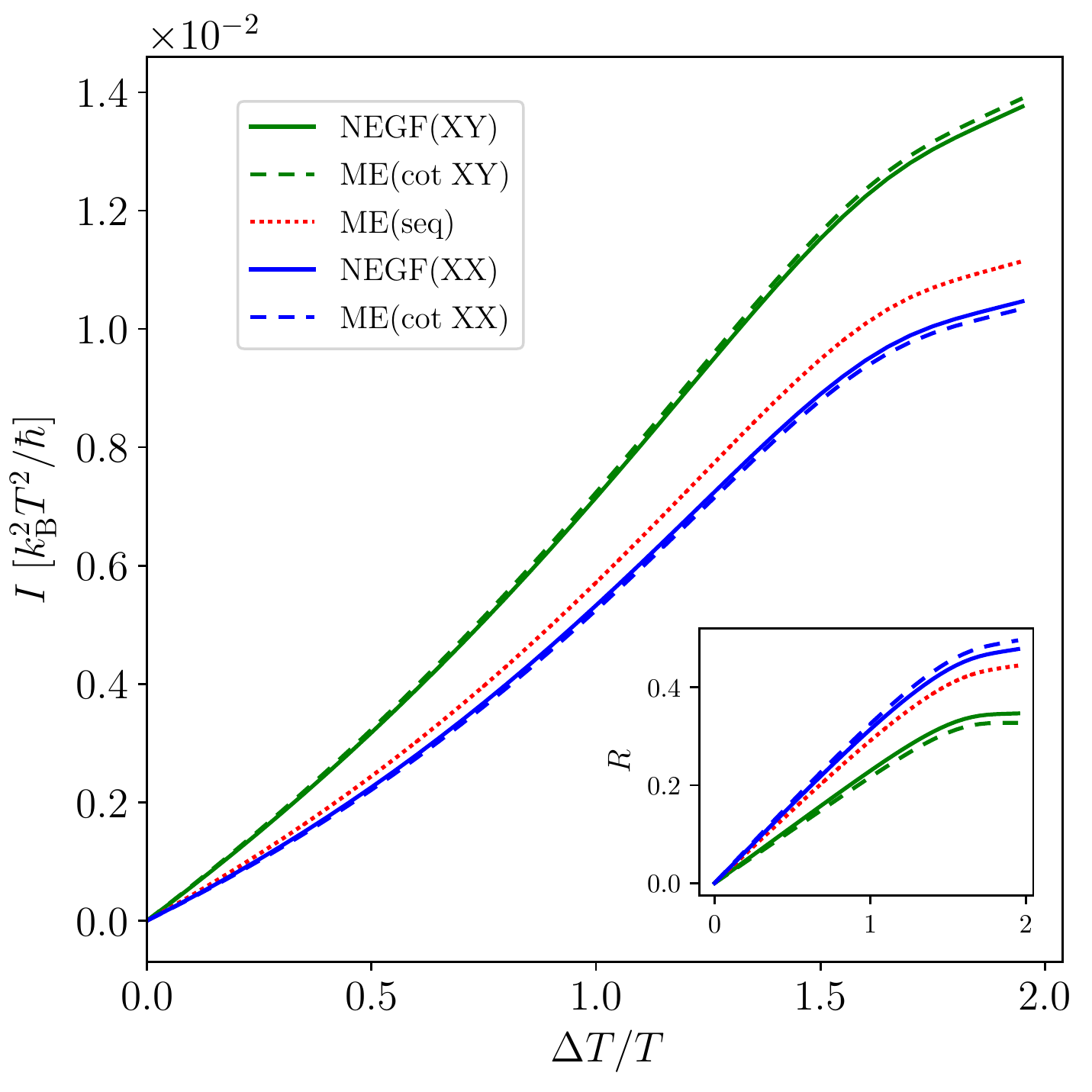}
	\vspace{0.05\columnwidth}
	\caption{Heat current and rectification (inset) as functions of the temperature bias $\Delta T$ for $\Delta=0.8 k_B T$ and all other parameters as in Fig.~\ref{fig:r_delta}. The color code of the curves is the same as for Figs.~\ref{fig:r_gamma} and \ref{fig:r_delta}.}
	\label{vsDT}
\end{figure}

We now study, in Fig.~\ref{vsDT}, the behavior of the currents, calculated with the three methods, and rectification (inset) as a  function of the temperature bias.
As far as the currents are concerned, all the curves show that the increase of $I$ with $\Delta T$ slightly deviates from the linear behavior already for small values of temperature bias, and all the curves have the same qualitative behavior.
We also notice that NEGF and ME(cot) methods give essentially the same results in both in the XX and XY cases.
In particular, the current in the XY (XX) case is larger (smaller), in the whole range of $\Delta T$ considered, with respect to the result obtained accounting for sequential processes only.
On the other hand, $R$ shows a nearly linear behavior up to $\Delta T/T=1$.
The curves relative to $R$ calculated with the two methods [NEGF and ME(cot)] show an increasing relative deviation with $\Delta T$, while the value of rectification reaches around 0.35 for the XY case and almost 0.5 for the XX case.
\begin{figure}[t]
	\centering
	\includegraphics[width=0.99\columnwidth]{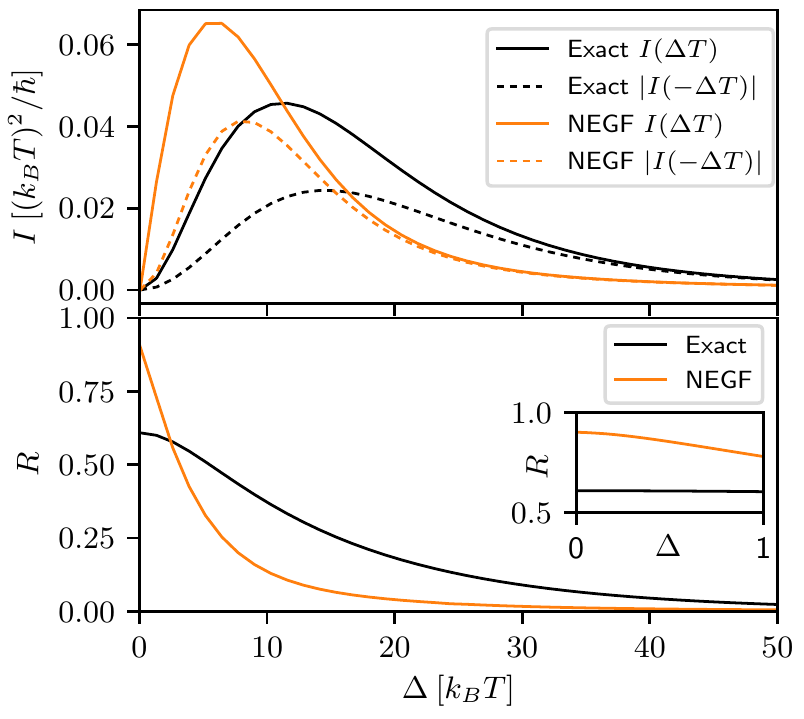}
\caption{Heat current (upper panel) and rectification coefficient (lower panel) plotted as functions of the qubit gap $\Delta$ in the strong coupling regime (in the XX coupling case) computed using the two methods indicated in the legend. The NEGF calculation includes the Lamb shift. The parameters are: $K_\text{L} = 0.49$, $K_\text{R} = 0.01$, $\epsilon_\text{C} = 100 k_\text{B}T$ and $\Delta T/T=1.9$.}
	\label{fig:exact_green}
\end{figure}

In Fig.~\ref{fig:exact_green} we plot the heat current (upper panel) and the rectification coefficient (lower panel) as a function of  $\Delta$. We compare the analytic results obtained in the XX case using the NEGF method including the Lamb-shift (see Section~\ref{subsec:negf}) with the exact calculation obtained using the Feynman-Vernon 
path integral approach (see Section~\ref{subsec:exact}). In doing so, we are constrained to fixing the coupling strength as $K_\text{L}+K_\text{R}=1/2$. The exact and NEGF calculations for the heat current, although in qualitative agreement, give quantitatively different results. The NEGF method tends to overestimate the magnitude of the heat current for values of $\Delta/(k_\text{B}T) \lesssim 10$, while it underestimates the heat current for larger values of $\Delta/(k_\text{B}T)$.
A similar trend is followed by the rectification coefficient: the NEGF calculation overestimates $R$, with respect to the exact one, when $\Delta/(k_\text{B}T) \lesssim 3$. For large values of $\Delta$, the rectification coefficient predicted by both methods tends to $0$.
Note that the qualitative agreement was not a priori expected, considering that the NEGF method  is perturbative in the coupling strength (valid, strictly speaking, only for $K_\text{L}$ and $K_\text{R} \ll 1$).

\begin{figure}[!tb]
	\centering
	\includegraphics[width=0.99\columnwidth]{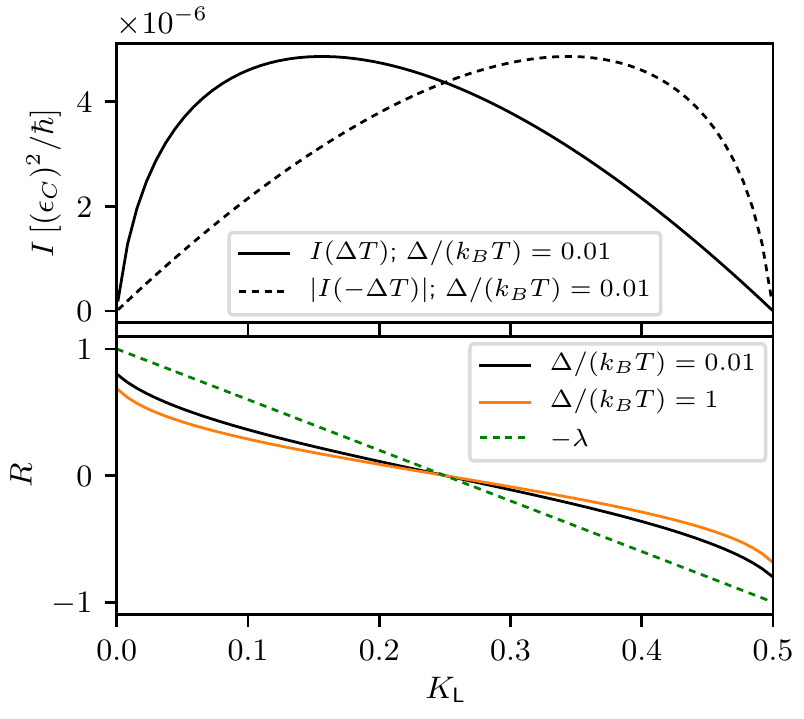}
\caption{Heat current and rectification, obtained with the exact calculation, as a function of $K_\text{L}$ setting $K_\text{R}=1/2-K_\text{L}$. The parameters not shown on the figure are $\Delta/\epsilon_\text{C}= 0.005$ and $\Delta T/T= 1.9$.}
	\label{fig:kl}
\end{figure}

In Fig.~\ref{fig:kl} we plot the heat current (top panel) and the rectification (bottom panel) obtained with the exact calculation as functions of the coupling constant $K_\text{L}$.
Because of the constraint $K_\text{R}=1/2-K_\text{L}$, here we can explore a very asymmetric coupling condition (one bath strongly coupled and the other weakly coupled), which cannot be treated by using approximate analytical approaches and it is difficult to address even numerically.
In the top panel the solid curve refer to positive bias, while the dashed curve refer to the negative bias, for a fixed value of $\Delta=0.01 k_BT$. Notice that the two curves are symmetric with respect to the point $K_\text{L}=1/4$.
Both currents vanish for $K_\text{L}=0$ and $K_\text{L}=1/2$, since no current can flow when one of the two coupling strengths is zero. The maximum occurs at around $K_\text{L}=0.15$ for the current relative to the positive bias. 
In the bottom panel the two solid curves of $R$ correspond to different values of $\Delta$. It is worth stressing that they differ by a little extent even though $\Delta$ spans two orders of magnitude. This is peculiar of the asymmetric coupling condition, with one bath
strongly coupled, and significantly differs from the behavior observed for smaller couplings reported in Fig.~\ref{fig:r_delta}. 
The rectification vanishes when the coupling strengths are equal,  $K_\text{L}=1/4$, and it is maximum for $K_\text{L}=0$ or 1/2, as expected from the fact that this is the most asymmetric situation.
We observe that the rectification fulfills the bound $|R|< |\lambda|$  [Eq.~(\ref{eq:r_similar_bound})] derived for the weak coupling, as shown by the green dashed curve which represents $(-\lambda)$ as a function of $K_\text{L}$.

\begin{figure}[!tb]
	\centering
	\includegraphics[width=0.99\columnwidth]{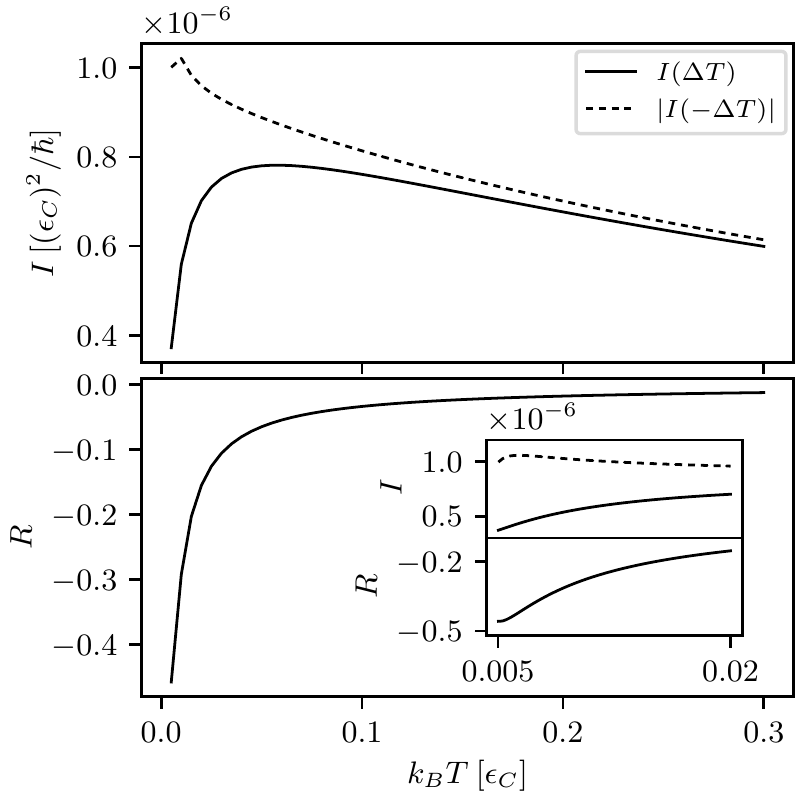}
\caption{Plot of the heat currents (top panel) and rectification (bottom panel) as functions of $k_BT$, obtained with the exact calculation.
The inset in the bottom panel shows a zoom of the curves in a range of small temperatures (from $k_BT = k_B\Delta T/2=0.005\epsilon_\text{C}$).
Rectification is clearly maximized at low temperatures. The parameters are: $K_\text{L}=0.49$, $K_\text{R} = 0.01$, $\Delta= 0.05\epsilon_\text{C}$ and $k_B\Delta T=0.01\epsilon_\text{C}$.}
	\label{fig:kl2}
\end{figure}

The dependence on the (average) temperature of the currents and rectification is analyzed in Fig.~\ref{fig:kl2} under asymmetric
coupling conditions ($K_L=0.49$, $K_R=0.01$) for fixed values of $\Delta T$ and $\Delta$.
As shown in the top panel, for $k_BT\gtrsim 0.05 \epsilon_\text{C}$ both currents (obtained with the exact calculation) decrease, as one can expect from the fact that the ratio $\Delta T/T$ decreases.
On the other hand, for the smallest temperatures, with $k_BT\lesssim\Delta$, both currents show an increase (the current for positive bias being maximum when $k_B T \approx \Delta$).
The (absolute value of) rectification, instead, monotonously decreases with $T$, taking its maximum value when the
weakly coupled bath (R) is at zero temperature ($T_R=0$, i.~e. $T=\Delta T/2$).
This can be explained by the fact that the ratio $\Delta T/T$ is maximum in this situation, thus maximizing the asymmetry between the two baths.
For increasing $T$, $R$ reduces tending to zero.

Analogously to what we did in the sequential tunneling limit (Section~\ref{subsubsec:weak_sigma}), also here it is interesting to understand the case of a generic coupling. We focus on the $\Delta/(k_\text{B}T) \to 0$ 
limit where heat transport is determined by higher order coherent processes (sequential tunneling component being vanishingly small, as discussed above).
We therefore consider the coupling Hamiltonian, given in Eq.~(\ref{eq:h_int_generic2}), with an arbitrary coupling to the left bath, i.e. $\theta_\text{L} = \theta$ and $\phi_\text{L} = \phi$, 
but with fixed $\sigma_x$ coupling to the right lead, i.e $\theta_\text{R} = \pi/2$ and $\phi_\text{R} = 0$. 
The XX and XY cases can be recovered, respectively, by setting $\theta=\pi/2$ and $\phi=0$, or $\theta=\pi/2$ and $\phi=\pi/2$. Notice that, by considering a coupling 
with $\theta \neq \pi/2$, we are including also a $\sigma_z$ coupling to the left lead. We recall that, in the weak coupling regime, the $\sigma_z$ coupling does not 
contribute to the heat current. In order to isolate the impact on rectification of different spin couplings, we consider the case of identical spectral densities for the two baths, 
i.e. $\Gamma_\text{L}(\epsilon)=\Gamma_\text{R}(\epsilon)$. Therefore, the only asymmetry in the coupling, which can give rise to rectification, is given by the different 
directions described by $\vec{u}_{\text{L}}$ and $\vec{u}_{\text{R}}$.

\begin{figure}[!tb]
	\centering
	\includegraphics[width=0.9\columnwidth]{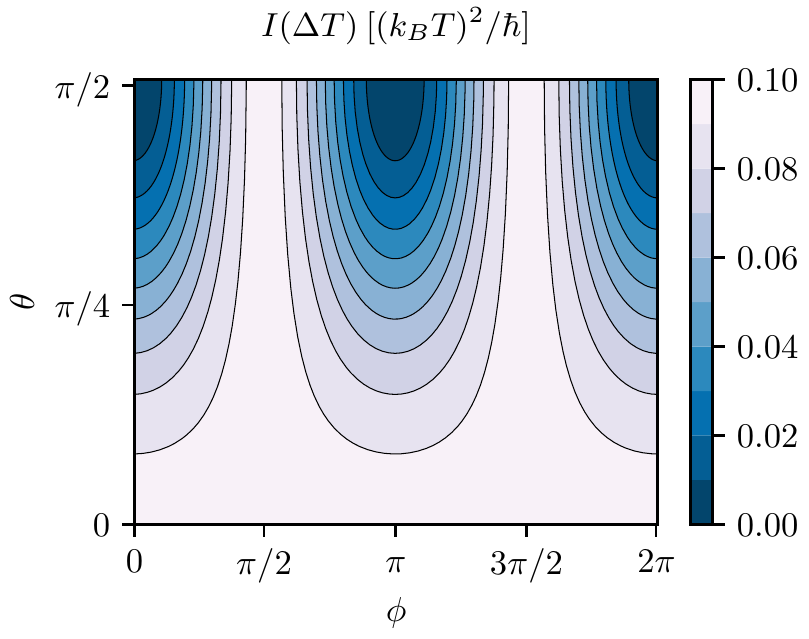}
\caption{Contour plot of $I(\Delta T)$, calculated with the NEGF approach, as a function of $\theta$ and $\phi$. The parameters are: $\epsilon_\text{C}=80 k_\text{B}T$,  $K_\text{L}=K_\text{R}=0.06$, $\Delta=0$ and $\Delta T/T = 1.9$.}
	\label{fig:j0202}
\end{figure}

In Fig.~\ref{fig:j0202} we show a contour plot of the heat current $I(\Delta T)$, at fixed temperatures and for equal Ohmic spectral densities [i.e.
 $K_\text{L}=K_\text{R}$, see Eq.~(\ref{eq:negf_spectral})], as a function of the two angles $\theta$ and $\phi$ in the small gap limit, i.e. for $\Delta/(k_\text{B}T)\to 0$. 
 For simplicity, we neglected the Lamb shift. Strikingly, the heat current is \textit{maximum} when the left lead is coupled through $\sigma_z$, i.e. for $\theta=0$ (lower 
 part of Fig.~\ref{fig:j0202}). This is surprising for two reasons: first, in the weak coupling limit the heat current at $\theta=0$ would be null even for finite 
 values of $\Delta$, since $\sigma_z$ does not contribute to the heat currents. Second, regardless of the coupling strength, a single bath coupled to S 
 through $\sigma_z$ cannot transfer heat to the system, since the Hamiltonian of S would commute with the total Hamiltonian (and thus it would be a conserved quantity). 
 In this case, the $\sigma_z$ coupling would only produce dephasing in the qubit state. We can therefore qualitatively describe transport in this regime as a direct 
 transfer of heat from one bath to the other. As $\theta$ increases, and therefore as the $\sigma_z$ component decreases, the heat current decreases monotonously, to 
 the point that it is null in the XX case ($\phi =0$), while it remain constant in the XY case (along $\phi=\pi/2$).

\begin{figure}[!tb]
	\centering
	\includegraphics[width=0.9\columnwidth]{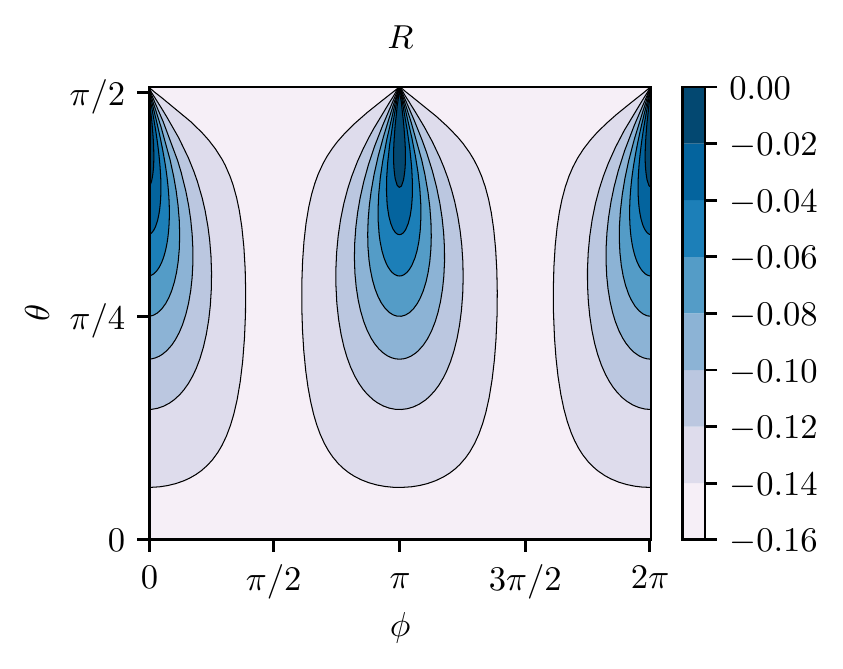}
\caption{Contour plot of $R$ as a function of $\theta$ and $\phi$. The parameters are the same as in Fig.~\ref{fig:j0202}.}
	\label{fig:r0202}
\end{figure}
Interestingly, also the rectification coefficients roughly follows a similar trend, i.e. it is maximum where also the heat currents are maximum. This can be seen 
in Fig.~\ref{fig:r0202}, where $R$ is contour-plotted as a function of $\theta$ and $\phi$ for the same parameters as in Fig.~\ref{fig:j0202}. Indeed, $R$ is maximum 
for $\theta=0$, i.e. when the left lead is coupled only through $\sigma_z$. As $\theta$ increases, the modulus of $R$ decreases monotonically along $\phi =0$, just as the heat current itself.
However, it remain constant along $\phi=\pi/2$, while for intermediate values of $\phi$ it displays a non-monotonic behavior.

We can therefore conclude that the optimal operational points in the $\Delta/(k_\text{B}T)\to 0$ limit are the XY and the XZ coupling cases. These couplings 
simultaneously maximize the magnitude of the heat current and of the rectification coefficient. We emphasize that the heat current, which in this limit is solely due to coherent 
quantum processes, behaves in the opposite way with respect to what would be expected from weak coupling calculations (the heat currents should be zero both because 
$\Delta=0$ and because $\sigma_z$ does not contribute to the heat current).

\section{Rectification at finite $U$}
\label{sec:nonlin}
How does the picture described so far changes when several levels come into play in the rectification process?
In this section we will study a non-linear resonator
defined by the Hamiltonian~(\ref{eq:sys_res}) at finite $U$, coupled to bosonic leads by the Hamiltonian~(\ref{eq:lin_coup}).
For the sake of convenience, in our numerical procedures we use the Ohmic spectral density in Eq.~(\ref{eq:negf_spectral}) in a form with a sharp cut-off, namely
\begin{equation}
\Gamma_\alpha(\epsilon)=\pi K_\alpha \epsilon\; \theta(\epsilon)\theta(\epsilon_{\rm C}-\epsilon).
\end{equation}

The most important effects are expected in the non-perturbative regime.
To this end we will employ the Keldysh non-equilibrium Green's function technique, with the retarded Green's function for the system defined as
$
G_{b;b}^r(t,t')=-i\theta (t-t')\left\langle \left[b(t),b^\dagger(t')\right]\right\rangle
$.
By following Ref.~\onlinecite{meir1991}, we use the equation of motion (EOM) decoupled to the second order (see App.~\ref{app:nonlin} for details)
to obtain the following analytic expression for the retarded Green's function
\begin{equation}
G_{b;b}^r(\epsilon)=\frac{1+2A(\epsilon)\left\langle n\right\rangle}{\epsilon-\Delta-\Sigma^{(0)}(\epsilon)+2A(\epsilon)\left(\Sigma^{(2)}(\epsilon)+\Sigma^{(3)}(\epsilon)\right)},
\label{Gr_bos_main}
\end{equation}
where $n(t)=b^\dagger (t) b(t)$ is the number operator, $\left\langle n\right\rangle$ is the expectation value of the occupation given by
\begin{equation}
\left\langle n\right\rangle=\sum_\alpha\int \frac{d\epsilon}{2\pi}G_{b;b}^r(\epsilon)n_\alpha(\epsilon)\Gamma_\alpha(\epsilon)G_{b;b}^{r^*}(\epsilon),
\label{n_bos}
\end{equation}
and
\begin{equation}
A(\epsilon)/U=\left(\epsilon-\Delta-U\left\langle n\right\rangle-\left(2\Sigma^{(0)}(\epsilon)+\Sigma^{(1)}(\epsilon)\right)\right)^{-1},
\end{equation}
$n_\alpha$ being the Bose-Einstein distribution relative to bath $\alpha$.

In order to calculate the Green's function $G^r_{b;b}(\epsilon)$, one has to solve Eqs.~(\ref{Gr_bos_main}) and (\ref{n_bos}) self-consistently.
In Eq.~(\ref{Gr_bos_main}), the  embedded self-energy is given by
\begin{equation}
\Sigma^{(0)}(\epsilon)
=\sum_\alpha \int d\omega \left[\frac{\Gamma_{\alpha}(\omega)}{\epsilon-\omega+i\eta}\right],
\end{equation}
while the other self energies are given by
\begin{eqnarray}
\Sigma^{(1)}(\epsilon)&=&\sum_\alpha \int \frac{d\omega}{2\pi} \left[\frac{\Gamma_{\alpha}(\omega)}{\epsilon+\omega-2\Delta-2U\left\langle n\right\rangle-U+i\eta}\right],\nonumber \\
\Sigma^{(2)}(\epsilon)&=&\sum_\alpha \int \frac{d\omega}{2\pi} \left[\frac{n_\alpha(\omega)\Gamma_{\alpha}(\omega)}{\epsilon-\omega+i\eta}\right],\nonumber \\
\Sigma^{(3)}(\epsilon)&=&\sum_\alpha \int \frac{d\omega}{2\pi} \left[\frac{\Gamma_{\alpha}(\omega)n_\alpha(\omega)}{\epsilon+\omega-2\Delta-2U\left\langle n\right\rangle-U+i\eta}\right].\nonumber
\end{eqnarray}
The inclusion of the self-energies defined by the expressions above ensures that the onsite correlations are correctly captured~\cite{meir1991}.
The advantage of this approach is that it describes both weak and strong coupling regimes and keeps the processes involving virtual states of the system~\cite{meir1993}.
It is worth stressing that in deriving Eq.~(\ref{Gr_bos_main}), we have neglected terms involving correlation in the baths by setting $\left\langle\left[b^\dagger(t)b_{\alpha k}(t)b_{\alpha k}(t),b^\dagger(t')\right]\right\rangle=0$ and $\left\langle b^\dagger b_{\alpha k}\right \rangle=\left\langle b b_{\alpha k}^\dagger\right \rangle=0$. The contributions from these terms become significant for very strong coupling~\cite{meir1991}. The lesser and greater Green's functions used to calculate the heat currents [see Eq.~(\ref{eq:j_green})] are obtained from the following relation
\begin{equation}
G_{b;b}^{\lessgtr}(\epsilon)=G_{b;b}^{r}(\epsilon)\Sigma^{(0)<}(\epsilon)G_{b;b}^{r^*}(\epsilon).
\end{equation}
The mean field (MF) approximation, on the other hand, is obtained by decoupling the EOM for the retarded Green's function to the first order, so that the latter takes the simple form
\begin{equation}
G_{b;b}^{r,MF}(\epsilon)=\left[\epsilon-\Delta-U\langle n\rangle-\Sigma^{(0)}(\epsilon)\right]^{-1} .
\label{green_mf}
\end{equation}
This expression makes clear that the MF approximation renormalizes the energy of the resonator, while leaving unchanged the self-energy compared to the non-interacting case.
This means that higher order onsite correlations are not taken into account.
The MF approximation was employed in Ref.~\onlinecite{ruokola2009} to calculate the rectification in nonlinear quantum circuits.
Considering an average thermal energy $k_{\rm B}T$ of the order of $\Delta$, we checked that the onsite correlation effects become significant when $U$ is of the order of the spectral density, i.~e. $U\approx  \pi K_\alpha \Delta$.
For $ U\ll \pi K_\alpha \Delta$, MF approximation and EOM give similar results~\cite{ruokola2009}.
In the absence of the non-linearity, Eqs.~(\ref{Gr_bos_main}) and (\ref{green_mf}) reduce to the same expression for exact Green's function of a non-interacting harmonic resonator, which yields no rectification.

\begin{figure}[!htb]
	\centering
	\includegraphics[width= 0.9\columnwidth]{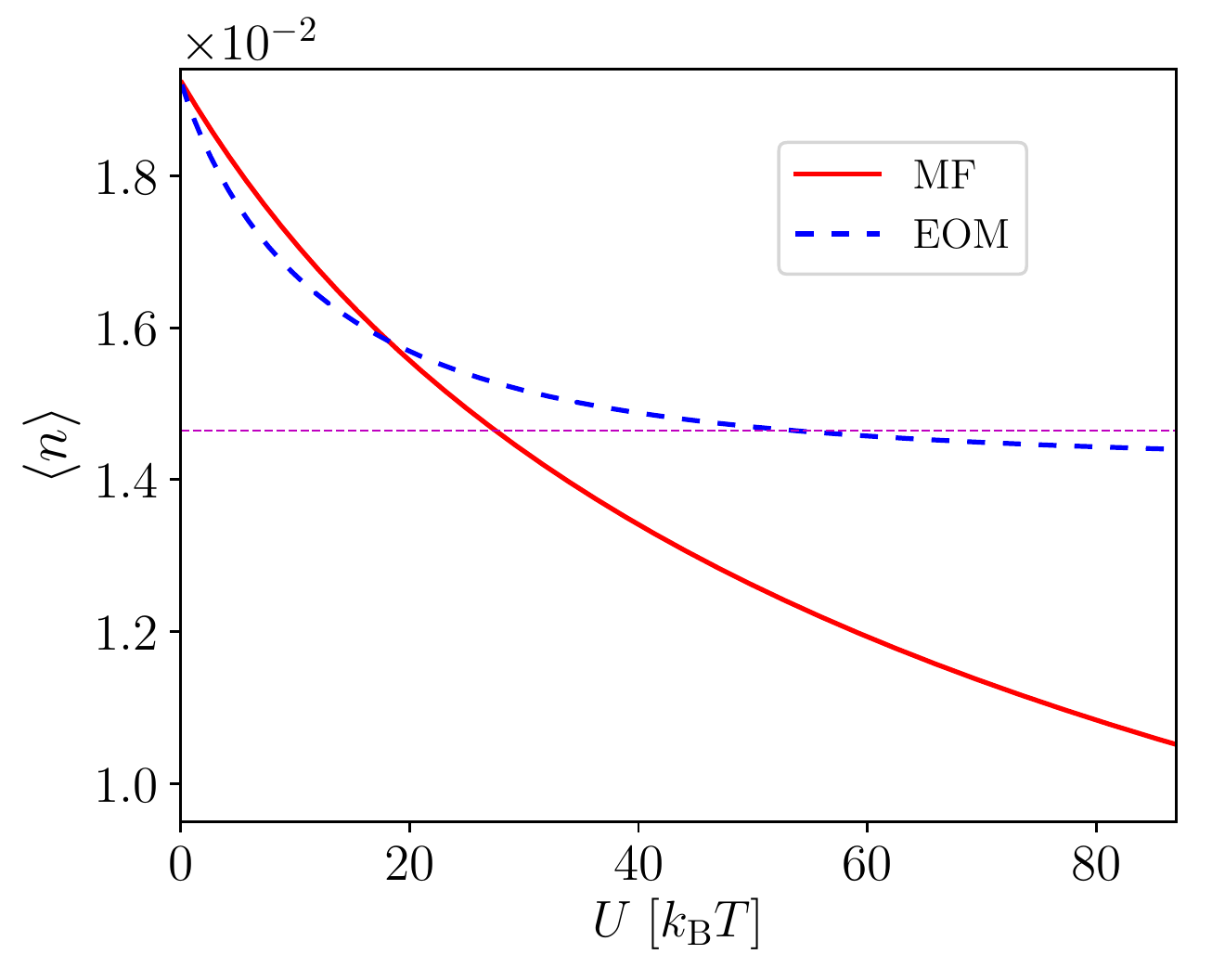}
	\caption{Occupation as a function $U$ for the following parameters $K_{\rm L}=0.06$, $K_{\rm R}=0.003$, $\epsilon_C=100 U$, $\Delta = 5k_B T$, and $\Delta T/T = 0.4$. The blue curve is calculated using the EOM, while the red curve refers to the MF approximation. The thin magenta horizontal line is the occupation of the excited level for a qubit with the same parameters, calculated accounting for sequential tunneling processes only.}
	\label{nvsU}
\end{figure}
In Fig.~\ref{nvsU} we plot the occupation $\langle n\rangle$, calculated using the EOM (blue dashed curve) and in the MF approximation (red curve), as a function of the non-linearity parameter $U$.
For small values of $U$ the two curves are close to each other, but they start to significantly depart from $U\simeq 20 k_BT$.
We also checked that MF and EOM results nearly coincide for $U<0.1k_BT$ and $\Delta\approx k_BT$.
The two curves of $\langle n\rangle$ are monotonically decreasing, as expected from the fact that with increasing $U$ the number of levels available in the resonator for transport decreases, and so does the expectation value of the occupation.
For $U$ very large one reaches the situation corresponding to a qubit.
Indeed, the curve relative to the EOM approaches the magenta line which represent the occupation of the excited level for a qubit, characterized by the same parameters, calculated accounting for sequential tunneling processes only.

\begin{figure}[!htb]
	\centering
	\includegraphics[width= 0.8\columnwidth]{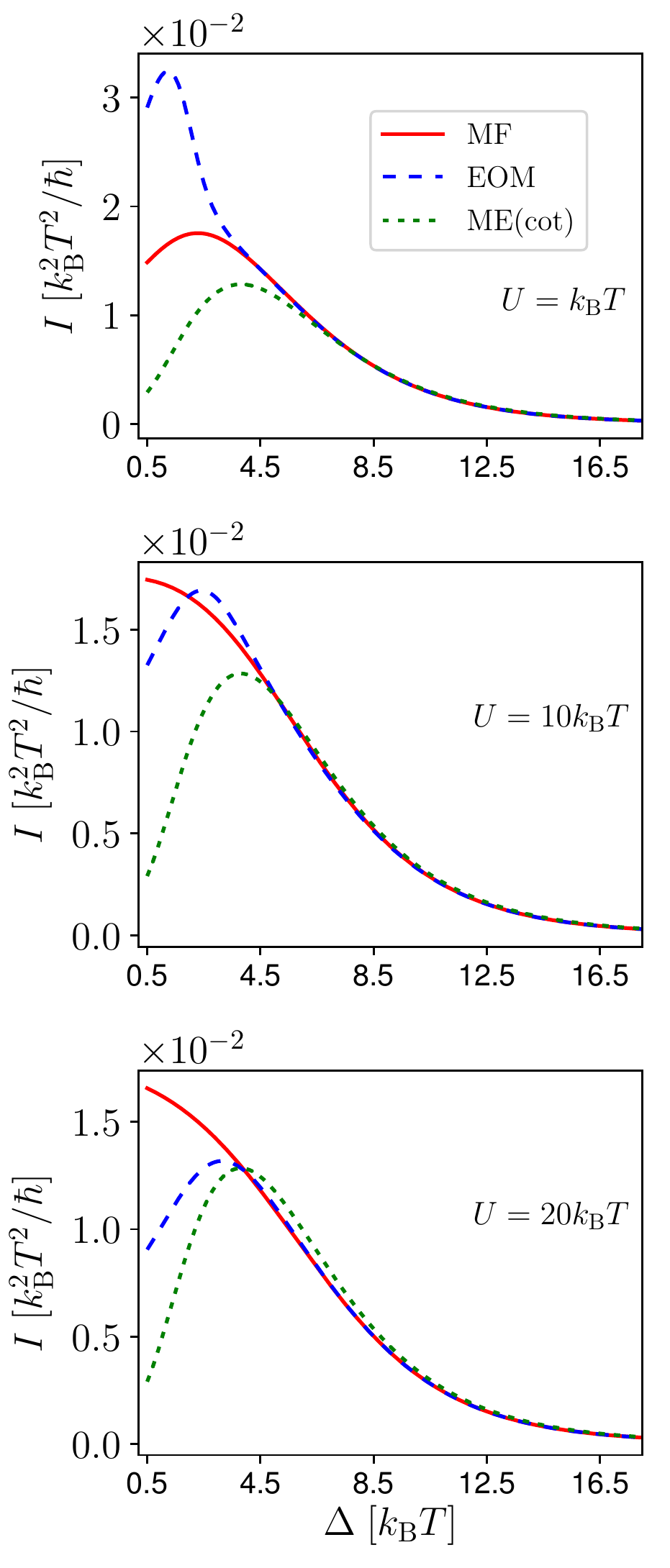}
	\vspace{0.05\columnwidth}
	\caption{Heat current calculated with three different methods as a function of resonator energy $\Delta$ for $\Delta T/T= 8/5$, $\epsilon_C=100 U$, $K_{\rm L}=0.06$ and $K_{\rm R}=0.003$. The three panels refer to different values of $U$.}
	\label{fig:boscur}
\end{figure}
In Fig.~\ref{fig:boscur}, we plot the heat current as a function of the resonator energy $\Delta$ for a few values of the non-linearity parameter $U$.
We fix the coupling strength to be within the weak coupling regime, i.~e. $K_{\rm L}=0.06$ and $K_{\rm R}=0.003$.
The heat current is calculated using both methods: i) the MF approximation (red curve), whose Green's function is Eq.~(\ref{green_mf}), and ii) the EOM method to second order (dashed blue curve), whose Green's function is Eq.~(\ref{Gr_bos_main}).
In addition, as a reference we include the heat current relative to the qubit (i.~e. $U\to\infty$) calculated using the master equation formulation taking into account the co-tunneling contributions (dotted green curve).

Fig.~\ref{fig:boscur} shows that, for all values of $U$ considered, the three curves coincide starting from $\Delta \geq 5 k_{\rm B}T$. This means that for $\Delta \geq 5 k_{\rm B}T$ the system behaves as a qubit, whereas for $\Delta \leq 5 k_{\rm B}T$ the non-linear resonator acts as a multi-level quantum system. At $U=10k_BT$ (middle panel) the curve for ME departs from the other two curves for a slightly smaller value of $\Delta$, with respect to the case $U=k_BT$. 
For $U=k_BT$, EOM and MF curves start deviating for $\Delta \leq 4 k_{\rm B}T$, and the EOM method predicts a much higher current.
Both MF and EOM curves display a maximum for small values of $\Delta$.
When $U$ reaches $10k_BT$ (middle panel) the heat current calculated with EOM reduces by roughly a half, tending to agree more with the MF result (which though does not display a maximum).
Nearly no changes are observed for the MF curve by increasing $U$ up to $20k_BT$, while the heat current predicted by the EOM method gets further reduced.
We note that the heat current obtained for the qubit case, green curve, gets vanishingly small for $\Delta=0$, whereas the MF and EOM calculations predict a finite heat current.
Interestingly, for larger values of $U$ (lower panel), the heat current computed with the EOM method approaches the ME(cot) curve even for small values of $\Delta$, as one would expect.

\begin{figure}[!htb]
	\centering
	\includegraphics[width= 0.8\columnwidth]{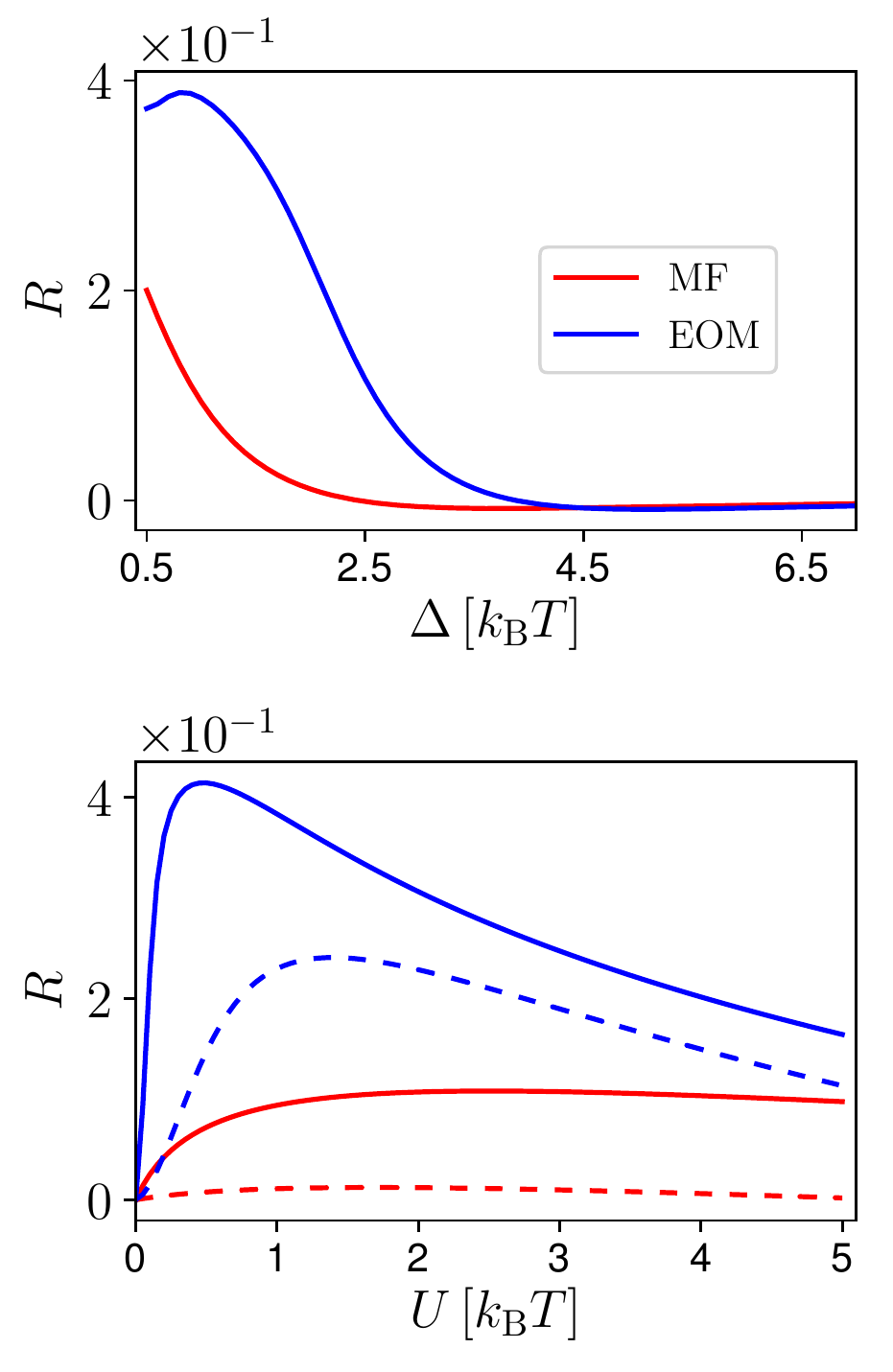}
	\vspace{0.05\columnwidth}
	\caption{Rectification as a function of the resonator energy $\Delta$ (top panel) for $U = k_B T$ and of the non-linear parameter $U$ (bottom panel) for $\Delta=k_B T$ (solid curve) and $\Delta=2k_BT$ (dashed curve). Other parameters are the same as in Fig.~\ref{fig:boscur}.}
	\label{fig:bosintDel}
\end{figure}
In Fig.~\ref{fig:bosintDel} we plot the rectification coefficient $R$ (calculated using both EOM and MF methods) as a function of the resonator energy $\Delta$ (top panel) for fixed thermal bias and coupling strengths, and of the non-linearity parameter $U$, for two values of $\Delta$ (bottom panel).
The top panel shows that the two methods agree for $\Delta>4.5k_BT$, while for smaller values of $\Delta$ EOM predicts always larger rectification, with respect to the MF method, which displays a maximum at $\Delta\simeq k_BT$.
The largest value of rectification found is around 40\%.
In the bottom panel, we first notice that the two methods do not agree over the whole range of values of $U$ considered, since we have chosen small resonator energies ($\Delta=k_BT$ and $\Delta=2k_BT$).
The EOM method, for both values of $\Delta$, predicts a larger rectification with respect to the MF method.
Furthermore, the EOM curve presents a maximum for $U$ of order $k_BT$, then $R$ steadily decreases with increasing $U$.
The MF curves, on the other hand, are also non-monotonous with $U$, but with a very broad maximum.

\section{Conclusions}
\label{sec:conclusions}
In this paper we have presented a systematic theoretical study of thermal rectification in a paradigmatic multi-level quantum system, namely a non-linear harmonic resonator, coupled to two thermal baths kept at different temperatures $T_{\rm H}$ and $T_{\rm C}$, corresponding to a thermal bias $\Delta T=T_{\rm H}-T_{\rm C}$.
Thermal rectification, consisting in an asymmetric heat conduction occurring under reversal of the thermal bias, is possible in the presence of an asymmetry in the coupling between system and baths and of interactions.
On a general perspective, both in the fermionic and bosonic cases, our aim was to explore in such a quantum system under which conditions thermal rectification can be induced and maximized, to find fundamental bounds to the maximum rectification, and to assess the impact of coherent processes on rectification.
To this end, we have first focused on the case where the strength of the non-linear term $U$ in the Hamiltonian [Eq.~(\ref{eq:sys_res})] is so large that the system behaves as a qubit (a two-level system with energy spacing, gap, $\Delta$).
In this case we have considered different transport regimes, depending on the strength and on the nature of the coupling between system and baths, and have used different theoretical approaches.
To the lowest order in the coupling, we have employed a master equation to calculate the heat current and we have determined, in different conditions, the behavior and fundamental bounds of the rectification coefficient $R$ in terms of the temperatures $T_{\rm H}$ and $T_{\rm C}$ and of the asymmetry in the system-bath couplings ($|R|\leq 1$ and $R=0$ for no rectification).
In particular, under the only assumption that the dissipation rates of the two baths, as a function of temperature, are equal up to a gap-dependent prefactor, we have found that:
\begin{itemize}
\item the rectification ratio $R$ is monotonous with respect to $\Delta T$, if the dissipation rates are monotonous in temperature (for example when the baths are bosonic);
\item $R$ is a linear function of $\lambda$;
\item the modulus of $R$ is upper bounded by $\lambda$ ($|R|\leq |\lambda|$);
\item $R$ is larger when the temperature dependence of the dissipation rates is stronger.
\end{itemize}
Here $\lambda$ is the asymmetry between the spectral densities of the two baths.
In particular, for the case of bosonic baths we have found that $R$ is a decreasing function of the energy spacing $\Delta$ and, in the limit of small $\Delta$,
\begin{itemize}
\item $|R|$ is upper bounded by $\lambda (T_{\rm H}-T_{\rm C})/(T_{\rm H}+T_{\rm C})$ in the case of linear coupling;
\item $|R|$ is upper bounded by $\lambda (T^2_{\rm H}-T^2_{\rm C})/(T^2_{\rm H}+T^2_{\rm C})$ in the case of non-linear coupling.
\end{itemize}
On the other hand, we have found that when the dissipation rates of the two baths are arbitrary the rectification can be stronger.
For example, in the case of bosonic baths and assuming the qubit to be linearly coupled to the left and non-linearly coupled to the right we have found that $R=(T_{\rm H}-T_{\rm C})/(T_{\rm H}+T_{\rm C})$ in the limit of small gap. This means that rectification can be made arbitrarily large simply by choosing a large temperature difference, regardless of $\lambda$.
We have then considered the case of arbitrary spin operators involved in the coupling Hamiltonian Eq.~(\ref{eq:h_int_gen1}) to the left ${\cal H}_{\rm L,S}$ and to the right ${\cal H}_{\rm R,S}$. In particular, with XX coupling we specify that both left and right baths are coupled to the system through $\sigma_x$, while with XY coupling we specify that the left bath is coupled through $\sigma_x$ and right bath is coupled through $\sigma_y$.
We have found that $R$ only depends on the angle between the coupling term and the qubit Hamiltonian (proportional to $\sigma_z$).
Finally, we have assessed how the Lamb-shift can give rise to an enhancement of the rectification when the baths have a gap in their density of states.

Next, we have investigated what happens in the regime beyond the weak coupling by making use of three different techniques which allow us to describe increasingly stronger coupling between system and baths and the effect of quantum coherence.
Namely, first we have included co-tunneling effects in the master equation approach, then we have used a perturbative approach based on non equilibrium Green's function theory and, finally, we have performed an exact calculation employing the Feynman-Vernon path integral approach which accounts for general spectral densities and coupling conditions.
All these approaches allow us to conclude that the rectification can be enhanced by going beyond the weak coupling regime, even violating the bounds found in the first-order coupling regime.
In particular, we have found that:
\begin{itemize}
\item co-tunneling processes enhance rectification when heat transport is dominated by sequential tunneling;
\item $R$ is in general non-monotonous with the coupling strength and depends on the spin operators involved in the coupling Hamiltonian;
\item $R$ increases with increasing coupling strength in the XX coupling case;
\item $R$ increases with $\lambda$ faster that linear in the XX coupling case;
\item $R$ increases with decreasing $\Delta$ in the XX coupling case, and is non-monotonous for the XY coupling;
\item the heat currents calculated with non equilibrium Green's function method and with the exact method are qualitatively in agreement even for large system bath couplings;
\item in the limit of small $\Delta$, where the coherent (higher order) contributions to the heat current dominates, heat currents and $R$ are maximized in the XY and the XZ coupling cases.
\end{itemize}

Finally, we have considered the case in which the non-linear term $U$ is finite.
Employing the Keldysh non-equilibrium Green's function technique we have calculated the heat current using the equation of motion (EOM) decoupled to the second order and in the mean field approximation. We have discussed the behavior of the heat current and of the rectification as functions of the resonator energy and of $U$.
We have found that the rectification, using the EOM approach, is not monotonous in both $\Delta$ and $U$ and reaches a maximum of 40\%, much larger that what is predicted employing the mean field approximation.

The paper is enriched with a number of appendices which contain the details of the calculations described in the main text.

To conclude, we believe that our results can be very valuable for the design and interpretation of experiments on thermal rectifiers based on qubits and non-linear harmonic resonators.

\section{Acknowledgements}
We thank Jukka Pekola for many stimulating discussions and for his comments on the draft.
R.F. research has been conducted within the framework of the Trieste Institute for Theoretical Quantum Technologies (TQT).
E.P. acknowledeges hospitality of ICTP where
part of this work has been carried out.
We acknowledge support from the SNS-WIS joint lab QUANTRA, and E.P. acknowledges support by the University of Catania, Piano di Incentivi per la Ricerca di Ateneo 2020/2022, proposal Q-ICT and by the CNR- QuantERA grant SiUCs.

\begin{appendix}

\section{Most Generic System-Bath Coupling in the qubit case}
\label{app:qubit_master_eq}
In this appendix we prove that the system bath interaction described by Eq.~(\ref{eq:h_int_gen1}) is indeed the most generic system-bath interaction in the qubit case. 

The most generic Hermitian operator acting on the tensor product space between S (a two-dimensional Hilbert space) and the baths (an arbitrary dimensional Hilbert space) can be expanded on the product basis of the two Hilbert spaces. We therefore consider a basis $\{\mathcal{B}_i\}_i$ of Hermitian operators acting on the space of the bath, and the specific basis $\vec{\sigma}_j \equiv \{\mathbb{1}, \sigma_x, \sigma_y, \sigma_z\}$ of Hermitian operators acting on the qubit space. This yields
\begin{equation}
	\mathcal{H}_{\alpha, \text{S}} = \sum_{i,j} a_{ij} \, \mathcal{B}_i \otimes \sigma_j = \sum_j B_j\otimes \sigma_j,
	\label{eq:h_gen_app}
\end{equation}
where $B_j = \sum_i a_{ij} \mathcal{B}_i$ is an Hermitian operator acting on the bath space. Using the relations
\begin{equation}
\begin{aligned}
	\sigma_x &= \sigma^+ + \sigma^- \\
	\sigma_y &= i\sigma^+ -i\sigma^-,
\end{aligned}
\label{eq:sigma_relations}
\end{equation}
we obtain Eq.~(\ref{eq:h_int_gen1}), where $B_\alpha = B_x + i B_y$.

\section{Rectification in the weak coupling regime and qubit case}
\label{app:rect_weak}
We now compute the heat current flowing out of the leads in the weak coupling regime, valid when $\mathcal{H}_{\alpha, \text{S}}$ is ``small enough''~\cite{breuer2002}.
 Under these conditions the evolution of the reduced density matrix $\rho_\text{S}$ of the qubit obeys a Lindblad master equation. Furthermore, when the qubit is not degenerate (i.e. when $\Delta\neq 0$), the Lindblad master equation can be cast in the form of a rate equation for the occupation probabilities of the qubit, defined by $P_1 = \Tr{\rho_\text{S} \sigma^+\sigma^-}$ and $P_0 = 1-P_1$. Only the terms in $\mathcal{H}_{\alpha,\text{S}}$ proportional to $\sigma^+$ and $\sigma^-$ contribute to the rate equation.
Indeed, rewriting $\mathcal{H}_{\alpha, \text{S}}$ as in Eq.~(\ref{eq:h_gen_app}), the rate equation only depends on the following matrix elements of the $\sigma_j$ operators~\cite{breuer2002}:
\begin{equation}
	\mel*{0}{\sigma_j}{1} .
\end{equation}
where $\{\ket{0},\ket{1}\}$ are the eigenstates of the qubit. 
Since $\mel*{0}{\sigma_j}{1}=0$ for $\sigma_j = \mathbb{1}, \sigma_z$, the only terms that determine the populations are the ones proportional to $\sigma_x$ and $\sigma_y$, and therefore to $\sigma^+$ and $\sigma^-$.

Neglecting for the moment the Lamb shift, the probabilities satisfy~\cite{breuer2002}
\begin{equation}
\frac{\partial}{\partial t}
\begin{pmatrix}
	P_0 \\ P_1
\end{pmatrix}
=
\begin{pmatrix}
	-\Upsilon^+(\Delta) & \Upsilon^-(\Delta) \\
	\Upsilon^+(\Delta) & -\Upsilon^-(\Delta)
\end{pmatrix}
\begin{pmatrix}
	P_0 \\ P_1
\end{pmatrix},
\label{eq:master}
\end{equation}
where $\Upsilon^\pm(\Delta) = \Upsilon^\pm_\text{L}(\Delta,T_\text{L}) + \Upsilon^\pm_\text{R}(\Delta,T_\text{R}) $, and $\Upsilon^\pm_\alpha(\Delta,T)$, for $\alpha = \text{L},\text{R}$, are derived in App.~\ref{app:tunn_rates}. Using Eq.~(\ref{eq:master}) and  $P_0+P_1=1$, we can find the steady state populations
\begin{equation}
\begin{aligned}
	P_0 &= \frac{\Upsilon^-(\Delta)}{\Upsilon^-(\Delta) + \Upsilon^+(\Delta)},\quad & P_1 = \frac{\Upsilon^+(\Delta)}{\Upsilon^-(\Delta) + \Upsilon^+(\Delta)}.
	\label{eq:populations}
\end{aligned}
\end{equation}
The heat current flowing out of bath $\alpha$ at temperature $T_\alpha$ [see Eq.~(\ref{eq:j_gen_def})] can then be computed as
\begin{equation}
	I_\alpha(T_\alpha) =  \Delta \left( P_0 \Upsilon^+_\alpha(\Delta,T_\alpha) - P_1 \Upsilon^-_\alpha(\Delta,T_\alpha)  \right).
	\label{eq:j_def}
\end{equation}
Obviously, the steady-state heat current and, as a consequence, the rectification ratio
within the weak-coupling regime only arise from the terms proportional to $\sigma^+$ and $\sigma^-$.
Since $\Upsilon^+_\alpha(\Delta,T_\alpha)$ and $\Upsilon^-_\alpha(\Delta,T_\alpha)$ are related by the detailed balance equation, $\Upsilon^+_\alpha(\Delta,T_\alpha)=e^{-\Delta/(k_BT_\alpha)} \Upsilon^-_\alpha(\Delta,T_\alpha)$, we can express them as $\Upsilon_\alpha^\pm(\Delta,T_\alpha) = \Upsilon_\alpha(\Delta,T_\alpha) f(\pm  \Delta/(k_\text{B}T_\alpha))$,
where $f(x) = (1+e^x)^{-1}$.
Using Eqs.~(\ref{eq:populations}) and (\ref{eq:j_def}), the heat current is given by
\begin{multline}
	I(\Delta T) = \Delta \frac{\Upsilon_\text{L}(\Delta,T_\text{L})\Upsilon_\text{R}(\Delta,T_\text{R}) }{\Upsilon_\text{L}(\Delta,T_\text{L})+\Upsilon_\text{R}(\Delta,T_\text{R})} \\
	\times \left[ f(\Delta/(k_\text{B}T_\text{L})) - f(\Delta/(k_\text{B}T_\text{R})) \right],
\label{eq:heat_currrents_weak}
\end{multline}
where $T_\text{L} = T+\Delta T/2$ and $T_\text{R} = T-\Delta T/2$.
In conclusion, under weak coupling and neglecting the Lamb shift, the
heat current only depends on the tunneling rates $\Upsilon_\alpha(\Delta, T_\alpha)$ which can be explicitly evaluated for the
models considered
in Section~\ref{sec:weak_coupling}.
By plugging Eq.~(\ref{eq:heat_currrents_weak}) into Eq.~(\ref{eq:r_def}) we find the general expression for $R$, i.e. Eq.~(\ref{eq:r_general}).

\section{Tunneling Rates}
\label{app:tunn_rates}
In this section we prove Eq.~(\ref{eq:gamma_def}) following Ref.~\onlinecite{breuer2002}. To this end we consider the system-bath Hamiltonian as written in Eq.~(\ref{eq:h_gen_app}), such that all operators are Hermitian. As shown in App.~\ref{app:rect_weak}, the term proportional to $\sigma_z$ does not contribute to the heat current, therefore we consider
\begin{equation}
	\mathcal{H}_{\alpha, \text{S}} = B_{\alpha x} \otimes \sigma_x + B_{\alpha y}\otimes \sigma_y,
	\label{eq:app_hint}
\end{equation}
where
\begin{equation}
\begin{aligned}
	B_{\alpha x} &= \frac{B_\alpha^\dagger + B_\alpha}{2}, &
	B_{\alpha y} &= \frac{i(B_\alpha^\dagger - B_\alpha)}{2}.
\end{aligned}
\label{eq:app_bx_b}
\end{equation}
Using the results of Ref.~\onlinecite{breuer2002} with $\mathcal{H}_{\alpha, \text{S}} $ given by Eq.~(\ref{eq:app_hint}), the dissipation rate
induced by bath $\alpha$ is given by
\begin{multline}
	\Upsilon^-_\alpha(\Delta, T) = \sum_{i,j = \{x,y\}} \gamma_{ij}\mel*{1}{\sigma_i}{0}\mel*{0}{\sigma_j}{1} = \\ =\gamma_{xx} + \gamma_{yy} + i\gamma_{yx} -i\gamma_{xy} = \int\limits_{-\infty}^{+\infty}dt\, e^{i\Delta t}\ev*{B_\alpha(s)B^\dagger_\alpha(0)},
	\label{eq:app_gamma_m}
\end{multline}
where 
\begin{equation}
	\gamma_{ij} = \int_{-\infty}^{+\infty} dt\, e^{i\Delta t}\ev*{B^\dagger_{\alpha i}(t)B_{\alpha j}(0)}.
	\label{eq:gamma_ij}
\end{equation}
Here $B_{\alpha i}(t)$ is the time evolution of $B_{\alpha i}$ in the interaction picture, i.e. $B_{\alpha i}(t)$ is the Heisenberg picture operator evolved solely according to Hamiltonian of the bath $\mathcal{H}_\alpha$. In the last step of Eq.~(\ref{eq:app_gamma_m}) we used Eq.~(\ref{eq:app_bx_b}) to express $B_{\alpha x}$ and $B_{\alpha y}$ in terms of $B_\alpha$ and $B_\alpha^\dagger$. This concludes the proof.

\section{Tunneling Rates in Specific Models}
\label{sec:tunneling_rates}
In this section we derive the expression for $\Upsilon_\alpha(\Delta,T)$ in various models.

\subsection{Fermionic baths with linear (tunnel) couplings}
\label{subsec:fermionic_rates}
In this subsection we consider a fermionic bath $\mathcal{H}_\alpha^\text{(F)}$, as defined in Eq.~(\ref{eq:h_bath}). Since we consider the case of equal chemical potentials, we define the energies $\epsilon_{\alpha k}$ in Eq.~(\ref{eq:h_bath}) as measured respect to the common chemical potential $\mu$. Therefore, the energies $\epsilon_{\alpha k}$ are defined in the interval $[-\infty,+\infty]$. Plugging the linear coupling Hamiltonian, given in Eq.~(\ref{eq:qubit_lin_coupling}), into Eq.~(\ref{eq:gamma_def}) yields
\begin{equation}
	\Upsilon^-_\alpha(\Delta,T_\alpha) = \sum_{k,k^\prime} V_{\alpha k} V_{\alpha k^\prime}^* \int\limits_{-\infty}^{+\infty} dt\, e^{i\Delta t}  \ev{c_{\alpha k}(t) c_{\alpha k^\prime}^\dagger}.	\label{eq:app:gamma_m}
\end{equation}
In the interaction picture, time-evolved bath operators $\hat{O}$ satisfy (with $\hbar=1$)
\begin{equation}
	\frac{d\hat{O}(t)}{dt} = i\left[ \mathcal{H}_\alpha, \hat{O}(t) \right].
\end{equation}
Using the fact that $[\mathcal{H}_\alpha, c_{\alpha k}] = -\epsilon_{\alpha k} c_{\alpha k}$, we find
\begin{equation}
c_{\alpha k}(t) = e^{-i\epsilon_{\alpha k} t} c_{\alpha k}.
\label{eq:ck_t}
\end{equation}
Plugging Eq.~(\ref{eq:ck_t}) into Eq.~(\ref{eq:app:gamma_m}) yields
\begin{multline}
	\Upsilon^-_\alpha(\Delta,T_\alpha) = 2\pi \sum_k |V_{\alpha k}|^2 \ev*{c_{\alpha k}c^\dagger_{\alpha k}} \delta(\Delta-\epsilon_{\alpha k}) = \\2\pi \sum_k |V_{\alpha k}|^2 [1-f(\beta_\alpha \epsilon_{\alpha k})] \delta(\Delta-\epsilon_{\alpha k}),	\label{eq:app_gamma_step1}
\end{multline}
where $f(x) = (\exp(x)+1)^{-1}$. Recognizing the spectral function, defined in Eq.~(\ref{eq:spectraldensity}), we have that
\begin{equation}
\Upsilon^-_\alpha(\Delta, T_\alpha) = \Gamma_\alpha(\Delta) [1-f(\Delta/(k_\text{B}T_\alpha))].
\end{equation}
Using the detailed balance condition, we find that
\begin{equation}
	\Upsilon_\alpha(\Delta,T) = \Gamma_\alpha(\Delta),
\end{equation}
which implies $g(\Delta,T)=1$. This proves Eq.~(\ref{eq:fermionic_rates}).

\subsection{Bosonic baths with linear (tunnel-like) coupling}
\label{subsec:bosonic_tunnel_rates}
In this section we consider a bosonic bath $\mathcal{H}_\alpha^\text{(B)}$, as defined in Eq.~(\ref{eq:h_bath}), and a linear coupling as in Eq.~(\ref{eq:qubit_lin_coupling}). As in the fermionic case, we have that  $[\mathcal{H}_\alpha^\text{(B)}, b_{\alpha k}] = -\epsilon_{\alpha k} b_{\alpha k}$, so also in this case we have that the interaction picture destruction operator is given by
\begin{equation}
	b_{\alpha k}(t) = e^{-i\epsilon_{\alpha k} t} b_{\alpha k}.
	\label{eq:bk_t}
\end{equation}
Performing the same steps as in the fermionic case, we end up with Eq.~(\ref{eq:app_gamma_step1}) with $\ev*{b_{\alpha k}b^\dagger_{\alpha k}}$ instead of $\ev*{c_{\alpha k}c^\dagger_{\alpha k}}$, which leads to having $1+n(\epsilon_{\alpha k}/(k_\text{B}T_\alpha))$ instead of $1-f(\epsilon_{\alpha k}/(k_\text{B}T_\alpha))$, where $n(x)\equiv (\exp(x)-1)^{-1}$. We therefore find
\begin{equation}
	\Upsilon^-_\alpha(\Delta, T_\alpha) = \Gamma_\alpha(\Delta) [1+n(\Delta/(k_\text{B}T_\alpha))],
\end{equation}
which, using the detailed balance condition, leads to
\begin{equation}
	\Upsilon_\alpha(\Delta,T_\alpha) = \Gamma_\alpha(\Delta) \coth{(\Delta/(2k_\text{B}T_\alpha))}. 
\end{equation}
This proves Eq.~(\ref{eq:bosonic_tunnel_rates}).

\subsection{Bosonic baths with non-linear coupling}
\label{app:bosonic_nonlin}
In this section we consider a bosonic bath $\mathcal{H}_\alpha^\text{(B)}$, as defined in Eq.~(\ref{eq:h_bath}), and a non-linear coupling as in Eq.~(\ref{eq:nonlin_coupling}).
Using Eq.~(\ref{eq:gamma_def}), we have that
\begin{equation}
	\Upsilon^-_\alpha(\Delta,T_\alpha) = \sum_{k,k^\prime} V_k V_{k^\prime}^* \int\limits_{-\infty}^{+\infty} dt\, e^{i\Delta t}  \ev{b_k^2(t) (b_{k^\prime}^\dagger)^2}.
\end{equation}
Using Eq.~(\ref{eq:bk_t}), we can compute the time integral, finding
\begin{equation}
	\Upsilon^-_\alpha(\Delta,T_\alpha) = 2\pi\sum_{k} |V_{\alpha k}|^2 \ev{b_{\alpha k}^2 (b_{\alpha k}^\dagger)^2} \, \delta(\Delta-2\epsilon_{\alpha k}).
	\label{eq:app_nonlin_s1}
\end{equation}
We therefore need to compute the expectation value $\ev*{b_{\alpha k}^2 (b_{\alpha k}^\dagger)^2}$. Using the commutation relations, we have that
\begin{equation}
	\ev*{b_{\alpha k}^2 (b_{\alpha k}^\dagger)^2} = \ev{n_{\alpha k}^2} + 3n(\epsilon_{\alpha k}/(k_\text{B}T_\alpha)) +2,
\end{equation}
where $\ev*{n_{\alpha k}^2} = \ev*{(b_{\alpha k}^\dagger b_{\alpha k})^2}$ is the thermal expectation value of the square number. The calculation of $\ev*{n_{\alpha k}^2}$ is performed in App.~\ref{sec:thermal_average}. Using Eq.~(\ref{eq:n2_boson}), we have that
\begin{equation}
	\ev*{b_{\alpha k}^2 (b_{\alpha k}^\dagger)^2} =2\left[n(\epsilon_{\alpha k}/(k_\text{B}T_\alpha))+1\right]^2.
\end{equation}
Plugging this into Eq.~(\ref{eq:app_nonlin_s1}), recalling that $\delta(\Delta-2\epsilon_k) = \delta(\Delta/2 - \epsilon_k)/2$ (which can be proven changing variables), and recognizing the spectral function, we find
\begin{equation}
	\Upsilon^-_\alpha(\Delta,T_\alpha) = \Gamma_\alpha(\Delta/2) [1+n(\Delta/(2k_\text{B}T_\alpha))]^2.
\end{equation}
Finally, using the detailed balance condition we find
\begin{equation}
	\Upsilon_\alpha(\Delta,T_\alpha) =\frac{1}{2} \Gamma_\alpha(\frac{\Delta}{2})\,[1+\coth^2(\Delta/(4k_\text{B}T_\alpha))].
\end{equation}
Replacing Eq.~(\ref{eq:similar_rates}) with $\Upsilon_\alpha(\Delta,T) = \Gamma_\alpha(\Delta/2) g(\Delta,T)/2$, 
$g(\Delta, T)$ is given by  Eq.~(\ref{eq:rates_nonlin_bosonic}). 
This implies that, for the results of Section~\ref{subsec:boson_nonlin}, $\Gamma_\alpha(\Delta)$ has to be replaced with $\Gamma_\alpha(\Delta/2)$ in the definition of $\lambda$ [Eq.~(\ref{eq:l_def})].

\subsection{Arbitrary baths with different $\sigma$ couplings}
\label{app:different_sigma}
In this subsection we consider arbitrary baths coupled to the qubit via Eq.~(\ref{eq:h_int_generic2}). In fact, in this appendix we consider a more general case given by
\begin{equation}
	\mathcal{H}_{\alpha, \text{S}} = \sum_{i=x,y,z} (u_{\alpha,i} \sigma_i) \otimes (B_\alpha+B_\alpha^\dagger),
\end{equation}
where $\vec{u}_\alpha = (\sin\theta_\alpha\cos\phi_\alpha, \sin\theta_\alpha\sin\phi_\alpha, \cos\theta_\alpha)$ is  a unit vector, and $B_\alpha$ is an arbitrary bath operator. As discussed in Section~\ref{sec:weak_coupling}, the term proportional to $\sigma_z$ does not contribute to the heat current, so we can neglect it. The term that matters is
\begin{equation}
	[\vec{u}_{\alpha}]_x \sigma_x + [\vec{u}_{\alpha}]_y\sigma_y = \sin\theta_\alpha \left( e^{i\phi_\alpha}\sigma^+ + e^{-i\phi_\alpha}\sigma^- \right).
\end{equation}
Assuming that $\Delta >0$, and that $B_\alpha^\dagger$ produces excitations in bath $\alpha$ with positive energy, also the terms proportional to $B^\dagger \sigma^+$ and $B \sigma^-$ vanish. The relevant terms of the interacting Hamiltonian thus become
\begin{equation}
	\mathcal{H}_{\alpha, \text{S}} =  \sigma^+ \otimes \tilde{B}_\alpha + \sigma^-\otimes \tilde{B}_\alpha^\dagger,
\end{equation}
where we define
\begin{equation}
	\tilde{B}_\alpha = \sin\theta_\alpha e^{i\phi_\alpha} B_\alpha.
\end{equation}
This interacting Hamiltonian is now of the form of Eq.~(\ref{eq:h_int_gen1}). Therefore, the tunneling rates can be computed from Eq.~(\ref{eq:gamma_def}), yielding
\begin{equation}
	\Upsilon^-_\alpha(\Delta,T_\alpha) = \sin^2\theta_\alpha h(\Delta,T_\alpha),
\end{equation}
where $h(\Delta,T_\alpha) = \int dt\, e^{i\Delta t} \ev*{B_\alpha(t)B_\alpha^\dagger(0)}_\alpha$ only depends on the bath through the temperature, and it does not depend on $\theta_\alpha$ nor $\phi_\alpha$. Using the detailed balance condition we find
\begin{equation}
	\Upsilon_\alpha(\Delta,T_\alpha) = (\sin^2\theta_\alpha) h(\Delta,T_\alpha)(1+e^{-\Delta/(k_BT_\alpha)}).
\end{equation}
This situation therefore corresponds to the {\it equal-g} case [$g_{\rm L}=g_{\rm R}\equiv g$ in Eq.~(\ref{eq:similar_rates})], where $\Gamma_\alpha(\Delta) = \sin^2\theta_\alpha$, and $g(\Delta,T) = h(\Delta,T)(1+e^{-\Delta/(k_BT)})$.
\section{Thermal Averages}
\label{sec:thermal_average}
In this section we show how to compute the expectation value $\ev*{n_{\alpha k}^2}$ for the bosonic bath. Let us define the inverse temperature $\beta_\alpha = 1/(k_BT_\alpha)$ The partition function $Z_\alpha$ is given by
\begin{equation}
Z = \sum_{\{n_i\}=0}^{+\infty} p(\{n_i\}),
	\label{eq:z}
\end{equation}
where the sum is over each $n_i$ from $0$ to $+\infty$, and where
\begin{equation}
	p(\{n_i\}) = e^{-\beta_\alpha \sum_j n_{j} \epsilon_{\alpha j}}
	\label{eq:p}
\end{equation}	
is the canonical probability of finding the bath in a Fock state with occupation numbers $\{n_i\}$. Using these two definitions, and recalling that $\ev{n_{\alpha k}^m} \equiv \sum p(\{n_i\})\, n^m_k$, it is easy to prove that
\begin{align}
-\frac{1}{\beta} \frac{\partial \ln Z}{\partial \epsilon_{\alpha k}} &= \ev*{n_{\alpha k}}, \label{eq:nk} \\
	\frac{1}{\beta^2} \frac{\partial^2 \ln Z}{\partial \epsilon_{\alpha k}^2} &= \ev*{n_{\alpha k}^2} - \ev*{n_{\alpha k}}^2. \label{eq:nk2}
\end{align}
Plugging Eq.~(\ref{eq:p}) into (\ref{eq:z}), and recognizing that we can perform all the sums as geometric series, we can express the logarithm of the bosonic partition function as
\begin{equation}
	\ln Z = -\sum_k \ln{(1- e^{-\beta \epsilon_{\alpha k}})}.
	\label{eq:z_bosonic}
\end{equation}
Plugging Eq.~(\ref{eq:z_bosonic}) into (\ref{eq:nk}), we find the well know result that $\ev{n_{\alpha k}} = n(\beta_\alpha\epsilon_{\alpha k})$. Plugging Eq.~(\ref{eq:z_bosonic}) into (\ref{eq:nk2}), we find 
\begin{equation}
	\ev*{n_{\alpha k}^2} = 2n^2(\beta_\alpha\epsilon_{\alpha k})  + n(\beta\epsilon_{\alpha k}).
	\label{eq:n2_boson}
\end{equation}

\section{Lamb shift}
\label{sec:lamb_shift}
In this appendix we compute the Lamb shift of the qubit gap induced by the bath in the weak coupling regime, and we derive Eq.~(\ref{eq:hls}). In order to use the results of Ref.~\onlinecite{breuer2002}, we consider a coupling Hamiltonian as written in Eq.~(\ref{eq:app_hint}).
As shown in Ref.~\onlinecite{breuer2002}, we have that
\begin{equation}
	\tilde{\mathcal{H}}  = \sum_{\substack{\epsilon=\{0,\pm\Delta\} \\ i,j = \{ x,y \}}} S_{xy}(\epsilon) \sigma_i^\dagger(\epsilon)\sigma_j(\epsilon),	\label{eq:hls_gen}
\end{equation}
where
\begin{equation}S_{ij}(\epsilon) = \frac{1}{2\pi}\mathcal{P} \int_{-\infty}^{+\infty} \frac{\gamma_{ij}(\omega)}{\epsilon-\omega} d\omega \equiv \mathcal{S}_\epsilon [ \gamma_{ij}(\omega) ],	\label{eq:sij_def}
\end{equation}
with
\begin{equation}
	\gamma_{ij}(\omega) = \int_{-\infty}^{+\infty} dt\, e^{i\omega t}\ev*{B^\dagger_i(t)B_j(0)}
	\label{eq:gammaij_omega}
\end{equation}
defined exactly as in Eq.~(\ref{eq:gamma_ij}), where $\Delta$ is replaced with $\omega$, and where
\begin{equation}
	\sigma_i(\epsilon) = \sum_{\epsilon^\prime-\epsilon^{''} = \epsilon} \ket{\epsilon{''}}\bra{\epsilon{''}}\sigma_i \ket{\epsilon^\prime}\bra{\epsilon^\prime}.
	\label{eq:sigmai_omega}
\end{equation}
Notice that the functional $\mathcal{S}_\epsilon[\dots]$, defined in Eq.~(\ref{eq:sij_def}), is linear, and that $\epsilon''$ and $\epsilon^\prime$ run over the two eigenvalues of the qubit, $-\Delta/2,\Delta/2$. For ease of notation, we identify the excited state of the qubit with $\ket{1} = \ket{\Delta/2}$, and the ground state with $\ket{0}=\ket{-\Delta/2}$. Expanding the sum in Eq.~(\ref{eq:sigmai_omega}), we have
\begin{equation}
\begin{aligned}
	\sigma_i(\Delta) &=  \mel*{0}{\sigma_i}{1}\, \ket{0}\bra{1}, \\
	\sigma_i(-\Delta) &=  \mel*{1}{\sigma_i}{0}\, \ket{1}\bra{0}, \\
	\sigma_i(0) &=  \sum_{k=0,1}\mel*{k}{\sigma_i}{k}\, \ket{k}\bra{k} =0, 
\end{aligned}
\end{equation}
where we used the fact that both $\sigma_x$ and $\sigma_y$ have only zeros on the diagonal in the last equality. Therefore, the non-null elements are given by
\begin{equation}
\begin{aligned}
	\sigma_x(\Delta) &=  \sigma^-, &  \sigma_x(-\Delta) &=  \sigma^+,\\
	\sigma_y(\Delta) &= -i\sigma^-, &  \sigma_y(-\Delta) &=  i\sigma^+.
\end{aligned}
\end{equation}
Plugging these results into Eq.~(\ref{eq:hls_gen}), using the anti-commutation relation $\{\sigma^-,\sigma^+\} = \mathbb{1}$, and neglecting the terms proportional to the identity, we find
\begin{multline}
	\tilde{\mathcal{H}} = \sigma_z \left[ S_{xx}(\Delta) + S_{yy}(\Delta) -iS_{xy}(\Delta) + iS_{yx}(\Delta) \right] \\
	-\sigma_z \left[ S_{xx}(-\Delta) + S_{yy}(-\Delta) +iS_{xy}(-\Delta) - iS_{yx}(-\Delta)  \right].
\end{multline}
Expressing $S_{ij}(\pm\Delta)$ in terms of the functional $\mathcal{S}_\epsilon$ yields
\begin{multline}
	\tilde{\mathcal{H}} = \sigma_z\mathcal{S}_\Delta \left[ \gamma_{xx}(\omega) + \gamma_{yy}(\omega) -i\gamma_{xy}(\omega) + i\gamma_{yx}(\omega) \right] \\
	-\sigma_z \mathcal{S}_{-\Delta}\left[ \gamma_{xx}(\omega) + \gamma_{yy}(\omega) +i\gamma_{xy}(\omega) - i\gamma_{yx}(\omega)  \right].
	\label{eq:hls_1}
\end{multline}
Using the definition of $\gamma_{ij}(\omega)$ in Eq.~(\ref{eq:gammaij_omega}), and expressing $B_x$ and $B_y$ in terms of $B$ and $B^\dagger$ through Eq.~(\ref{eq:app_bx_b}), it can be shown that
\begin{equation}
\begin{aligned}
	\gamma_{xx}(\omega) + \gamma_{yy}(\omega) -i\gamma_{xy}(\omega) + i\gamma_{yx}(\omega) = \Upsilon^-(\omega), \\
	\gamma_{xx}(\omega) + \gamma_{yy}(\omega) +i\gamma_{xy}(\omega) - i\gamma_{yx}(\omega) = \Upsilon^+(-\omega),
\end{aligned}
\label{eq:gamma_sum}
\end{equation}
where $\Upsilon^-(\omega)$ and $\Upsilon^+(\omega)$ are the rates introduced in Eq.~(\ref{eq:master}). 
Plugging Eq.~(\ref{eq:gamma_sum}) into Eq.~(\ref{eq:hls_1}) yields
\begin{equation}
	\tilde{\mathcal{H}}  = \sigma_z\left( \mathcal{S}_\Delta\left[\Upsilon^-(\omega)\right] - \mathcal{S}_{-\Delta}\left[ \Upsilon^+(-\omega)  \right] \right)
\end{equation}
Using Eq.~(\ref{eq:sij_def}), it can be shown that the operator $\mathcal{S}_\Delta[\dots]$ satisfies the general property $\mathcal{S}_{-\Delta}[f(-\omega)] = - \mathcal{S}_\Delta[f(\omega)]$. Therefore, we find
\begin{equation}
	\tilde{\mathcal{H}}  = \sigma_z \mathcal{S}_\Delta\left[\Upsilon^-(\omega) + \Upsilon^+(\omega)\right].
\end{equation}
Finally, recalling that $\Upsilon^{\pm}(\omega) = \Upsilon^{\pm}_\text{L}(\omega,T_\text{L}) + \Upsilon^{\pm}_\text{R}(\omega,T_\text{R})$, we have that
\begin{equation}
	\Upsilon^-(\omega) + \Upsilon^+(\omega) = \Upsilon_\text{L}(\omega,T_\text{L}) + \Upsilon_\text{R}(\omega,T_\text{R}).
\end{equation}
Therefore, we find 
\begin{equation}
	\tilde{\mathcal{H}}  = \sigma_z \left( \mathcal{S}_\Delta\left[\Upsilon_\text{L}(\omega,T_\text{L}) \right] + \mathcal{S}_\Delta\left[\Upsilon_\text{R}(\omega,T_\text{R}) \right]\right),
\end{equation}
which proves Eq.~(\ref{eq:hls}).

\section{Co-tunneling calculation}
\label{app:cotunn}
In this appendix we derive Eq.~(\ref{eq:cot}), i.e. the contribution to the heat current of co-tunneling processes.
We will focus on the XX and XY coupling cases, defined in Section~\ref{sec:qubit_strong}. For simplicity, in this appendix we express the system bath Hamiltonian $\mathcal{H}_{\alpha,\text{S}}$ as
\begin{equation}
\begin{aligned}
	\mathcal{H}_{\text{L}, \text{S}}&=(\sigma^+ + \sigma^-) \otimes \sum_{k}V_{\alpha k}(b_{\alpha k}+b_{\alpha k}^\dagger), \\
	\mathcal{H}_{\text{R}, \text{S}}&=(q \sigma^++q^* \sigma^-) \otimes \sum_{k}V_{\alpha k}(b_{\alpha k}+b_{\alpha k}^\dagger),
\end{aligned}
\label{concot}
\end{equation}
where $q$ is a complex coefficient given by $q = 1$ in the XX case (since $\sigma_x = \sigma^+ + \sigma^-$) and by $q=i$ in the XY case (since $\sigma_y = i \sigma^+-i \sigma^-$). 

Co-tunneling is a second-order process where a state of the uncoupled system evolves into another state of the uncoupled system passing through a ``virtual state'' by interacting twice with $\mathcal{H}_{\alpha, \text{S}}$.  Since $\mathcal{H}_{\alpha, \text{S}}$ contains the operators $\sigma^+$ and $\sigma^-$, and since co-tunneling rates are obtained by acting twice with $\mathcal{H}_{\alpha, \text{S}}$, the state of the qubit remains unaltered during a co-tunneling process. This property, which is denoted as ``elastic co-tunneling'', implies that co-tunneling rates do not enter the master equation for the probabilities.

We start by considering all processes which transfer an excitation from the left to the right bath while the qubit is in the ground state. Let us denote with $\ket{0}$ and $\ket{1}$ the ground and excited state of the qubit, and with $\ket{n_\alpha}_{k}$ a Fock state with $n_\alpha$ excitations in mode $k$ of bath $\alpha$. The initial $\ket{i}$, final $\ket{f}$, and possible intermediate states $\ket{\nu_i}$ involved in the co-tunneling process are respectively given by
\begin{equation}
\begin{aligned}
	\ket{i} &= \ket{0}\otimes \ket{n_\text{L}}_{k} \otimes \ket{n_\text{R}}_{k^\prime}, \\
	\ket{f} &= \ket{0}\otimes \ket{n_\text{L}-1}_{k} \otimes \ket{n_\text{R}+1}_{k^\prime}, \\
	\ket{\nu_1} &= \ket{1}\otimes \ket{n_\text{L}-1}_{k} \otimes \ket{n_\text{R}}_{k^\prime}, \\
	\ket{\nu_2} &= \ket{1}\otimes \ket{n_\text{L}}_{k} \otimes \ket{n_\text{R}+1}_{k^\prime}, 
\end{aligned}
\label{statescot}
\end{equation}
for all choices of $k$ and $k^\prime$. Using the Fermi golden rule, the rate of transition from the initial state $\ket{i}$ to the final state $\ket{f}$ is given by
\begin{equation}
\Upsilon_{i\to f}=\frac{2\pi}{\hbar}\left|A_{if}\right|^2 \delta(\epsilon_i-\epsilon_f),
\label{ratecot}
\end{equation}
where $\epsilon_{i/f}$ is the energy of the initial/final state in the absence of the system-bath interaction, and
\begin{equation}
A_{if}=\sum_j \frac{\langle f|\sum_\alpha\mathcal{H}_{\alpha,\text{S}}|\nu_j\rangle\langle \nu_j|\sum_\alpha\mathcal{H}_{\alpha,\text{S}}|i\rangle}{\epsilon_i-\epsilon_{\nu_j}+i\eta},
\label{aa}
\end{equation}
$\eta$ being an infinitesimal positive quantity and $\epsilon_{\nu_j}$ the energy of $\ket{\nu_j}$. Using Eqs.~(\ref{concot}) and (\ref{statescot}), we have that the non-null matrix elements are
\begin{equation}
\begin{aligned}
	&\mel*{f}{\mathcal{H}_{\text{R}, \text{S}}}{\nu_1} = \mel*{\nu_2}{\mathcal{H}_{\text{R}, \text{S}}}{i} =q V_{\text{R}k_\text{R}}\sqrt{n_\text{R}+1}, \\
	&\mel*{\nu_1}{\mathcal{H}_{\text{L}, \text{S}}}{i} = \mel*{f}{\mathcal{H}_{\text{L}, \text{S}}}{\nu_2} = \,V_{\text{L}k_\text{L}}\sqrt{n_\text{L}}.
\end{aligned}
\label{eq:inter}
\end{equation}
The co-tunneling heat current $I^{\rm cot(0)}_{\text{L}\to \text{R}}$ that accounts for the transfer of an excitation from left to right, while the qubit is in state $\ket{0}$, is obtained by performing a weighed sum, according to the equilibrium probabilities, over all the initial and final states of the quantity $\Upsilon_{i\to f}$ multiplied by the transferred energy.
Combining Eqs.~(\ref{eq:inter}), (\ref{aa}) and (\ref{ratecot}), and using for simplicity $\epsilon_k = \epsilon_{\text{L}k}$ and $\epsilon_{k^\prime} = \epsilon_{\text{R}k^\prime}$, we have that 
\begin{multline}
 I^{\rm cot(0)}_{\text{L}\to \text{R}}=\frac{2\pi}{\hbar}\sum_{k k^\prime} \epsilon_k\, |V_{\rm Lk}|^2|V_{\rm Rk^\prime}|^2n_{\rm L}(\epsilon_{\rm k})[1+n_{\rm R}(\epsilon_{{\rm k^\prime}})]\\
\times  \left|\frac{q^*}{\Delta +\epsilon_{\rm k}+i\eta}+\frac{q}{\Delta-\epsilon_{\rm k}+i\eta}\right|^{2}\delta(\epsilon_{\rm k}-\epsilon_{\rm k^\prime}).
 \label{eqn:ltor}
\end{multline}
As usual, we assume that the energies in the leads form a continuum, so we can replace the sum with an integral. Performing some calculations, and recalling that $|q|^2=1$ both in the XX and XY case, we have that
\begin{multline}
 I^{\rm cot(0)}_{\text{L}\to \text{R}}=\int_0^{+\infty}\frac{d\epsilon}{2\pi\hbar}\,\epsilon\, \Gamma_\text{L}(\epsilon) \Gamma_\text{R}(\epsilon)  n_{\rm L}(\epsilon)[1+n_{\rm R}(\epsilon)]\\
 \times \left|\frac{1}{\Delta +\epsilon+i\eta}+\frac{q}{q^*}\frac{1}{\Delta-\epsilon+i\eta}\right|^{2}.
 \label{eq:cot_rect_s1}
\end{multline}
Note that the term $q/q^*$ is respectively $1$ and $-1$ in the XX and XY cases. 

The co-tunneling heat current $I^{\rm cot(0)}_{\text{R}\to \text{L}}$ transferring an excitation from right to left bath when the qubit is in 
$|0\rangle$ is given by Eq.~(\ref{eq:cot_rect_s1}) exchanging $\text{L} \leftrightarrow \text{R}$.  We thus find that the net heat
current while the qubit is in the ground state $I^{\rm cot(0)} \equiv I^{\rm cot(0)}_{\text{L}\to \text{R}} - I^{\rm cot(0)}_{\text{R}\to \text{L}}$ is given by
\begin{multline}
 I^{\rm cot(0)}=\int_0^{+\infty}\frac{d\epsilon}{2\pi\hbar}\,\epsilon\, \Gamma_\text{L}(\epsilon) \Gamma_\text{R}(\epsilon)  [n_{\rm L}(\epsilon) - n_{\rm R}(\epsilon)]\\
 \times \left|\frac{1}{\Delta +\epsilon+i\eta}+\frac{q}{q^*}\frac{1}{\Delta-\epsilon+i\eta}\right|^{2}.
 \label{eq:cot_rect_s2}
\end{multline}

Repeating the same derivation assuming that the qubit is in the excited state, it can be shown that the net heat
current  $I^{\rm cot(1)}$ is the same, i.e. $I^{\rm cot(0)}=I^{\rm cot(1)}$. Therefore, the heat current due to co-tunneling
processes is  $I^{\rm cot} \equiv I^{\rm cot(0)}=I^{\rm cot(1)}$ given by Eq.~(\ref{eq:cot_rect_s2}) which corresponds to Eq.~(\ref{eq:cot})
in the main text.

\section{Non equilibrium Green's
function calculation}
\label{app:negf}
In this appendix we will consider a qubit in contact with bosonic baths. In all our calculations, we fix the coupling on the left hand side to have only the $\hat{\sigma}_x$ component, i.e $u_{{\rm L},y}=u_{{\rm L},z}=0$ and $u_{{\rm L},x}=1$. The total Hamiltonian in terms of spin operators
\begin{equation}
H=\frac{\Delta}{2}\hat{\sigma}_z+\sum_{k,\alpha}\epsilon_{\alpha k}\hat{b}_{\alpha k}^\dagger \hat{b}_{\alpha k}+\sum_{j}u_{{\rm R},j}\hat{\sigma}_{j}\hat{B}_{{\rm R}}+\hat{\sigma}_{x}\hat{B}_{{\rm L}}.
\end{equation}
Spin operators do not satisfy the usual Wick's theorem. The usual Feynman diagram techniques applied to obtain Dyson equations cannot be used. In order to overcome this difficulty, one can undergo Majorana fermion transformation of spin operators using the following relations
\cite{liu,schad}:
\begin{equation}
\hat{\sigma}_{x}={-i}\hat{\eta}_{y}\eta_{z};~\hat{\sigma}_{y}={-i}\hat{\eta}_{z}\hat{\eta}_{x};~\hat{\sigma}_{z}={-i}\hat{\eta}_{x}\eta_{y}.
\label{spintomaj}
\end{equation}
The total Hamiltonian in terms of Majorana fermions reads
\begin{align}
H = &-\frac{i\Delta}{2}\hat{\eta}_{x}\hat{\eta}_{y}+\sum_{k,{\alpha}}\epsilon_{{\alpha}k}b_{{\alpha}k}^{\dagger}b_{{\alpha}k}-i\Big[u_{{\rm R},x}\hat{\eta}_y \hat{\eta}_z \hat{B}_{\rm R}\nonumber \\
&+u_{{\rm R},y}\hat{\eta}_z \hat{\eta}_x \hat{B}_{\rm R} + u_{{\rm R},z}\hat{\eta}_x \hat{\eta}_y \hat{B}_{\rm R}\Big]-i\hat{\eta}_y\hat{\eta}_z \hat{B}_{\rm L}.
\label{hamilmaj}
\end{align}
We write the Green's function for spin operators as:
\begin{eqnarray}
\hat{G}^<_{l,l^{\prime}}(t,t^{\prime})&=&-i \langle \hat{\sigma}_{l^{\prime}}(t^{\prime}) \hat{\sigma}_l(t) \rangle ,\nonumber \\
\hat{G}^r_{l,l^{\prime}}(t,t^{\prime})&=&-i \Theta(t-t^{\prime}) \langle \left[ \hat{\sigma}_l(t),\hat{\sigma}_{l^{\prime}}(t^{\prime}) \right] \rangle.
\end{eqnarray}
The relations between the Green's function in the Majorana representation and the Green's function in spin representation are given by~\cite{langreth,makhlin},
\begin{align}
&\hat{G}^{</>}_{l,l^{\prime}}(t,t^{\prime})=\mp\hat{\Pi}^{</>}_{l,l^{\prime}}(t,t^{\prime})\nonumber\\
&\hat{G}^{r}(t,t')=\theta(t-t')\left[\hat{\Pi}^{>}(t,t')+\hat{\Pi}^{<}(t,t')\right],
\label{allcon}
\end{align}
where $\hat{\Pi}^{<}_{l,l^{\prime}}(t,t^{\prime})= i \langle \hat{\eta}_{l^{\prime}}(t^{\prime})\hat{\eta}_{l}(t) \rangle$ and $\hat{\Pi}^{>}_{l,l^{\prime}}(t,t^{\prime})= -i \langle \hat{\eta}_{l}(t)\hat{\eta}_{l^{\prime}}(t^{\prime}) \rangle$ are the lesser and greater Green's functions for Majorana operators, respectively. The heat current flowing from the lead ${\rm L}$ to the system is given by
\begin{align}
I(t)&=\frac{i}{\hbar}\left\langle\left[H_{\rm L}(t),H(t)\right]\right\rangle \nonumber \\
&=-\frac{2}{\hbar}\sum_k \epsilon_{L k}V_{L  k}{\rm Re}\left[G_{x,L k}^<(t,t)\right],
\end{align}
where $G_{x,L k}^<(t,t')=-i\left\langle \hat{b}_{L k}^\dagger (t') \hat{\sigma}_x (t)\right\rangle$. Following standard Keldysh NEGF treatment using Langreth theorem, the steady state heat current  as defined in Eq.~(\ref{eq:j_gen_def}) can be written as:
\begin{equation}
I(\Delta T)=-\frac{2}{\hbar}\int d\epsilon \;\epsilon\; {\rm Re}\left[G_{xx}^r(\epsilon)\Sigma_{L }^<(\epsilon)+G_{xx}^<(\epsilon)\Sigma_{L }^a(\epsilon)\right],
\end{equation}
where $\Sigma_L(\epsilon)=\sum_{k}|V_{L k}|^2 g_{L k}(\epsilon)$ is the self energy of the bath L and $g_{L k}(\epsilon)$ is the Green's function for the uncoupled bath L. Applying the relations of Eq.~(\ref{allcon}), the heat current can be computed as
\begin{equation}
I(\Delta T)=-\int_0^\infty \frac{d\epsilon}{2\pi\hbar}\epsilon \left[\Pi_{xx}^>(\epsilon)\Sigma_{L}^<(\epsilon)+\Pi_{xx}^<(\epsilon)\Sigma_{L}^>(\epsilon)\right],
\label{eq:heatcurr_gr}
\end{equation}
where the self energies due to system bath coupling, $\Sigma_L^<(\epsilon)$ and $\Sigma_L^>(\epsilon)$ are defined in Sec~\ref{sec:heat_rect}. In order to evaluate the heat currents, one needs to calculate the lesser and greater components of the Majorana Green's function.
\subsection{Derivation of Green's function}
In this section we will derive the Green's functions in Majorana representation. Normal ordering for Majorana fermions is not defined. It is useful to write the Majorana operators in terms of Dirac operators~\cite{langreth}
\begin{equation}
\hat{\eta}_{x}=\hat{f} + \hat{f}^{\dagger};~~\hat{\eta}_{y}=i(\hat{f}^{\dagger}-\hat{f});~~\hat{\eta}_{z}=\hat{g}+\hat{g}^{\dagger}.
\end{equation}
The fermionic nature of $\hat{f}$ is consistent with,
\begin{equation}
\hat{f}=\frac{\hat{\eta}_{x}+i\hat{\eta}_{y}}{2};~~\hat{f}^{2}=0;~~\hat{f}^{\dagger^2}=0;~~\left\{\hat{f},\hat{f}^{\dagger}\right\}=1,
\end{equation}
and should hold for $g$ as well.
The Majorana representation does not suffer from vertex problem~\cite{langreth} and the constraints on spins are naturally imposed on Majorana operators~\cite{schad2}. The Hamiltonian for the qubit gets transformed to
\begin{equation}
{H}_{\infty} =\frac{\Delta}{2}(1-2\hat{f}^{\dagger}\hat{f}),
\end{equation}
whereas the contact Hamiltonians are
\begin{multline}
H_{\rm R,S}=\sum_k V_{Rk}\Big[u_{{\rm R},x}\,(f^\dagger-f)\hat{\eta}_z-iu_{{\rm R},y}\,\hat{\eta}_z(f+f^\dagger)\\
+u_{{\rm R},z}(1-2f^\dagger f)\Big]\hat{B}_R,
\label{hamildirac}
\end{multline}
and 
\begin{equation}
H_{\rm L,S}=\sum_k{V_{Lk}}(\hat{f}^\dagger-\hat{f})\hat{\eta}_{z}\hat{B}_{\rm L}.
\end{equation}
Note that we consider general spin coupling in the right lead whereas a fixed $\sigma_x$ coupling in the left. The contour ordered Green's function for the Majorana operators can be written as:
\begin{equation}
\hat{\Pi}_{xx}(\tau,\tau')=
\begin{bmatrix}
\hat{\Pi}_{xx}^{t}(t,t') & \hat{\Pi}_{xx}^<(t,t')\vspace{0.03\columnwidth} \\
\hat{\Pi}_{xx}^>(t,t') & \hat{\Pi}_{xx}^{\bar{t}}(t,t')
\end{bmatrix}
\end{equation}
We also define the Green's function for Dirac fermions $f$ in the Bogoliubov-Nambu representation, $\hat{\psi}\equiv (\hat{f}, \hat{f}^{\dagger})^{T}$ and $\hat{\psi}^{\dagger}\equiv ( \hat{f}^{\dagger},\hat{f})$, such that $\hat{G}_{\psi}(\tau,\tau')=-i\left\langle\mathcal{T}\hat{\psi}(\tau)\hat{\psi}(\tau')\right\rangle $. On expansion in the Keldysh contour,
\begin{multline}
\hat{G}_{\psi}(\tau,\tau')=\\
\begin{bmatrix}
{G}_{ff^{\dagger}}^{t}(t,t') & {G}_{ff}^{t}(t,t') & {G}_{ff^{\dagger}}^{<}(t,t') & {G}_{ff}^{<}(t,t')\vspace{0.03\columnwidth} \\
{G}_{f^{\dagger} f^{\dagger}}^{t}(t,t') & {G}_{f^{\dagger}f}^{t}(t,t') & {G}_{f^{\dagger} f^{\dagger}}^{<}(t,t') & {G}_{f^{\dagger}f}^{<}(t,t') \vspace{0.03\columnwidth}\\
{G}_{ff^{\dagger}}^{>}(t,t') & {G}_{ff}^{>}(t,t') & {G}_{ff^{\dagger}}^{\bar{t}}(t,t') & {G}_{ff}^{\bar{t}}(t,t')\vspace{0.03\columnwidth} \\
{G}_{f^{\dagger} f^{\dagger}}^{>}(t,t') & {G}_{f^{\dagger}f}^{>}(t,t') & {G}_{f^{\dagger} f^{\dagger}}^{\bar{t}}(t,t') & {G}_{f^{\dagger}f}^{\bar{t}}(t,t') \vspace{0.03\columnwidth} 
\end{bmatrix},
\label{realtime}
\end{multline}
where for instance, $G_{ff^{\dagger}}(\tau,\tau')=-i\left\langle\mathcal{T}\hat{f}(\tau)\hat{f}^{\dagger}(\tau ')\right\rangle$. For more clarification, see Eqs. (A2) and (A3) in Ref.~\onlinecite{agarwalla}. The lesser and greater Green's function in Majorana representation are
\begin{align}
&\Pi_{xx}^{<,>}(t,t')= \begin{bmatrix}
1 & 1
\end{bmatrix}\hat{G}_{\psi}^{<,>}(t,t')
\begin{bmatrix}
1 \\
1
\end{bmatrix}\nonumber \\
&\Pi_{yy}^{<,>}(t,t')= \begin{bmatrix}
1 & -1
\end{bmatrix}\hat{G}_{\psi}^{<,>}(t,t')
\begin{bmatrix}
1 \\
-1
\end{bmatrix}
\label{majdir}
\end{align}

\subsection{Calculation of Dyson equation}
In order to obtain a Dyson equation for $\hat{\psi}$, we need to do perturbation expansion in terms of the contact Hamiltonian for Dirac fermions $f$, namely
\begin{multline}
G_{ff^{\dagger}}(\tau,\tau')=G_{ff^{\dagger}}^{0}(\tau,\tau')+\frac{i}{2}\sum_\alpha \\
 \int d\tau_1  d\tau_2\left\langle \mathcal{T}\left[ \hat{H}_{\alpha, \rm S}(\tau_{1})\hat{H}_{\alpha, \rm S}(\tau_{2}) \hat{\tilde{f}}(\tau)\hat{\tilde{f}}^{\dagger}(\tau')\right]\right\rangle + \cdots
\label{diracpert}
\end{multline}
After a long but straightforward calculation, we obtain
\begin{multline}
\hat{G}_{\psi}(\tau,\tau')=\hat{G}_{\psi}^{0}(\tau,\tau')\\
+\int d\tau_1 d\tau_2\, \hat{G}_{\psi}(\tau,\tau_1)\hat{\Sigma}_{\psi}(\tau_1,\tau_2)\hat{G}_{\psi}^{0}(\tau_2,\tau'),
\label{dyson}
\end{multline}
where $\hat{\Sigma}_{\psi}=\hat{\Sigma}_{\psi,{\rm L}}+\hat{\Sigma}_{\psi,{\rm R}}$,
\begin{multline}
\hat{\Sigma}_{\psi,{\rm R}}(\tau_1,\tau_2)=iD_{R}(\tau_1,\tau_2)\Big(u_{{\rm L},x}^2\Pi_{z,z}^{0}(\tau_1,\tau_2)\hat{\lambda}\\
+u_{{\rm R},y}^2\Pi_{z,z}^{0}(\tau_1,\tau_2)\hat{{1}}+4u_{{\rm R},z}^2
\begin{bmatrix}
G_{ff^\dagger}^0(\tau_1,\tau_2) & 0 \\
0 & G_{f^\dagger f}^0(\tau_1,\tau_2) 
\end{bmatrix}\Big),
\label{eq:selfL}
\end{multline}
and 
\begin{equation}
\hat{\Sigma}_{\psi,{\rm L}}(\tau_1,\tau_2)=iD_{L}(\tau_1,\tau_2)\Pi_{z,z}^{0}(\tau_1,\tau_2)\hat{\lambda},
\label{eq:selfR}
\end{equation}
where $\hat{{1}}$ is the matrix of ones, the embedded self energy $
D_{\alpha}(\tau_1,\tau_2)=-i\sum_{k}|{V_{\alpha k}|^{2}}\left\langle\mathcal{T}\left[B_{k\alpha}(\tau_1)B_{k\alpha}(\tau_2)\right]\right\rangle$, and 
\[\hat{\lambda}=\begin{bmatrix}
1 & -1 \\
-1 & 1
\end{bmatrix}.\]
Writing the equation of motion for $\hat{G}_{\psi}^{0}$, we get
\begin{align}
\hat{G}_{\psi}^{0}(\tau,\tau')(-i\overleftarrow{\partial}_{\tau'}+\Delta\hat{\sigma}_{z})=\delta(\tau-\tau')\hat{\mathbb{1}},
\label{dyson_psi0}
\end{align}
where $\hat{\mathbb{1}}$ is a unit matrix. The retarded and advanced self energies due to coupling to the bath are given by
\begin{align}
D_{\alpha}^{r/a}(\epsilon)&=\sum_{k}\left|V_{\alpha k}\right|^2\left(\frac{1}{\epsilon-\epsilon_{\alpha k}\pm i\eta}-\frac{1}{\epsilon+\epsilon_{\alpha k}\pm i\eta}\right)\nonumber \\
&={\delta{\Delta}_\alpha(\epsilon)}\mp \frac{i}{2}\left(\Gamma_\alpha(\epsilon)-\Gamma_\alpha(-\epsilon)\right),
\label{eq:selfra}
\end{align}
where $\delta{\Delta}_\alpha(\epsilon)$ is the lamb shift defined as
\begin{equation}
\label{lamb}
\delta{\Delta}_\alpha(\epsilon)=\mathcal{P}\int_{-\infty}^\infty \frac{d\epsilon'}{2\pi}\left(\frac{\Gamma_\alpha(\epsilon')}{\epsilon-\epsilon'}-\frac{{\Gamma}_\alpha(\epsilon')}{\epsilon+\epsilon'}\right),
\end{equation}
The lesser and greater components of self energy take the form
\begin{eqnarray}
D_\alpha^<(\epsilon)&=&-in_\alpha(\epsilon)\left({\Gamma}_\alpha(\epsilon)-\Gamma_\alpha(-\epsilon)\right),\nonumber\\
D_\alpha^>(\epsilon)&=&-i(1+n_\alpha(\epsilon))\left({\Gamma}_\alpha(\epsilon)-\Gamma_\alpha(-\epsilon)\right).
\label{eq:selflg}
\end{eqnarray}
The integration for the Lamb shift can be simplified to
\begin{equation}
\delta{\Delta}_{\alpha}(\epsilon)=\frac{\Gamma_{\alpha}
}{2\pi}\left(\epsilon\;e^{-\epsilon/\epsilon_{\rm C}}\mathcal{E}\left[\frac{\epsilon}{\epsilon_{\rm C}}\right]-\epsilon \;e^{\epsilon/\epsilon_{\rm C}}\mathcal{E}\left[\frac{-\epsilon}{\epsilon_{\rm C}}\right]-2\epsilon_{\rm C}\right),
\end{equation}
where
\begin{equation} 
\mathcal{E}[\epsilon]=-\mathcal{P}\int_{-\epsilon}^\infty e^{-t}/t\; dt
\end{equation}
is a well known exponential integral function. Note that, we used $\Gamma_\alpha(\epsilon)=0$ for $\epsilon<0$.
Moreover, we have
\begin{multline}
\hat{\Sigma}_{\psi,{\rm R}}(\epsilon)=i\int \frac{d\epsilon'}{2\pi}D_{R}(\epsilon-\epsilon')\Big(u_{{\rm R},x}^2\Pi_{z,z}^{0}(\epsilon')\hat{\lambda}\\
+u_{{\rm R},y}^2\Pi_{z,z}^{0}(\epsilon')\hat{{1}}+4u_{{\rm R},z}^2
\begin{bmatrix}
G_{ff^\dagger}^0(\epsilon') & 0 \\
0 & G_{f^\dagger f}^0(\epsilon') 
\end{bmatrix}\Big),
\end{multline}
\begin{equation}
\hat{\Sigma}_{\psi,{\rm L}}(\epsilon)=i\int \frac{d\epsilon'}{2\pi}D_{L}(\epsilon-\epsilon')\Pi_{z,z}^0(\epsilon')\hat{\lambda}.
\end{equation}
Following Ref.~\onlinecite{jauho} for
\begin{equation}
\Sigma(\tau_1,\tau_2)=A(\tau_1,\tau_2)B(\tau_1,\tau_2)
\label{eq:jauho}
\end{equation}
the Langreth rules are given by
\begin{eqnarray}
&&\Sigma^{<}(\tau_1,\tau_2)=A^{<}(\tau_1,\tau_2)B^{<}(\tau_1,\tau_2),\nonumber \\
&&\Sigma^{r}(\tau_1,\tau_2)=A^{<}(\tau_1,\tau_2)B^{r}(\tau_1,\tau_2)+A^{r}(\tau_1,\tau_2)B^{<}(\tau_1,\tau_2)\nonumber\\
&&~~~~~~~~~~~~~~~+A^{r}(\tau_1,\tau_2)B^{r}(\tau_1,\tau_2).
\label{eq:lan_rel}
\end{eqnarray}
Since both Eqs.~(\ref{eq:selfL}) and (\ref{eq:selfR}) have the form of Eq.~(\ref{eq:jauho}), one can obtain the lesser $(\Sigma_\psi^<)$, greater $(\Sigma_\psi^>)$, retarded $(\Sigma_\psi^r)$ and advanced $(\Sigma_\psi^a)$ self energies in terms of different components of embedded self energy and the free Green's function for the system using Eq.~(\ref{eq:lan_rel}). For instance,
\begin{multline}
\hat{\Sigma}_{\psi,{\rm L}}^<(\epsilon)=i\int \frac{d\epsilon'}{2\pi}D_{L}^<(\epsilon-\epsilon')\Big(u_{{\rm L},x}^2\Pi_{z,z}^{0,<}(\epsilon')\hat{\lambda}\\
+u_{{\rm L},y}^2\Pi_{z,z}^{0,<}(\epsilon')\hat{{1}}+4u_{{\rm L},z}^2
\begin{bmatrix}
G_{ff^\dagger}^{0,<}(\epsilon') & 0 \\
0 & G_{f^\dagger f}^{0,<}(\epsilon') 
\end{bmatrix}\Big),
\end{multline}
\begin{equation}
\hat{\Sigma}_{\psi,{\rm R}}^<(\epsilon)=i\int \frac{d\epsilon'}{2\pi}D_{R}^<(\epsilon-\epsilon')\Pi_{z,z}^{0,<}(\epsilon')\hat{\lambda}.
\end{equation}
Eqs.~(\ref{eq:selfra}) and (\ref{eq:selflg}) give all the components of the embedded self energies of the baths. The only unknowns are the free Green's functions of the system which we will discuss below.
The free dynamics of the system Hamiltonian can be easily computed to obtain
\begin{align}
G_{ff^\dagger}^{0,r}(\epsilon)&=\mathcal{P}\left\{\frac{1}{\epsilon+\Delta}\right\}-i\pi \delta(\epsilon+\Delta),\nonumber \\
G_{f^\dagger f}^{0,r}(\epsilon)&=\mathcal{P}\left\{\frac{1}{\epsilon - \Delta}\right\}-i\pi \delta(\epsilon-\Delta).
\end{align}
We can use the relation, $G^r-G^a=G^>-G^<$ to write
\begin{eqnarray}
G_{f^\dagger f}^{0,>}(\epsilon)-G_{f^\dagger f}^{0,<}(\epsilon)&=&-2i\pi \delta(\epsilon-
\Delta),\nonumber \\
G_{ff^\dagger}^{0,>}(\epsilon)-G_{ff^\dagger}^{0,<}(\epsilon)&=&-2i\pi \delta(\epsilon+\Delta).
\end{eqnarray}
Using the fluctuation dissipation relation~\cite{jauho}, $G^{0,<}(\epsilon)=-f(\epsilon)\left(G^{0,>}(\epsilon)-G^{0,<}(\epsilon)\right)$ and $G^{0,>}(\epsilon)=\left(1-f(\epsilon)\right)\left(G^{0,>}(\epsilon)-G^{0,<}(\epsilon)\right)$ where $f(\epsilon)$ is the Fermi distribution of the system defined at average temperature of the two baths, we can write
\begin{align}
G_{f^\dagger f}^{0,</>}(\epsilon)&=\pm 2 i \pi f(\pm \epsilon)\delta (\epsilon-\Delta),\nonumber \\
G_{ff^\dagger }^{0,</>}(\epsilon)&=\pm 2 i \pi f(\pm \epsilon)\delta (\epsilon+\Delta).
\end{align}
The retarded and advanced Green's function for the system in the Majorana notation are
\begin{equation}
{\Pi}_{z,z}^{0,r/a}(\omega)=\frac{2}{\omega\pm i\eta},
\end{equation}
such that
\begin{equation}
{\Pi}_{z,z}^{0,r}(\omega)-{\Pi}^{0,a}(\omega)={\Pi}_{z,z}^{0,>}(\omega)-{\Pi}_{z,z}^{0,<}(\omega)=-4i\pi\delta(\omega).
\label{lessmgrt}
\end{equation}
If we take the effective temperature of the Majorana fermions to be given by $\beta_{\rm eff}(\beta_L,\beta_R)$, we have from the fluctuation-dissipation theorem for the ordinary fermionic system in equilibrium~\cite{jauho}:
\begin{equation}
{\Pi}_{z,z}^{{0},>}(\omega)+\Pi_{z,z}^{0,<}(\omega)=\left({\Pi}_{z,z}^{0,r}(\omega)-{\Pi}_{z,z}^{0,a}(\omega)\right)\tanh\left(\frac{\beta_{\rm eff}\omega}{2}\right)
\label{lessbgrt}
\end{equation}
Using Eqs.~(\ref{lessmgrt}) and (\ref{lessbgrt}), one can find the lesser and greater Green's function for the Majorana operators.

Similarly, the time ordered and anti-time ordered self energies are obtained from
\begin{eqnarray}
\hat{\Sigma}_\psi^t(\epsilon)+\hat{\Sigma}_\psi^{\bar{t}}(\epsilon)&=&\hat{\Sigma}_\psi^>(\epsilon)+\hat{\Sigma}_\psi^<(\epsilon),\nonumber \\
\hat{\Sigma}_\psi^t(\epsilon)-\hat{\Sigma}_\psi^{\bar{t}}(\epsilon)&=&\hat{\Sigma}_\psi^a(\epsilon)+\hat{\Sigma}_\psi^r(\epsilon).\nonumber \\
\end{eqnarray}
Substituting Eq.~(\ref{realtime}) in Eq.~(\ref{dyson}) and undergoing Fourier transform, we obtain:
\begin{equation}
\hat{G}_\psi^{-1}(\epsilon)=\hat{G}_{\psi}^{0^{-1}}(\epsilon)-\hat{\Sigma}_{\psi}(\epsilon),
\end{equation}
where the bare system Green's function,
\begin{equation}
\hat{G}_{\psi}^{0,t^{-1}}(\epsilon)=-\hat{G}_{\psi}^{0,\bar{t}^{-1}}(\epsilon)=\epsilon \hat{\mathbb{1}}+\Delta\hat{\sigma}_z,
\end{equation}
such that
\begin{widetext}
\begin{equation}
\hat{\sigma}_k\hat{G}_\psi(\epsilon)
=\begin{bmatrix}\vspace{0.02\columnwidth}
&\epsilon+\Delta-[\Sigma_\psi^t(\epsilon)]_{11} & -[\Sigma_\psi^t(\epsilon)]_{12} & -[\Sigma_\psi^<(\epsilon)]_{11} & -[\Sigma_\psi^<(\epsilon)]_{12} \\ \vspace{0.02\columnwidth}
&-[\Sigma_\psi^t(\epsilon)]_{21} & \epsilon-\Delta-[\Sigma_\psi^t(\epsilon)]_{22} & -[\Sigma_\psi^<(\epsilon)]_{21} & -[\Sigma_\psi^<(\epsilon)]_{22} \\ \vspace{0.02\columnwidth}
& [\Sigma_\psi^>(\epsilon)]_{11} & [\Sigma_\psi^>(\epsilon)]_{12} & \epsilon+\Delta+[\Sigma_\psi^{\bar{t}}(\epsilon)]_{11} & [\Sigma_\psi^{\bar{t}}(\epsilon)]_{12}\\
& [\Sigma_\psi^>(\epsilon)]_{21} & [\Sigma_\psi^>(\epsilon)]_{22} & [\Sigma_\psi^{\bar{t}}(\epsilon)]_{21} & \epsilon-\Delta+[\Sigma_\psi^{\bar{t}}(\epsilon)]_{22}
\end{bmatrix}^{-1},
\label{eq:Gpsi}
\end{equation}
\end{widetext}
where $\hat{\sigma}_k=\text{diag}(1,1,-1,-1)$ is introduced to keep the appropriate sign for two different branches of the Keldysh contour~\cite{agarwalla}. Using Eq.~(\ref{majdir}) along with Eq.~(\ref{eq:Gpsi}), one can obtain the lesser and greater Green's function in the Majorana representation. Substituting the Majorana Green's functions in Eq.~(\ref{eq:heatcurr_gr}), we obtain the final expression for current with general spin coupling in the right lead and a fixed spin coupling $\hat{\sigma}_x$ in the left lead. 

\subsection{Calculation of currents for simple models}
\subsubsection{The XX and XY case}
The current for the XX and the XY system-bath coupling can be calculated from Eq.~(\ref{eq:heatcurr_gr}) after calculating the Green's functions from Eq.~(\ref{eq:Gpsi}). Note that one has to properly choose $u_{{\rm R},j}$ to obtain the XX and XY case. 
Considering the zero dimensionality of the spin system we arrive at the following expression for the heat current in the XY case
\begin{equation}
I(\Delta T)=
\int \frac{d\epsilon}{2\pi\hbar}\epsilon\; \mathcal{T}_{XY}(\epsilon)[n_{\rm L}(\epsilon)-n_{{\rm R}}(\epsilon)],
\end{equation}
where 
\begin{equation}
\mathcal{T}_{XY}(\epsilon)=\frac{4\,\epsilon^2\,\Gamma_\text{L}(\epsilon)\Gamma_\text{R}(\epsilon)}{\left(\epsilon^2-\mathcal{X}(\epsilon)-\Delta^2\right)^2+\mathcal{Y}^{2}(\epsilon)},
\label{xytra}
\end{equation}
where $\mathcal{X}(\epsilon)$ and $\mathcal{Y}(\epsilon)$ are defined in Eqs.~(\ref{x1}) and (\ref{x2}), respectively. We will consider Ohmic spectral density for both baths with high frequency cut off given by $\epsilon_C$. In Eq.~(\ref{xytra}), when the Lamb shift term is neglected, the transmission probability has a Lorentzian form, whose width is determined by ${\Gamma}_\alpha(\epsilon)$. When the coupling is very weak, i.e. ${\Gamma}_\alpha(\epsilon)\ll \Delta \ll k_{\rm B}T_\alpha$, the Lorentzian effectively shrinks to vanishing width and peaked around $\Delta$. In this limit, one can write $\epsilon\approx 
\Delta$ giving:
\begin{equation}
\mathcal{T}_{XY}(\Delta)\approx\frac{4\Delta^2{\Gamma}_{\rm L}(\Delta){\Gamma}_{\rm R}(\Delta) }{
\xi^2(\Delta)}.
\label{seqtraxy}
\end{equation}
The above result corresponds to the one obtained using master equation in the sequential tunneling limit.
Following a similar calculation, the heat current for the XX case is given by
\begin{multline}
I(\Delta T)=
\int_{0}^{\infty}\frac{d\epsilon}{2\pi\hbar}\epsilon\; \mathcal{T}_{XX}(\epsilon)[n_{{\rm L}}(\epsilon)-n_{\rm R}(\epsilon)],
\end{multline}
where the transmission probability $\mathcal{T}_{XX}(\epsilon)$ is given by
\begin{equation}
	\mathcal{T}_\text{XX}(\epsilon)=
	\frac{4\,\Delta^2\Gamma_\text{L}(\epsilon)\Gamma_\text{R}(\epsilon)}{\left(\epsilon^2-2\epsilon\left(\delta {\Delta}_\text{L}(\epsilon)+
	\delta{ \Delta}_\text{R}(\epsilon)\right)-\Delta^2\right)
	^2+\xi^2(\epsilon)},
\label{rescur}
\end{equation}
where $\xi (\epsilon)=\epsilon\sum_\alpha{\Gamma}_\alpha(\epsilon)\Big(1+2n_\alpha(\epsilon)\Big)$. 
In Refs.~\onlinecite{agarwalla,boudjada2014} a similar form for the transmission function has been 
derived within the NEGF but without including the frequency renormalization expressed by the Lamb shift. For $\Gamma_\alpha(\epsilon) \ll \Delta \ll k_B T$, we obtain $\mathcal{T}_{XX}=\mathcal{T}_{XY}$ given by Eq.~(\ref{seqtraxy}).
In the low temperature and weak coupling regime ($\Gamma_\alpha (\epsilon)\ll k_{\rm B}T\leq \Delta$), the first order sequential processes are generally suppressed and the dominant contribution comes from second order co-tunneling processes. For the heat current in the XX case we obtain
\begin{equation}
I(\Delta T)\approx\int_{0}^\infty \frac{d\epsilon}{2\pi\hbar}\frac{4\epsilon\,{\Gamma}_{\rm L}(\epsilon){\Gamma}_{\rm R}(\epsilon)}{\Delta^2}[n_{\rm L}(\epsilon)-n_{\rm R}(\epsilon)].
\end{equation}
The above result corresponds to the co-tunneling contribution \cite{agarwalla} and matches with 
Eq.~(\ref{eq:cot}). 

\subsection{Exact calculation}
\label{app:exact}
In this section we derive the formal exact expressions for the dynamical susceptibility entering the transmission function
Eq.~(\ref{eq:transmission_exact})
\begin{equation}
\chi(t)= \frac{i}{\hbar} \Theta(t) \langle [ \sigma_x(t), \sigma_x(0)] \rangle 
\label{chi-def}
\end{equation}
within the path-integral approach to the spin-boson model \cite{weiss}.
To deal with a correlated initial state at time $t=0$, we assume that the system starts at a preparation
time $t_p <0$ in a factorized state (Feynman Vernon)
\begin{equation}
W_{tot} = \hat{\rho}_\text{L}(T_\text{L}) \otimes \hat{\rho}_\text{R}(T_\text{R}) \otimes \hat{\rho}(t_p),
\end{equation}  
where each bath is in the thermal equilibrium state described by the density matrix 
$\hat{\rho}_\alpha(T_\alpha)$ and $\hat{\rho}(t_p)$ is a general state of the qubit at the preparation time. Assuming that 
the system is ergodic, the response function will not depend on the chosen initial state when $t_p \to - \infty$. For 
the sake of simplicity, we assume that the qubit starts in a diagonal state (or sojourn) of the Pauli matrix which couples to
the bath coordinates, $\sigma_x$, $| \eta_p \rangle$, with 
$\eta_p=1$. 

It is easy to demonstrate that in the case of XX coupling the effect of the two baths on the qubit evolution is expressed by 
the influence functional
\begin{multline}
 \mathcal{F}[\sigma, \sigma^\prime; t_0] = \exp\bigg\{
 \int_{t_0}^t  dt^\prime 
 \int_{t_0}^{t^\prime} dt^{\prime \prime} 
 \sum_\alpha \Big[
 \dot{\xi}(t^\prime)\\
 {\rm Re}\left[Q_\alpha(t^\prime - t^{\prime \prime})\right] \dot{\xi}(t^{\prime \prime}) 
+ i 
\dot{\xi}(t^\prime) {\rm Im} \left[ Q_\alpha (t^\prime - t^{\prime \prime})\right] \dot{\eta}(t^{\prime \prime}) 
  \Big]
 \bigg\}
 \end{multline}
 where $Q_\alpha(t)= {\rm Re}[Q_\alpha (t)] + i {\rm Im}[Q_\alpha (t)]$ is the complex bath-$\alpha$ correlation function
  \begin{multline}
 Q_\alpha(t)= \int_0^\infty \frac{d \omega}{\pi \hbar} \frac{ 2 \Gamma_\alpha(\hbar \omega)}{\omega^2} 
 \big [ \coth \left( \frac{\hbar \omega}{2 k_\text{B} T_\alpha} \right) \\ 
 (1 - \cos(\omega t)) + i \sin(\omega t)  
 \big ] \, .
 \end{multline}
 The dynamical susceptibility, Eq. (\ref{chi-def}) is expressed as follows 
\begin{equation}
\chi(t) =  \frac{i}{\hbar} \Theta (t) \lim_{t_p \to - \infty} \, \sum_{\eta= \pm 1} \sum_{\xi_0= \pm 1} \eta \xi_0  
J(\eta, t; \xi_0, 0; \eta_p, t_p) \,,
\label{chi-formal}
\end{equation}
where $J(\eta, t; \xi_0, 0; \eta_p, t_p)$ is the conditional propagating function to find the qubit in the diagonal (sojourn) state  $\eta=\pm 1$ 
at time $t$, conditioned to having measured the system in off-diagonal (blip) state $\xi_0$ at
time $t=0$ and having prepared it in state $\eta_p$ at time $t_p$
We find
\begin{widetext}
\begin{eqnarray}
J(\eta, t; \xi_0, 0; \eta_p, t_p) &=& \eta \eta_p \sum_{m,n=1}^\infty \big (- \frac{\Delta^2}{4 \hbar^2}\big )^{m+n-1} \int_{t_p}^t \mathcal{D}_{2m-1,2n-1} \{t_j\} \sum_{\{\xi_j= \pm 1\}^\prime} \, G_{n+m-1}^L G_{n+m-1}^R  \, \sum_{\{\eta=\pm 1\}^\prime}  H_{n+m-1}^L H_{n+m-1}^R \nonumber \\
\label{Jcond}
\end{eqnarray}
\end{widetext}
where the integration paths consist of $2n-1$ transitions for $t_p < t^\prime <0$ 
and $2m-1$ transitions for $0 < t^\prime < t$ and we introduced the compact notation
 $\int_{t_p}^t \mathcal{D}_{k,l} \{t_j\} \times \cdots \equiv \int_{0}^{t} d t_{l+k} .... \int_0^{t_{l+2}}dt_{l+1} \int_{t_p}^0 dt_l  .. \int_{t_p}^{t_2}dt_1 \times \cdots$. The symbol $\{ \}^\prime$ reminds that the sum is over all sequences of blips and sojourns in accordance 
with the constraints indicated in the argument. The blip-sojourn interactions enter the $H_{i}$s, whereas
the $G_j$s include the blip-blip interactions and are given by
\begin{eqnarray}
H_{n+m-1}^{\alpha} & =& \exp{i \sum_{k=0}^{m+n-2} \sum_{j=0}^{m+n-1} \xi_j X_{j,k}^{\alpha} \eta_k } 
\label{blip-soj}
\end{eqnarray}
\begin{multline}
G_{n+m-1}^{\alpha}  =\exp{- \sum_{j=1}^{n+m}  {\rm Re }\left[Q^{\alpha}_{2j,2j-1}\right]}\\ 
\exp{- i \sum_{j=2}^{m+n} \sum_{k=1}^{j-1} \xi_j  \xi_k \Lambda_{j,k}^{\alpha} }
\label{blipblip}
\end{multline}
\begin{eqnarray}
 X_{j,k}^{\alpha} &=& {\rm Im}\left[ Q^{\alpha}_{2j,2k +1} +  Q^{\alpha}_{2j-1,2 k} -  Q^{\alpha}_{2j,2k } - Q^{\alpha}_{2j-1,2 k+1}\right],\nonumber \\
\Lambda_{j,k}^{\alpha} &=&  {\rm Re}\left[Q^{\alpha}_{2j,2 k-1} +  Q^{\alpha}_{2j-1,2 k} -  Q^{\alpha}_{2j,2 k} - Q^{\alpha}_{2j-1,2 k-1} \right].\nonumber \\
\end{eqnarray}
Inserting the conditional propagating function Eq.~(\ref{Jcond}) in the susceptibility Eq.~(\ref{chi-formal}) it is possible
to perform the sum over the sojourns leading to
\begin{widetext}
\begin{multline}
\chi(t) =  \frac{2}{\hbar} \lim_{t_p \to -\infty} \sum_{m=1}^\infty \sum_{n=1}^\infty 
\big (- \frac{\Delta^2}{2 \hbar^2}\big )^{m+n-1} \int_{t_p}^t \mathcal{D}_{2m-1,2n-1} \{t_j\} 
 \sum_{ \{\xi_j= \pm 1 \}} \xi_n  G_{n+m-1}^L G_{n+m-1}^R   \\ 
 \sin(\phi_{0,n+m-1}^L + \phi_{0,n+m-1}^R)  
 \Pi _{k=1}^{m+n-2} \cos(\phi_{k,n+m-1}^L + \phi_{k,n+m-1}^R)
 \label{chi-exact} 
\end{multline}
\end{widetext}
where
\begin{equation}
\phi_{k,m}^\alpha = \sum_{j=k+1}^m \xi_j X_{j,k}^\alpha \, .
\label{eq:phi}
\end{equation}
Eq. (\ref{chi-exact}) is the formal exact expression for the susceptibility for a qubit simultaneously coupled to two
harmonic baths at different temperatures for general spectral densities and temperatures.

\subsubsection{Ohmic baths and the case $K_\text{L} + K_\text{R}  =1/2$}
We now specialize to the case of two baths with Ohmic damping defined in Eq.~(\ref{eq:negf_spectral}) where
we assume identical dependence on the energies included in  $J(\epsilon)$.
The bath correlation functions take the form
\begin{eqnarray}
 Q_\alpha(t) &=& 2 K_\alpha \ln{  \Big \{ \Big (\frac{\epsilon_{\rm C}}{\pi k_\text{B} T_\alpha} \Big ) 
 \sinh \Big (\frac{\pi k_\text{B} T_\alpha |t|}{\hbar \beta_\alpha} \Big) \Big \}} \nonumber \\
 &+& i \pi K_\alpha {\rm sgn}(t) \, .
 \label{Q-ohmic}
\end{eqnarray}
The blip-sojourn interactions and the phases $\phi_{k,m}^\nu$, Eq.~(\ref{eq:phi}) simplify, taking the form
\begin{eqnarray}
&& X_{j,k}^\alpha = \pi K_\alpha  \;, \; {\rm for} \, j=k+1 \qquad   X_{j,k}^\alpha= 0 \;, \; {\rm for} \,  j \neq k+1 \nonumber \\
&& \phi_{k,n+m}^{\alpha} = \xi_{k+1} \pi K_\alpha \, .
\end{eqnarray}
The susceptibility Eq. (\ref{chi-exact}) becomes
\begin{widetext}
\begin{multline}
\chi(t)= \frac{2}{\hbar} \lim_{t_p \to -\infty} \sum_{m=1}^\infty \sum_{n=1}^\infty 
\big (- \frac{\Delta^2}{2 \hbar^2}\big )^{m+n-1} \int_{t_p}^t \mathcal{D}_{2m-1,2n-1} \{t_j\} 
 \sum_{ \{\xi_j= \pm 1 \}} \xi_1 \xi_n  G_{n+m-1}^L G_{n+m-1}^R \\
\sin(\pi (K_{\rm L}+K_{\rm R})) \cos(\pi(K_{\rm L}+K_{\rm R}))^{n+m-2}.
 \label{chi-ohmic} 
\end{multline}
\end{widetext}
We observe that dependence on the damping strengths $K_\alpha$ coming from the blip-sojourn interactions $X_{i,j}$, 
is in the simple form $K_{\rm L}+K_{\rm R}$.  
Thus the two Ohmic baths coupled to the qubit with strengths such that  $K_{\rm L} + K_{\rm R} = 1/2$
can be treated analogously to the standard spin-boson model at the Toulouse point. We remark that in Eq.~(\ref{chi-ohmic}) the coupling
strengths enter non linearly the blip-blip interactions, $ G_{n+m-1}^L G_{n+m-1}^R$, which include the temperatures
of the two baths. Therefore, the two baths at $K_{\rm L}+ K_{\rm R} = 1/2$  are not simply equivalent to a single bath at $K=1/2$ 
with an effective temperature.  

We proceed with the evaluation of Eq.~(\ref{chi-ohmic}) for $K_{\rm L} + K_{\rm R} = 1/2$.
We observe that all the terms in the sum, except for the first one $m=n=1$, have $n+m-2$ zeros from
$ \cos(\pi(K_{\rm L}+K_{\rm R}))^{n+m-2} $. They give a non-vanishing contribution if a proper divergency comes from the 
interaction terms between the system's transitions included in the $ G_{n+m-1}^L G_{n+m-1}^R $. This is the
typical case of a bath at $K=1/2$. 
In the case of two baths with $K_{\rm L}+K_{\rm R}= 1/2$  we have
\begin{multline}
\lim_{K_{\rm L}+K_{\rm R} \to 1/2} \Delta^2 \cos(\pi (K_{\rm L}+K_{\rm R})) \nonumber  \\ 
\times \int_0^\infty d \tau e^{- \lambda \tau} e^{-\sum_\alpha {\rm Re}\left[Q_\alpha (\tau) \right]}
 = \frac{\pi}{2} \frac{\Delta^2}{\epsilon_{\rm C}} \equiv \hbar \gamma \,.
\end{multline}
Such an integral describes a collapsed dipole which does not interact with any other dipole, having effectively
a zero dipole moment. This mechanism allows to sum the different terms of the sum in Eq.~(\ref{chi-ohmic}), leading
to
\begin{multline}
\chi(t) =  \frac{4}{\hbar^3} \frac{\Delta^2}{2 \gamma} \Theta(t) \, e^{- \gamma t/2}
\int_0^\infty d \tau \\
e^{- \sum_\alpha {\rm Re}[Q_\alpha(\tau)]} [ e^{- \gamma |t- \tau |/2} - e^{- \gamma (t+\tau)/2} ] \,.
\label{chi-1su4-final2} 
\end{multline}
This solution extends to the non-equilibrium case the dynamical susceptibility at the Toulouse limit of the spin-boson model.

Performing its Fourier transform and inserting it in the transmission function Eq.(\ref{eq:transmission_exact}) the  
heat current between two harmonic baths under the strong coupling condition $K_{\rm L} + K_{\rm R} = 1/2$ is obtained.
Equivalently, the heat current Eq.~(\ref{eq:j_green_prop}) with Eq.~(\ref{eq:transmission_exact}) and the Fourier transform of 
Eq.~(\ref{chi-1su4-final2}) can be written as
\begin{equation}
I = \frac{1}{\hbar}\frac{K_\text{L} K_\text{R}}{K_\text{L}+ K_\text{R}} \Im \int_{-\infty}^{+\infty}dt\, \chi(t)F(-t),
\end{equation}
where
\begin{multline}
	F(-t)=  (k_{\rm B} T_\text{R})^3 \psi^{(2)}\left(1 + \frac{k_{\rm B} T_\text{R}}{\epsilon_\text{C}} \left(1-i \frac{ \epsilon_\text{C}\,t}{\hbar}\right)\right) \\
	 -(k_{\rm B} T_\text{L})^3 \psi^{(2)}\left(1 + \frac{k_{\rm B} T_\text{L}}{\epsilon_\text{C}} \left(1-i \frac{ \epsilon_\text{C}\,t}{\hbar}\right)\right)
\end{multline}
and $\psi^{(2)}(z)$ denotes the second derivative of the digamma function.

\section{Green's functions for the non-linear resonator}
\label{app:nonlin}
We first define a generic retarded Green's function as
\begin{equation}
G_{A,B,C,...;b}^r(t,t')= -i \theta (t-t') \langle [A(t)B(t)C(t) ...,b^\dagger(t')]\rangle ,
\end{equation}
where $A$, $B$ and $C$ are operators such as $b$, $n$ or $b_{\alpha k}$, $n(t)=b^\dagger (t) b(t)$ being the number operator.
In the subscript ($A,B,C,...;b$), the operators appearing before the semicolon are taken at time $t$, while the ones appearing after the semicolon are taken at time $t'$.

The Hamiltonian for the non-linear resonator is given by Eq.~(\ref{eq:sys_res}).
The equation of motion for the retarded Green's function of the system can be written as
\begin{multline}
i\partial_t G_{b;b}^r(t,t')=\delta(t-t')+\Delta G_{b;b}^r(t,t')\\
+UG_{n,b;b}^r(t,t')+\sum_{k,\alpha}V_{\alpha k}G_{\alpha k;b}^r(t,t')
\label{deom1}
\end{multline}
where
\begin{equation}
 G_{n,b;b}^r(t,t')= -i \theta (t-t') \langle [n(t)b(t),b^\dagger (t')]\rangle ,
\end{equation}
and
\begin{equation}
G_{\alpha k;b}^r(t,t')=-i\theta (t-t')\left\langle \left[b_{\alpha k}(t),b^\dagger(t')\right]\right\rangle .
\end{equation}
Using the equation of motion for $G_{\alpha k;b}^r(t,t')$ we obtain
\begin{equation}
\left(\epsilon-\Delta-\Sigma^{(0)}(\epsilon)\right)G_{b;b}^r(\epsilon)=1+U G_{n,b;b}^r(\epsilon),
\end{equation}
where $\Sigma^{(0)}(\epsilon)$ is the usual self energy due to system-bath coupling defined as
\begin{equation}
\Sigma^{(0)}(\epsilon)=\sum_\alpha \Sigma_{\alpha}^{(0)}(\epsilon)=\sum_{k,\alpha}|V_{\alpha k}|^2 g^r_{\alpha k;\alpha k}(\epsilon),
\end{equation}
where $g^r_{\alpha k;\alpha k}$ is the retarded Green's function for the free bath, i.~e.
\begin{equation}
g^r_{\alpha k;\alpha k}(t,t')=-i\theta (t-t')\left\langle \left[b_{\alpha k}(t),b_{\alpha k}^\dagger(t')\right]\right\rangle .
\end{equation}
In order to evaluate Eq.~(\ref{deom1}), we need to evaluate $G^r_{n,b;b}(\epsilon)$ in terms of $G^r_{b;b}(\epsilon)$. 
Using the equation of motion we find
\begin{multline}
\left(\epsilon-\Delta\right)G^r_{n,b;b}(\epsilon)=2\left\langle n\right\rangle + U G^r_{n,n,b;b}(\epsilon)\\
+\sum_{\alpha k}V_{ \alpha k}\Big[2 G^r_{n,\alpha k;b}(\epsilon)- G^r_{b,b,\alpha k^\dagger;b}(\epsilon)\Big] .
\label{deom2}
\end{multline}
We decouple Eq.~(\ref{deom2}) to second order by approximating $G^r_{n,n,b;b}(\epsilon)=\left\langle n\right\rangle G^r_{n,b;b}(\epsilon)$ and we obtain
\begin{multline}
\left(\epsilon-\Delta-U\left\langle n\right\rangle\right)G^r_{n,b;b}(\epsilon)=2\left\langle n\right\rangle\\
+\sum_{\alpha k}V_{\alpha k}\Big[2 G^r_{n,\alpha k;b}(\epsilon)- G^r_{b,b,\alpha k^\dagger;b}(\epsilon)\Big]
\label{deom3}
\end{multline}
where
\begin{equation}
 G_{n,n,b;b}^r(t,t')= -i \theta (t-t') \langle [n(t)n(t)b(t),b^\dagger (t')]\rangle .
\end{equation}
We can again use equation of motion to evaluate $G^r_{n,\alpha k;b}(\epsilon)$ and obtain
\begin{multline}
(\epsilon-\epsilon_{\alpha k})G^r_{n,\alpha k;b}(\epsilon)=V_{\alpha k}\Big(G^r_{n,b;b}-n_{\alpha}(\epsilon_{\alpha k})G^r_{b;b}(\epsilon)\\
+\left\langle b^\dagger b_{\alpha k}\right \rangle G^r_{\alpha k;b}(\epsilon)\Big)+\left\langle b^\dagger b_{\alpha k}\right \rangle,
\label{deom4}
\end{multline}
and 
\begin{multline}
\left(\epsilon+\epsilon_{\alpha k}-2\Delta-2U\left\langle n\right\rangle -U\right)G^r_{b,b,\alpha k^\dagger;b}(\epsilon)=2\left\langle  b b_{\alpha k}^\dagger\right\rangle \\
-V_{\alpha k}G^r_{n,b;b}(\epsilon)+2V_{\alpha k}n_{\alpha}(\epsilon_{\alpha k}) G^r_{b;b}(\epsilon) .
\label{deom42}
\end{multline}
We do not take into account the terms involving correlation between the leads and the system, such that $\left\langle b^\dagger b_{\alpha k}\right \rangle=\left\langle b b_{\alpha k}^\dagger\right \rangle=0$~\cite{meir1991}.
Substituting Eq.~(\ref{deom4}) and Eq.~(\ref{deom42}) into Eq.~(\ref{deom3}) we obtain
\begin{equation}
G^r_{n,b;b}=\frac{2A(\epsilon)\left\langle n\right\rangle}{U}-\frac{2\left(\Sigma^{(2)}(\epsilon)+\Sigma^{(3)}(\epsilon)\right)A(\epsilon)}{U}G_{b;b}^r
\label{deom5}
\end{equation}
where
\begin{equation}
A(\epsilon)/U=\left(\epsilon-\Delta-U\left\langle n\right\rangle-\left(2\Sigma^{(0)}(\epsilon)+\Sigma^{(1)}(\epsilon)\right)\right)^{-1},
\end{equation}
$\Sigma^{(0)}(\epsilon)=\sum_{\alpha k} |V_{\alpha k}|^2\left(\epsilon-\epsilon_{\alpha k}+i\eta\right)^{-1}$, $\Sigma^{(1)}(\epsilon)=\sum_{\alpha k} |V_{\alpha k}|^2\left(\epsilon+\epsilon_{\alpha k}-2\Delta-2U\left\langle n\right\rangle-U+i\eta\right)^{-1}$, $\Sigma^{(2)}(\epsilon)=\sum_{\alpha k} |V_{\alpha k}|^2n_{\alpha}(\epsilon_{\alpha k})\left(\epsilon-\epsilon_{\alpha k}+i\eta\right)^{-1}$ and $\Sigma^{(3)}(\epsilon)=\sum_{\alpha k} |V_{\alpha k}|^2n_{\alpha}(\epsilon_{\alpha k})\big(\epsilon+\epsilon_{\alpha k}-2\Delta-2U\left\langle n\right\rangle -U\\
+i\eta\big)^{-1}$.
Substituting Eq.~(\ref{deom5}) in Eq.~(\ref{deom1}), we find the final expression for $G^r_{b;b}(\epsilon)$, i.e.
\begin{equation}
G^r_{b;b}(\epsilon)=\frac{1+2A(\epsilon)\left\langle n\right\rangle}{\epsilon-\Delta-\Sigma^{(0)}(\epsilon)+2A(\epsilon)\left(\Sigma^{(2)}(\epsilon)+\Sigma^{(3)}(\epsilon)\right)}.
\label{dGr_bos}
\end{equation}
 The self energies are given by
\begin{eqnarray}
\Sigma^{(1)}(\epsilon)&=&\sum_\alpha \int \frac{d\omega}{2\pi} \left[\frac{\Gamma_{\alpha}(\omega)}{\epsilon+\omega-2\Delta-2U\left\langle n\right\rangle-U+i\eta}\right],\nonumber \\
\Sigma^{(2)}(\epsilon)&=&\sum_\alpha \int \frac{d\omega}{2\pi} \left[\frac{n_\alpha(\omega)\Gamma_{\alpha}(\omega)}{\epsilon-\omega+i\eta}\right],\nonumber \\
\Sigma^{(3)}(\epsilon)&=&\sum_\alpha \int \frac{d\omega}{2\pi} \left[\frac{\Gamma_{\alpha}(\omega)n_\alpha(\omega)}{\epsilon+\omega-2\Delta-2U\left\langle n\right\rangle-U+i\eta}\right].\nonumber
\end{eqnarray}
For any function $g$ we can write
\begin{equation}
\int d\omega \frac{g(\omega)}{x-\omega+i\eta}=\mathcal{P}\int d\omega \left\{\frac{g(\omega)}{x-\omega}\right\}-\frac{i}{2}g(x),
\end{equation}
where the first term is the Cauchy-Hadamard principal value distribution.

\end{appendix}


\begin{thebibliography}{100}
\bibitem{footnote1}  In order to avoid clutter in the notation, in this section we use the shorthand symbol $\Upsilon^\pm_\alpha(T)$ instead of $\Upsilon^\pm_\alpha(\Delta,T)$.
\bibitem{footnote:majorana} In order to account for the counter-rotating terms in the coupling Hamiltonian, in the calculation of the Lamb-shift the spectral density $\Gamma_\alpha(\epsilon^\prime)$ must be extended to negative values according to $\Gamma_\alpha(-\epsilon^\prime) = -\Gamma_\alpha(\epsilon^\prime)$.
\end{thebibliography}
\end{document}